\RequirePackage{lineno} 

\documentclass[twocolumn,showpacs,preprintnumbers,amsmath,amssymb,nofootinbib]{revtex4}

\usepackage{graphicx}
\usepackage{dcolumn}
\usepackage{bm}

\usepackage{epstopdf}
\def\bea{\begin{eqnarray}}
\def\eea{\end{eqnarray}}

\def\pp{\mbox{$p$-$p$}}

\def\pa{\mbox{$p$-A}}

\def\auau{\mbox{Au-Au}}

\def\pbpb{\mbox{Pb-Pb}}
\def\ppb{\mbox{$p$-Pb}}
\def\pn{\mbox{$p$-N}}
\def\aa{\mbox{A-A}}
\def\nn{\mbox{N-N}}

\def\ee{\mbox{$e^+$-$e^-$}}
\def\ppbar{\mbox{$p$-$\bar p$}}

\def\pt{$p_t$}
\def\mt{$m_t$}
\def\yt{$y_t$}

\def\nch{$n_{ch}$}
\def\mmpt{$\bar p_t$}

\begin{document} 

\setlength{\pdfpagewidth}{8.5in}
\setlength{\pdfpageheight}{11in}

\setpagewiselinenumbers
\modulolinenumbers[5]

\preprint{version 1.8}

\title{Comparing the PYTHIA Monte Carlo to a two-component (soft + hard) model of hadron production in high-energy p-p collisions
}

\author{Thomas A.\ Trainor}\affiliation{CENPA 354290, University of Washington, Seattle, WA 98195}


\date{\today}

\begin{abstract}
The PYTHIA Monte Carlo (PMC), first introduced more than thirty years ago, remains a popular simulation tool both for analysis of $p$-$p$ collision dynamics and for detector design and calibration. The PMC assumes that almost all produced hadrons result from parton-parton scatterings (interactions) described by pQCD (a hard component), and that multiple parton interactions per collision event (MPIs) are a common occurrence. In contrast, a two-component (soft + hard) model (TCM) of high-energy collisions, inferred inductively from a variety of data formats, attributes a majority of final-state hadrons to a soft component (projectile-nucleon dissociation) and a minority to a hard component representing minimum-bias dijet production (corresponding to measured jet spectra and fragmentation functions). The hard-component hadron yield is precisely proportional to the square of the soft-component yield over an interval corresponding to 100-fold increase in dijet production. The two data descriptions appear to be in conflict. This study presents a detailed comparison of the two models and their relations to a broad array of collision data. The PMC appears to disagree with some data, whereas the TCM provides an accurate and comprehensive data description.
\end{abstract}


\maketitle

 \section{Introduction} \label{intro}
   
This study compares the PYTHIA Monte Carlo~\cite{pythia} to a two-component (soft + hard) model (TCM) of \pp\ hadron production near midrapidity~\cite{ppprd,ppquad}. The  two systems are confronted with a variety of data from several collision systems. The general context for \pp\ Monte Carlos is drawn from summaries reported in Refs.~\cite{pdgmcs,buckley}.
 
 
 General-purpose Monte Carlos (GPMCs) are intended  to explore physical models of high-energy \pp\ collisions and (with detector-model Monte Carlos) to design and calibrate detector systems. GPMCs  combine perturbative QCD (pQCD) at short distances and QCD-inspired phenomenological models at longer distances~\cite{pdgmcs}. 
 The main components of GPMCs are (a) a pQCD description of hard parton scattering and parton splitting cascades (showering), (b) a hadronization transition between partonic and hadronic final states and (c) a description of ``soft hadron physics'' said to include an underlying event, minimum-bias (MB) interactions and color reconnection.
   
QCD-inspired phenomenological models for item (b) include string models (e.g.\ as in PYTHIA) and cluster models (e.g.\ as in HERWIG). The string model or Lund model~\cite{lund} is based on strings (color flux tubes) joining color-connected energetic partons, which strings then fragment into hadrons upon elongation. The cluster model is said to ``...force `by hand' all gluons to split into quark-antiquark pairs at the end of the parton shower.'' Those ``excited mesons'' decay ``isotropically to two hadrons...''~\cite{pdgmcs}. The cluster scenario is consistent with the current TCM soft component and longitudinal projectile-nucleon dissociation (Sec.~\ref{tcmview}). GPMC hadronization models can be contrasted with measured fragmentation functions (FFs) described by pQCD energy evolution and applied to final-state hadrons~\cite{eeprd}.

Some models for item (c) -- soft QCD and underlying event physics -- are based on multiple parton interactions (MPIs), color reconnection (CR) and \pp\ impact parameter dependence.  The underlying event (UE) is defined as the complement to a hard scatter (dijet):  ``In  events containing a hard parton-parton interaction, the underlying event represents the additional activity which is not directly associated with that interaction''~\cite{buckley}. The UE as such is defined only in the context of a hard jet trigger. 

The dominant contribution to the UE is said to come from additional {\em color exchanges} (MPIs). The UE thus remains perturbative, the trigger hard scatter being an element of the high-\pt\ tail of the MPI distribution. The associated ``jet pedestal'' effect ``is interpreted as follows. When two hadrons collide at non-zero impact parameter, high-$p_\perp$ interactions can only take place inside the overlapping region. Imposing a hard selection cut therefore statistically biases the event sample toward more central collisions, which will also have more underlying activity. ... The shape of the pedestal...is therefore related to...modeling of the impact parameter dependence''~\cite{buckley}. The primary motivation for the CR mechanism is ensemble-mean $\bar p_t$ data: ``Without colour reconnections, the predicted $\langle p_\perp \rangle(N_{ch})$ [i.e.\ $\bar p_t(n_{ch})$] distributions appear to rise too slowly with $N_{ch}$''~\cite{buckley}.

In an alternative scenario inelastic scattering is modeled in terms of  cut {\em Pomerons} relating ``...diffractive and non-diffractive scattering that is absent in the MPI-based models.... [In the MPI formulation Pomeron exchange is restricted to a small number of diffractive events.] ...the picture now is one of both hard and soft pomerons, ideally with a smooth transition between the two''~\cite{buckley}. Pomerons have the quantum numbers of the vacuum and thus do not color connect collision partners (e.g.\ projectile protons or hard-scattered partons leading to jets).
    
 
The present study focuses on the PYTHIA Monte Carlo (PMC) wherein MPIs are assumed to be the dominant or exclusive mechanism for hadron production. The PMC contains no feature comparable to the TCM soft component. ``MPI modeling has traditionally been a hallmark of PYTHIA''~\cite{buckley}. Non-diffractive inelastic scattering is modeled in the PMC by extending the pQCD parton scattering cross section down to \pt\ = 0, with collision-energy-dependent soft cutoff parameter $p_{\perp0} \propto E_{CM}^\epsilon$. Cutoff parameter $p_{\perp0}$ ``is thus one of the main `tuning' parameters in such models''~\cite{buckley}. \pp\ centrality, described by the Glauber model with eikonal approximation, is a major feature of the model. Color reconnection is assumed so as to minimize the total string length (i.e.\ fragment number) resulting from multiple hard parton scatters (MPIs). The PMC is tuned to accommodate a specific subset of currently available data volumes and analysis methods, referred to as ``key experimental data''~\cite{sjostrand}. 


This study contrasts a {\em deductive} approach based on {\em a priori} assumptions (PMC) leading to predictions for certain preferred observables with an {\em inductive} approach based on a self-consistent phenomenological description (TCM) of all information derivable from all available data, expressed via the simplest algebraic formalism.

This article is arranged as follows:
Section~\ref{pythia}  introduces the PYTHIA Monte Carlo and MPIs in the context of the TCM.
Section~\ref{canonical}  describes the relation of the PMC to certain data features conventionally employed for design and testing of that Monte Carlo.
Section~\ref{comparison}  presents a PMC-TCM model comparison strategy.
Section~\ref{tcmstory}  reviews the TCM with illustrations from \pp\ spectrum data.
Section~\ref{mpimbjets}  summarizes the relation of minimum-bias (MB) jets to MPIs.
Section~\ref{uembjets}  considers the underlying event (UE) and its relation to MB dijets.
Sections~\ref{disc} and~\ref{summ} present discussion and summary.

\section{PYTHIA Monte Carlo and $\bf MPIs$} \label{pythia}

The PMC has been applied broadly to simulations of high-energy \pp\ and \ppbar\ collisions. It was first formulated in the mid eighties in response to high-energy data obtained from newly-constructed collider accelerators. The PMC was configured to describe and explain certain data features emerging from initial analysis of collider data and includes several novel assumptions about high-energy nuclear collisions. In this section PMC structure and development history are briefly reviewed in the context of the TCM.

\subsection{PYTHIA and related MC models}  \label{models}

Historical development of the PMC is summarized in Refs.~\cite{sjostrand2,sjostrand} which, taken together, provide some indication of its continuing development in recent years. The initially introduced PMC served as a response to information newly obtained from the intersecting storage rings (ISR) and the super proton-antiproton synchrotron (Sp\=pS) by the mid eighties, especially strong jet contributions to various collision manifestations. At that time three issues dominated: hard parton scattering to form eventwise-reconstructed dijets, soft (multi)Pomeron descriptions of the underlying event (UE, complement to a triggered hard dijet) and UA1 minijets~\cite{ua1jets}. 

Several outstanding {\em data features} drew attention: 
(a) a broad minimum-bias (MB) distribution $P(n_{ch})$ on charge multiplicity \nch, its width increasing substantially with collision energy, 
(b) ensemble-mean $\bar p_{t}(n_{ch})$ increasing strongly with \nch\ (whereas the opposite trend had been observed at lower energies), 
(c) correlated variation of the UE with increasing jet trigger energy -- the so called pedestal effect -- and
(d) strong ``forward-backward'' correlated fluctuations of \nch\ in separated pseudorapidity $\eta$ intervals.
Those effects could not be accommodated by a simple model of hard jet plus soft Pomeron exchange.

The PMC approach was based on a unified description applicable to triggered hard jets, UA1 minijets and the UE  based on multiple interactions~\cite{sjostrand2} or {\em multiparton interactions} (MPIs)~\cite{sjostrand} wherein almost all hadrons from almost all events arise from large-angle scattering of constituent partons as described by pQCD. 
A triggered hard jet is simply the most energetic MPI in an event, UA1 minijets correspond to lower-energy MPIs, and the UE is the complementary MPI spectrum extending down to low \pt. MPIs are said to arise naturally from the composite nature of projectile protons leading to large nonPoisson fluctuations and other notable data features. The model excludes any necessity for soft physics via Pomeron exchange except for a small minority of diffractive events (i.e.\ including no MPI) which are modeled by a cut Pomeron or pair of strings connecting valence quarks and diquarks in projectile protons. Given that system  ``...it [was] possible to obtain a quite reasonable description of essentially {\em all the key experimental data}...''~\cite{sjostrand} [emphasis added].  The PMC assumes exchange of colored objects, requiring complex color connections via strings, whereas momentum transfer could be dominated by soft and hard Pomerons~\cite{pomerons,levin}. The only color connection might  then be between scattered partons and their parent nucleons.

Further refinement of the PMC lead to the following: ``In summary, most if not all of MB and UE physics at collider energies is explained and reasonably well described once the basic MPI framework has been complemented by (a) a smooth turnoff of the [jet] cross section for $p_\perp \rightarrow 0$, (b) a requirement to have at least one MPI to get an event, (c) an impact-parameter dependence [based on the eikonal approximation applied to \pp\ collisions], and (d) a colour reconnection mechanism [labels added].'' Reference~\cite{sjostrand2} points out that ``...a sound understanding of multiple interactions [MPIs] is prerequisite for precision physics involving jets and/or the underlying event.''


Although data features (a) through (d)  were fairly well accommodated by the PMC there remained outstanding issues:  The  \pt\ spectrum was underestimated at lower \pt\ and ``...high-multiplicity $pp$ events have properties similar to those observed in heavy-ion $AA$ collisions''~\cite{sjostrand} such as (i) multistrange baryon enhancement, (ii) higher fraction of heavy hadrons, (iii) \mmpt\ larger for heavy hadrons, (iv) Lambda/kaon ratio has a peak near \pt\ = 2.5 GeV/c, (v) ridge on both sides of same-side jet peak, (vi) possible azimuthal flow $v_2$ similar to \aa\ data. It was suggested that ``...plausible explanations start out from a MPI picture and add some kind of collective behaviour [a QGP-like state within \pp\ collisions?] among the MPIs''~\cite{sjostrand}.

\vskip .2in

\subsection{Multiparton interactions -- $\bf MPIs$} \label{mpi}

A signature element of the PMC is multiparton interactions or MPIs, a concept motived as follows: Because of the composite structure of protons (especially low-$x$ partons) several pairs of partons may collide (within a \pp\ event), denoted by multiple interactions (scatterings).  ``Viewing hadrons as `bunches' of incoming partons, it is apparent that when two hadrons collide it is possible that several distinct pairs of partons collide with each other...''~\cite{sjostrand2}. ``...most inelastic [\pp] events...are guaranteed to contain several {\em perturbatively calculable} [sic] interactions''~\cite{sjostrand3}.  Thus, MPIs must exist and would lead to more ``activity'' in the UE as apparently required by data. ``The crucial leap of imagination is to postulate  that {\em all} particle production in inelastic hadronic collisions derives from the multiple-interactions [MPIs] mechanism. ...the starting point is perturbative''~\cite{sjostrand2}. 

The PMC is therefore a one-component model: almost all hadrons must emerge from a hard component consisting of MPIs (i.e.\ QCD jets). And for any non-single-diffractive (NSD) \pp\ event  ``...each event has to have at least one [partonic] interaction [MPI]...''~\cite{sjostrand2}.

The conventional pQCD jet-spectrum formula assumed to represent MPIs within the PMC is~\cite{sjostrand}
\bea \label{jettspec}
\frac{d\sigma}{dp_t^2} &=& \Sigma_{ijk} \int dx_1 dx_2 d\hat t\, f_i(x_1,Q^2) f_j(x_2,Q^2) 
\\ \nonumber
&&\times \frac{d\hat \sigma_{ij}^k}{d\hat t} \delta\left( p_t^2 - \frac{\hat t \hat u}{\hat s} \right).
\eea
Parton distribution functions (PDFs) $ f(x,Q^2)$ are presumably fixed mean-value distributions randomly sampled within the PMC, an approach that can be questioned based on available data as mentioned in Sec.~\ref{specific}.

If almost all final-state hadrons are to be represented by MPIs in the PMC the MPI spectrum must extend to very low parton $p_t \rightarrow p_{tmin}$ where a perturbative formulation may be questioned. One may ask how a very-low-momentum parton might fragment to hadrons in the context of measured fragmentation functions (FFs) from the HERA, LEP and Fermilab~\cite{eeprd}. Nevertheless, it is assumed that MPIs dominate \pp\ hadron production.

Introduction of the PMC represented a transition from a (multi)Pomeron-exchange description of soft events to a formulation of multiple pQCD interactions (MPIs) that combines MB (i.e.\ jet spectrum) and UE aspects: ``The $p_{\perp0}$ [$\sim p_{tmin}$] parameter has to be chosen accordingly small --- since now the concept of no-interaction [no MPI] low-$p_\perp$ events is gone...''~\cite{sjostrand2}. ``A hard-process event would just be the high-$p_\perp$ tail of the MB class, and a soft-process event just one where the hardest jet was too soft to detect as such [by eventwise reconstruction]''~\cite{sjostrand}.

The total cross section for MPIs is then the integral
\bea \label{sigint}
\sigma_{int}(p_{tmin}) &=& \int_{p_{tmin}^2}^{s/4} dp_t^2\, \frac{d\sigma}{dp_t^2},
\eea
and the mean number of MPIs per NSD \pp\ collision is
\bea
\bar n_{MPI}(p_{tmin}) &=& \frac{\sigma_{int}(p_{tmin})}{\sigma_{NSD}},
\eea
with the assumption $\bar n_{MPI}(p_{tmin}) \geq 1$ establishing a constraint (upper limit) on $p_{tmin}$.  ``At least one interaction [MPI] must occur when two hadrons pass by for there to be an event at all''~\cite{sjostrand}. It is assumed that MPIs are independent (Poisson distributed) except for energy conservation. 
 The adopted assumption that almost all hadrons proceed from MPIs suggests that $n_{ch} \propto \bar n_{MPI}(p_{tmin})$, but that leads to problems with observations that $\bar p_t(n_{ch})$ increases strongly with \nch\ (see Sec.~\ref{color} on color reconnection for proposed resolution). A review of MPI-related results from the LHC is presented in Ref.~\cite{mpiref}.

Other significant issues for initial versions of the PMC were 
(a) how to cut off the MPI spectrum: ``A sharp [\pt] cutoff, below which cross sections vanish [as for UA1 minijets at 5 GeV~\cite{ua1jets}], is not plausible''~\cite{sjostrand},%
\footnote{The 5 GeV cutoff reflects a measurement limitation, not an assumed limit to the physical jet spectrum. Spectrum data suggest that the effective lower limit for hadron jets is near 3 GeV~\cite{fragevo,jetspec2}.}
(b) how to deal with ``soft'' events seeming to have no MPIs, (c) how to introduce a \pp\ impact parameter, and (d) how to scale the model with \pp\ collision energy. Item (a) was resolved with a soft cutoff in which $p_t \rightarrow 0$ was permitted but with increasing deviation from Eq.~(\ref{jettspec}) below some cutoff parameter $p_{\perp0}$ adjusted to accommodate data. The supporting argument was based on the inability of lower-\pt\ gluons to resolve color charges. Item (c) was resolved by applying a geometric Glauber model with  eikonal approximation to \pp\ collisions: ``MPIs can be viewed as occurring simultaneously in different parts of the [\pp] overlap region''~\cite{sjostrand}. 

\subsection{Two-component model: alternative viewpoint} \label{tcmview}

As noted, the PMC is  essentially a {\em one-component} model (OCM, hard component only): almost all hadron production arises from a single mechanism---production of MPIs by pQCD-described large-angle parton scattering. In the same context the currently most-popular description of more-central \aa\ collisions at RHIC and the large hadron collider (LHC) is also a OCM (soft component only) in that almost all hadrons are said to emerge by ``freezeout'' from a locally-thermalized flowing dense medium or QGP. As a OCM the PMC can be contrasted with the two-component combination of hadron production within the TCM as applied within this study. 

A two-component model separately describing soft and hard processes was proposed in 1985 for Sp\=pS data~\cite{pancheri}, essentially concurrent with introduction of the PMC. A soft component representing the majority of produced hadrons was retained. However, it was intended to ``...separate the mini-jet contribution from the bulk of many-parton interactions [soft component]'' in response to the following issues: (a) KNO scaling violations, (b) rise of NSD cross section, (c) increase of \mmpt\ with \nch, (d) rise of the ``central plateau'' and hence $\log^2(s)$ scaling of $n_{ch}$. $\log^2(s)$ scaling (hard, gluon bremsstrahlung) was distinguished from $\log(s)$ scaling (soft, quark bremsstrahlung) observed at lower energies. 


With the startup of the hadron-electron ring accelerator (HERA) and large electron-positron collider (LEP) knowledge about jet production and nucleon structure increased rapidly after 1990.  In addition, operation of the alternating gradient synchrotron (AGS) and super proton synchrotron (SPS) in heavy-ion mode and preparations for operation of the relativistic heavy ion collider (RHIC) led to major advances in the description of nucleus-nucleus (A-A) collisions during the nineties. Included in that effort was production of the HIJING MC model for \aa\ collisions based on the \pp\ PMC coupled with a TCM for hadron production in \aa\ collisions~\cite{hijing}.  In Ref.~\cite{kn} the TCM was summarized in comparison with a gluon {\em saturation model} following the startup of RHIC.

The current TCM, with hard {\em and} soft components, is intermediate between two extremes. As demonstrated in Sec.~\ref{tcmstory} the TCM is a simple {\em inductive} model based on inference from a broad array of data. The PMC is a complex {\em deductive} model based on certain {\em a priori} assumptions and includes an array of parameters adjusted to accommodated a limited subset of data manifestations. 

A previous study of \pp\ collision dynamics~\cite{pptheory} focused on UE-related experimental methods and results in the context of the TCM without emphasis on theoretical Monte Carlos. The present study is essentially the complement: direct comparison of the PMC and TCM against a range of established and currently-available analysis methods and data.
The next section considers the PMC within the context of selected \pp\ data features and interpretations that motivated its development.

\vskip .2in

\section{Canonical data Interpretations}  \label{canonical}

Certain analysis methods and associated data features have been emphasized in design and application of the PMC and continue to be used for that purpose despite more-recent alternatives. As noted in Sec.~\ref{models}  data features include 
(a) increasing width of multiplicity distribution $P(n_{ch})$ with collision energy,
(b) ensemble-mean \mmpt\ increasing strongly with \nch,
(c) a triggered-jet pedestal effect on azimuth associated with the UE and
(d) long-range FB correlations indicating nonPoisson fluctuations.
``...the broadening multiplicity distribution and the strong forward-backward correlations offer ...evidence... strongly suggesting that the bulk of events have several [MPIs]. We are not aware of any realistic alternative explanations for either of the observables''~\cite{sjostrand2}.

\subsection{Multiplicity distributions $\bf P(n_{ch})$}  \label{pnch}

A major stimulus for introduction of MPIs as a basis for the PMC was the increasing width of \pp\ multiplicity distribution $P(n_{ch}) = dP/dn_{ch}$ as CM collision energies increased, especially with the introduction of collider accelerators. The increasing nonPoisson behavior was identified with so-called KNO scaling~\cite{kno}.  ``...allowing at most one interaction [MPI] in p\=p events [and assuming \ee\ hadronization] there is no (known) way to accommodate the experimental multiplicity distributions [KNO scaling].... Either hadronization is very different in hadronic events from \ee\ ones, or one must accept multiple interactions [MPIs] as a reality''~\cite{sjostrand2}.

KNO scaling is the hypothesis that \pp\ multiplicity $n$ distributions in the form $\bar n P(n) ~\text{vs}~ n/\bar n \rightarrow Q(z) ~\text{vs}~ z$ with $z = n / \bar n$ are approximately independent of collision conditions, reflecting nonPoisson fluctuations. While the KNO trend seemed to describe ISR data substantial deviations were noted at higher energies. The significance of apparent KNO trends in \pp\ data was questioned in Ref.~\cite{guidokno} where it was pointed out that KNO scaling is equivalent to invariance of statistical moments in the form $C_n = \overline{n^2} / \bar n^2$, and the trend may be accidental. 

Over a broad energy range $P(n)$ data are accurately described by the negative binomial distribution (NBD) with parameters $\mu = \bar n$ and $k$~\cite{aliceppmult}. Fluctuations can be represented in the NBD context by $\sigma^2_n / \bar n = 1 + \bar n / k$. 
Poisson fluctuations correspond to $1/k \rightarrow 0$. The KNO hypothesis corresponds to $C_2 - 1 \approx 1/ \bar n + 1/k \approx$ constant. The first term decreases strongly with collision energy. KNO scaling thus implies that the second term, representing nonPoisson fluctuations, must be rapidly increasing with collision energy in such a way that the sum is approximately constant. Reference~\cite{guidokno} also points out that \pp\ data at higher energies are consistent with $1/k \approx -0.104 + 0.058 \, \ln(\sqrt{s}) \approx 0.06 \ln(\sqrt{s} / \text{6 GeV})$. $1/k$ then dominates $C_2$ at higher energies and breaks KNO scaling which is apparently a coincidence within the  energy range of the ISR. However, the energy dependence of $1/k$ has important implications for jet systematics as described in Sec.~\ref{1overk}. In effect, NBD fluctuation measure $1/k$ can provide a constraint on MB jet spectra.

\subsection{Ensemble-mean $\bf p_t$ and color reconnection}  \label{color}

A second major influence on development of the PMC was the trend of  $\bar p_t$ vs \nch\ at higher collision energies. The trend at lower energies had been decrease, consistent with momentum conservation. The trend observed at and above ISR energies was strong increase for {\em untriggered} or MB events, whereas  events triggered with a (reconstructed) jet exhibited $\bar p_t \approx$ constant. Since the only mechanism for transport from longitudinal to transverse phase space in the PMC is MPIs a problem then arises in attempts to represent the increasing $\bar p_t(n_{ch})$ trend.
Suppose $P_t$ represents {\em total} \pt\ integrated within some acceptance $\Delta \eta$ which also includes total charge \nch. Then for hadron production from independent MPIs per the PMC
\bea
\bar P_t &\propto& \bar n_{MPI}(p_{tmin}) 
\\ \nonumber
n_{ch} &\propto& \bar n_{MPI}(p_{tmin}) 
\\ \nonumber
\bar p_t &\equiv& \bar P_t / n_{ch} ~\approx~ \text{constant}.
\eea 
If MPIs are independent systems then $\bar p_t(n_{ch})$ is necessarily constant within the PMC. An additional mechanism is required to reproduce the observed $\bar p_t(n_{ch})$ trend.

The response was color reconnection (CR): ``To obtain a rising $\bar p_t(n_{ch})$ it is therefore essential to have a mechanism to connect the different MPI subsystems in colour, not only at random but specifically so as to reduce the total string length [and hence hadrons per string] of the event, {\em more and more the more MPIs there are} [emphasis added]. Each further MPI on the average then contributes less \nch\ than the previous, while still the same (semi)hard \pt\ kick is to be shared between the hadrons, thus inducing the rising trend. This is precisely what [color reconnection] is intended to do. It is the first large-scale application of colour reconnection (CR) ideas....''
By introducing a CR mechanism ``Not only the slope but also the absolute value of $\langle p_\perp \rangle$ [\mmpt] is well reproduced, without any need to modify the fragmentation \pt\ width tuned to \ee\ data.''
The ``...CR [mechanism] was essential to obtain a rising $\bar p_t(n_{ch})$, and that has remained a constant argument over the years, still valid today: separate MPIs must be colour-connected in such a way that topologies with a reduced $\lambda$ measure...are favoured''~\cite{sjostrand}.

CR determines how MPIs will hadronize {\em in a correlated way} via a string mechanism: ``Interactions gg $\rightarrow$ gg [are configured] such that...each of the gluons is connected to one of the strings `already' present.'' A color connection ``...which minimizes the total increase in string length is chosen''~\cite{sjostrand2} ``...the $\lambda$ measure is used to pick such reconnections.... A {\em free strength parameter} [$\lambda$, emphasis added] is introduced to regulate the fraction of [scattered parton] pairs that are being tested in this way. With  this further mechanism at hand it now again becomes possible to describe $\bar p_t(n_{ch})$ data approximately''~\cite{sjostrand}. However, it is cautioned that ``Neither of these three [color connection rules] follow naturally from any colour flow rules...''~\cite{sjostrand}.

The combination of $p_{tmin}$ and CR represents an {\em ad hoc} reconfiguration of the MB parton spectrum per Eqs.~(\ref{jettspec}) and (\ref{sigint}) and the systematics of parton fragmentation (modeled within the PMC by a string fragmentation mechanism). But parton fragmentation is represented by measured fragmentation functions (FFs) as in Refs.~\cite{eeprd,jetspec2}, and  {\em effective} jet spectra (i.e.\ as manifested by detected jet fragments) for various collision systems have also been measured~\cite{jetspec2}. The PMC MPIs and CR mechanism as summarized above effectively redefine the MB jet spectrum and FFs (based on the {\em density of scattered partons}) with free parameters used to match \mmpt\ data. There is no guarantee that such a system is correct within a QCD context or in comparison to other data.


\subsection{Triggered-jet pedestal effect and the UE}

A third major influence on development of the PMC has been the so-called ``pedestal effect'' -- strong correlation between an imposed jet trigger condition and a feature of the underlying event or UE. Triggered events including a hard dijet show increased ``activity'' or particle and momentum production (the pedestal) near the midpoint between jet cones at $\phi_\Delta = \phi_1 - \phi_2 \approx \pi/2$ (the {\em trans region} or TR~\cite{fieldue}). The pedestal increases up to trigger condition $E_{jet} \approx 5$ GeV and then saturates.  ``Events containing a hard jet also have an above-average level of particle production {\em well away from the jet core} [emphasis added], the `pedestal effect'  ...The pedestal effect is well described, and explained [by the PMC  tuned to data]. The rise is caused by a shift in the composition of events, from one dominated by fairly peripheral collisions to one strongly biased toward central ones''~\cite{sjostrand}. 

A more detailed and recent theoretical description is as follows~\cite{ppcent2}. Hard particle production (i.e., large-angle gluon scattering to dijets) should be most probable at small \pp\ impact parameter because the transverse size of the low-$x$ gluon distribution in the proton inferred from DIS data is substantially smaller than the overall proton size. 
``Soft'' particle production (not associated with a triggered dijet, i.e.\ MPIs in the context of the PMC) should vary with $b$ over a large range. Specifically, {\em transverse multiplicity} $N_\perp$ (perpendicular to a trigger-particle momentum and therefore to a dijet axis) may be strongly correlated with $b$.
It is assumed that jet production is correlated with smaller $b$ and therefore larger $N_\perp$.

Indirect selection of jets may be established with a single trigger-particle $p_t$ condition denoted by $p_{t,trig}$. For sufficiently high $p_{t,trig}$ 
$b$ should be relatively small and nearly independent of the trigger condition. Equivalently, transverse multiplicity $N_\perp$ should be nearly independent of $p_{t,trig}$ and substantially larger than for NSD \pp\ collisions.
Reference~\cite{ppcent2} then poses the question (given the several assumptions): above what critical $p_{t,trig}$ value is hadron production dominated by ``hard'' parton-parton interactions in more-central \pp\ collisions? 

The assumption that an azimuth interval exists ``well away from the jet core[s]'' -- equivalent to the assumption of ``zero yield at minimum'' or ZYAM invoked for some analyses of triggered-jet azimuth correlations~\cite{zyam} -- is questionable~\cite{tzyam}. As demonstrated in Sec.~\ref{notr}, although a high-energy dijet may {\em appear} to be concentrated in two well-separated ``cones'' it must contribute substantially to all azimuth regions. The relevance of centrality to \pp\ collisions and the suggestion that protons appear {\em smaller} in diameter at lower $x$ whereas the opposite is expected from Gribov diffusion~\cite{gribov} can be questioned: reference~\cite{sjostrand} notes that ``low-$x$ partons should diffuse out in $b$ [$r$] during the evolution [splitting cascade] down from higher-$x$ ones.'' UE analysis is discussed further in Sec.~\ref{uembjets}. The \pp\ pedestal effect and other interpretations relating to the UE are also considered in Ref.~\cite{pptheory}.

\subsection{Long-range FB correlations on pseudorapidity}  \label{fbcorr}

A fourth major influence on development of the PMC was forward-backward (FB) correlations on pseudorapidity $\eta$. ``Long-range'' FB correlations are measured by Pearson's normalized covariance~\cite{pearson,inverse} $b_{FB} \equiv \sigma^2_{n_F n_B} / \sqrt{\sigma^2_{n_F} \sigma^2_{n_B}} $, where $ \sigma^2_{n_F n_B} = \overline{n_F n_B} - \bar n_F \bar n_B$ is the covariance of fluctuating charges $n_F$ and $n_B$ in two $\eta$ bins nominally symmetric about midrapidity and separated by some $\eta$ interval (gap). (The $\sigma^2_{nX}$ are variances within the individual bins.) Observed non-Poisson FB correlations represent angular correlations spanning substantial intervals on $\eta$ and therefore requiring a global source mechanism. Within the PMC context the source of such global fluctuations is attributed to MPIs: ``$\bar n_{MPI}$ is a kind of global quantum number of the event''~\cite{sjostrand}.  FB correlations are described as surprisingly large for $\Delta \eta$ separations (gaps) over several units.

Statistical measure $b_{FB}$ as defined above represents a subset of all \pp\ angular correlations which have by now been studied in considerable detail~\cite{porter2,porter3,ppquad}. A more general measure is $\Delta \rho / \sqrt{\rho_{ref}} \sim  \sigma^2_{n_F n_B} / \sqrt{\bar n_{F} \bar n_B} $~\cite{anomalous,ppquad,inverse} where the statistical reference is represented by Poisson values for the two variances. That covariance density distribution can be determined on a 2D binned system of {\em difference variables} $\eta_\Delta = \eta_1 - \eta_2$ and $\phi_\Delta = \phi_1 - \phi_2$.
Several correlation components are then resolved and may be attributed to distinct hadron production mechanisms~\cite{axialci,anomalous,ppquad}.  Even when extended to multiple $\eta_\Delta$ values the 1D projection $b_{FB}(\eta_\Delta)$ from 2D $(\eta_\Delta,\phi_\Delta)$ is unable to establish such distinctions. While MB dijets do represent a substantial contribution to angular correlations details of measured angular-correlation structure appear inconsistent with the hypothesis of multiple MPIs per event and its implementation in the PMC~\cite{jetspec,jetspec2,ppquad}.

\section{Model Comparison Strategy} \label{comparison}

According to Refs.~\cite{sjostrand,sjostrand2} the PMC is a \pp\ OCM  assuming that most hadron production arises from scattered partons fragmenting to jets (MPIs, hard component) and any soft component is negligible. The flow-QGP OCM for \aa\ collisions asserts that almost all hadron production is soft (freezeout from a thermalized  QCD medium) and any jet contribution comprises a small minority. The TCM with soft and hard components offers a comprehensive intermediate description. This section presents a strategy for further comparisons.

\subsection{Specific model differences} \label{specific}

The PMC by assumption includes no element(s) comparable to the TCM soft component. The scattered-parton \pt\ spectrum extends down to zero, and the detailed spectrum shape for smaller \pt\ is tuned (via some $p_{\perp0}$) to match data within the PMC set of assumptions. In order to describe \mmpt\ vs \nch\ data a CR mechanism is introduced that includes free parameter $\lambda$ adjusted to accommodate such data. The CR mechanism in effect creates a freely adjustable FF ensemble.

The PMC further assumes that \pp\ centrality is relevant and can be modeled via a geometric Glauber model based on the eikonal approximation. \pp\ centrality variation coupled with MPIs as the dominant hadron source provide the principal source of fluctuations. The relation of MPIs to \pp\ centrality and  hadron production should lead to coupling between an imposed jet trigger and UE production, as evidenced by the so-called pedestal effect.

 The TCM, inferred from observed \nch\ dependence of \pt\ spectra and two-particle correlations, describes hadron production in terms of a greater source (soft component) arising from projectile-nucleon dissociation and a lesser source (hard component) arising from MB dijets. The soft component has universal properties including a L\'evy shape and slope parameter $T \approx 145$ MeV~\cite{jetspec2,ppquad}. The hard component (MB fragment distribution) is predicted by a convolution of measured FFs and measured jet spectra {\em not} adjusted to accommodate \pp\ \pt\ spectrum data. The effective lower bound of the jet spectrum is the single free parameter: spectrum data require a lower bound near 3 GeV for all currently-accessible \pp\ collision energies~\cite{fragevo,alicetomspec} consistent with analysis of jet spectra~\cite{jetspec2}.

The precisely-determined quadratic relation between TCM soft and hard components, persistent over a large range of parton densities, precludes any role for collision centrality. In any \pp\ collision all participant partons may freely interact in any combination. There is no restricted ``overlap region,'' and the eikonal approximation is not relevant to parton-parton interactions. Given the noneikonal quadratic soft/hard relation and no centrality (impact parameter $b$) variation, fluctuations must arise from soft-component production within {\em individual projectile protons}, suggesting that the mechanism is varying depth on momentum fraction $x$ of parton splitting cascades, i.e.\ strong eventwise fluctuation of nucleon PDFs.

\subsection{Relations among several forms of data}

Certain relations among analysis methods and data formats are relevant to the PMC-TCM comparison. The TCM for \mmpt\ data~\cite{tommpt} is simply related to the TCM for \pt\ spectra~\cite{ppprd,ppquad}. The TCM spectrum hard component is in turn directly and quantitatively related to measured FFs~\cite{eeprd,jetspec2} and measured MB jet spectra~\cite{jetspec2} via a convolution integral~\cite{fragevo}. Two-particle correlations~\cite{porter2,porter3}, both $(y_t,y_t)$ correlations and 2D angular correlations on $(\eta,\phi)$, are described quantitatively by the TCM~\cite{ppquad}. Soft and hard components of $(y_t,y_t)$ correlations correspond directly to \yt\ spectrum TCM (projection from 2D to 1D). 

TCM soft and hard components are clearly distinguished by \nch\ dependence, \pt\ dependence, $\sqrt{s}$ dependence and correlation structure as demonstrated in several studies~\cite{ppprd,porter3,ppquad,alicetomspec,tommpt}. The hard components of spectra and two-particle correlations, corresponding directly to MB dijets, represent a minority fraction of total hadron production. The majority fraction must then be a {\em nonjet} contribution, i.e.\ the TCM soft component. 

\subsection{Improved comparison strategy} 

This study emphasizes two themes:  (a) reexamine mid-eighties collider data trends invoked as motivating or supporting the PMC at its inception and (b) present evidence from more-recent data and interpretations against basic assumptions of the PMC. 
Certain data features described in Sec.~\ref{models}, inferred from ISR and Sp\=pS data and more recently from Fermilab and referred to as ``key experimental data''~\cite{sjostrand}, have been favored for subsequent development and support of the PMC. A large body of additional data and methods relating to \pp, \pa\ and \aa\ collisions and emerging in the intervening period at the RHIC and LHC appear to be underemployed. The present study extends model comparisons to better utilize available data.
Several topics are emphasized: 
(a) quantitative relations between TCM soft and hard components,  
(b) quantitative understanding of MB dijets in various manifestations, 
(c) UE systematics and the pedestal effect,
(d) \pp\ collision geometry -- relevance thereof -- and
(e) fluctuations and their sources. 


Section~\ref{tcmstory} introduces details of the TCM and its relation to data to provide context for further comparisons. The relation between the \pp\ TCM in isolation and a \pn\ TCM within more-complex \pa\ collisions is emphasized.

Section~\ref{mpimbjets} compares arguments for conjectured MPIs to experimentally-observed MB dijets. Measured jet spectra and fragmentation functions are linked directly and quantitatively via convolution integral to \pt\ spectrum hard components. That connection places a lower bound on jet energy spectra that contradicts PMC assumptions. It also challenges assumptions about the conjectured role of a CR mechanism and implicit variation of FFs with scattered-parton density. MB dijet systematics and the relation between TCM spectrum soft and hard components conflict with a PMC assumption about \pp\ centrality. MB dijet cross sections imply that a majority of \pp\ events (soft) include no significant jet structure (MPIs), consistent with the measured rate of double parton scattering. The relation of the \mmpt\ statistic to \pt\ spectrum structure is evident. Variation of \mmpt\ with \nch\ is then simply explained within the TCM context in terms of MB dijet properties and noneikonal \pp\ collisions.

Section~\ref{uembjets} considers assumed access to the UE and its inferred properties in the context of the TCM and MB dijets. In support of the PMC it is argued that response of UE activity to a jet trigger signals the presence of MPIs and relevance of \pp\ centrality. However, measured properties of MB dijets reveal a strong triggered-jet contribution to the  azimuth TR, i.e.\ the increased particle production referred to in the PMC context. Both the $N_\perp$ yield within the TR and the $dN_\perp / dp_t$ spectrum are accurately predicted by the TCM based on measured MB dijet properties and \pt\ spectra. A notable result: application of a jet \pt\ trigger does not change the soft component, arguably the actual UE. There is no coupling between an applied jet trigger and \pp\ centrality. The role of fluctuations is considered briefly in Sec.~\ref{disc}.

\section{TCM for $\bf p$-$\bf p$ and $\bf p$-N collisions} \label{tcmstory}

As an introduction to the TCM the \nch\ dependence of \pt\ spectra from 200 GeV \pp\ and 5 TeV \ppb\ collision systems is analyzed differentially.  Spectrum systematics are related to two-particle correlations on $(y_t,y_t)$ and $(\eta_\Delta,\phi_\Delta)$ to support interpretation of soft and hard TCM components.  Hard/soft ratio trends reveal a quadratic relation between MB dijet production and soft hadron production.  In App.~\ref{tcmenergy} a TCM parametrization directly related to fundamental QCD processes spans a range of collision systems to describe all data within their uncertainties.  In App.~\ref{etadist} the hadron density distribution on $\eta$ within $|\eta| < 1$ is decomposed into soft and hard components.  Some TCM results relevant to the PMC are: (a) a TCM soft component representing a majority of produced hadrons is required by data, (b) \pp\ centrality is not relevant and (c) spectrum hard components are isolated for comparison with measured jet properties.












\subsection{TCM description of p-p $\bf p_t$ or $\bf y_t$ spectra} \label{aa1}

The TCM for \pp\ collision data emerged from analysis of 200 GeV \pt\ spectrum data. Systematic analysis of the \nch\ dependence of \pt\ spectra from 200 GeV \pp\ collisions described in Ref.~\cite{ppprd} led to a compact phenomenological TCM with {approximate factorization} of multiplicity \nch\ and {transverse-rapidity} \yt\ dependence in the form
\bea \label{a1x}
\frac{d^2n_{ch}}{y_t dy_t d\eta} &\approx& \bar \rho_0(y_t) = S_{pp}(y_t,n_{ch}) + H_{pp}(y_t,n_{ch})
\\ \nonumber
&\approx& \bar \rho_s(n_{ch}) \hat  S_0(y_t) + \bar \rho_h(n_{ch}) \hat H_0(y_t)
\eea
with mean angular densities $\bar \rho_x = n_x / \Delta \eta$. Transverse rapidity $y_t \equiv \ln[(p_t + m_t)/m_h]$ with transverse mass defined by $m_t^2 = p_t^2 + m_h^2$ provides improved visual access to spectrum structure at lower \pt\ or \yt\ (for unidentified hadrons pion mass $m_h = m_\pi$ is assumed). Unit-integral soft-component model $\hat S_0(m_t)$ is consistent with a L\'evy distribution on \mt, while peaked hard-component model  $\hat H_0(y_t)$ is well approximated by a Gaussian on \yt\ centered near $y_t \approx 2.65$ ($p_t \approx 1$ GeV/c) with exponential (on \yt) tail reflecting an underlying power-law (on \pt) jet spectrum~\cite{jetspec2}. Transformation from \pt\ or \mt\ to \yt\ is simply accomplished with Jacobian $p_t m_t/  y_t$.

\begin{figure}[t]
	\includegraphics[width=1.65in,height=1.65in]{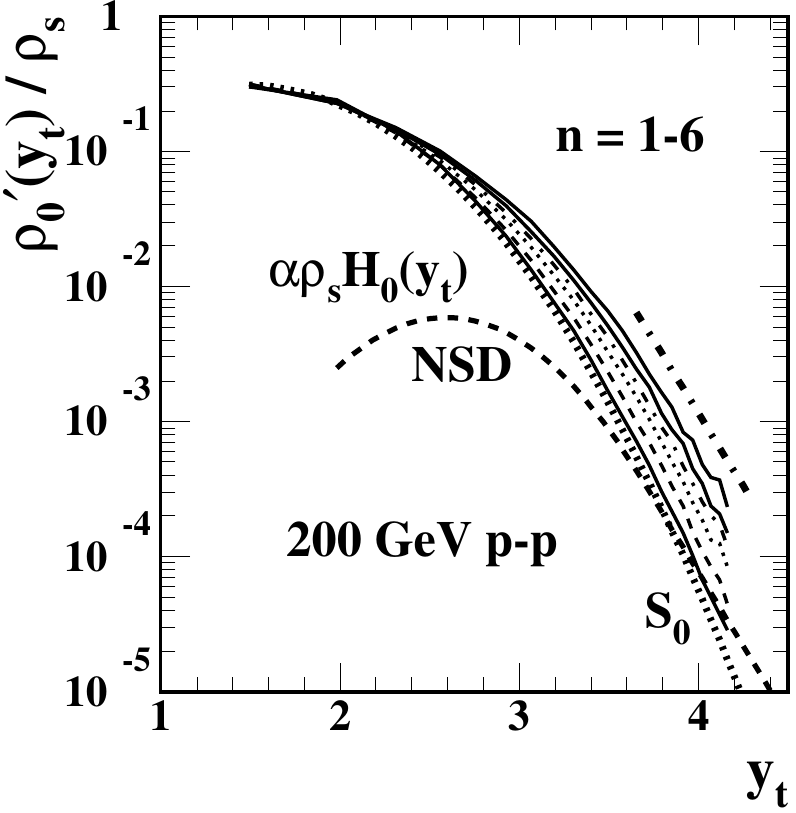}
	\includegraphics[width=1.65in,height=1.65in]{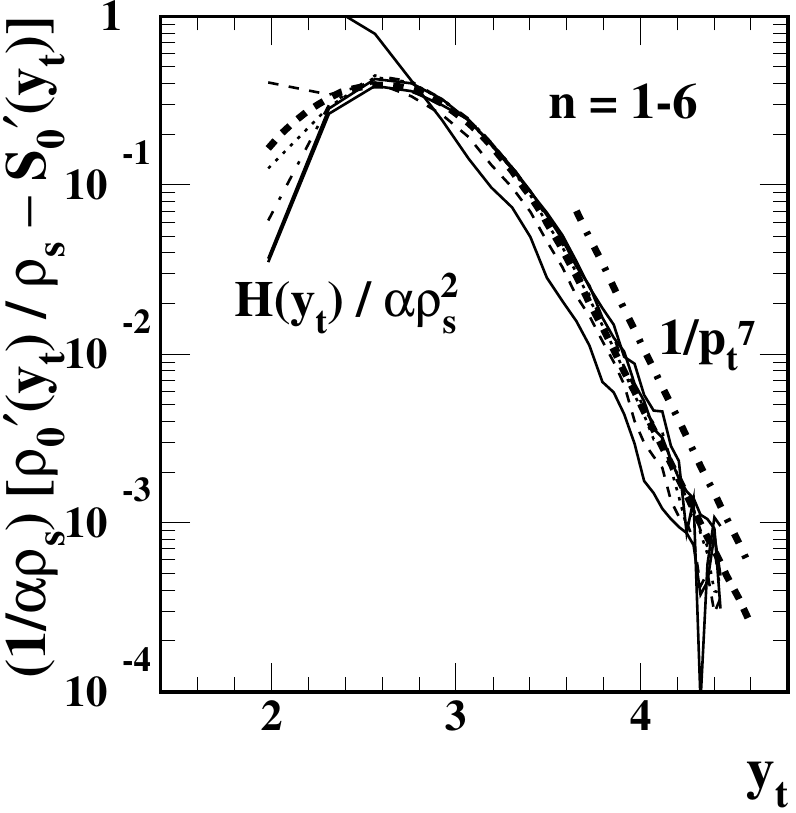}
	\caption{\label{pp1}
		Left: Hadron \yt\ spectra for six multiplicity classes of 200 GeV \pp\ collisions (thin curves) compared to fixed reference $\hat S_0(y_t)$ (bold dotted)~\cite{ppquad}. Spectra $\bar \rho_0'$ are uncorrected for low-\pt\ inefficiencies ($y_t < 2$). The bold dashed curve is the corresponding hard-component model $\alpha\bar \rho_s \hat H_0(y_t)$ for NSD \pp\ collisions.
		Right: Hard-component distributions inferred from spectra at left in the form $H(n_{ch},y_t)/\bar \alpha \rho_s^2$ (several line styles) compared to fixed reference $\hat H_0(y_t)$ (bold dashed).  The dash-dotted line in each panel indicates a power-law trend $\approx 1/p_t^7$ corresponding to the underlying 200 GeV jet energy spectrum.
	} 
\end{figure}

Integration of Eq.~(\ref{a1x}) over \yt\ results in the angular-density TCM $\bar \rho_0 = \bar \rho_s + \bar \rho_h$. Spectrum~\cite{ppprd,alicetomspec} and angular-correlation~\cite{ppquad} data reveal that soft and hard angular densities are related by $\bar \rho_h \approx \alpha \bar \rho_s^2$ with $\alpha \approx 0.006$ within $\Delta \eta = 2$ at 200 GeV. The two relations are equivalent to a quadratic equation that uniquely defines $\bar \rho_s$ and $\bar \rho_h$ in terms of $\bar \rho_0$ (when corrected for inefficiencies). That quadratic relation is valid over a $\bar \rho_s$ interval corresponding to {\em 100-fold variation} of MB dijet production~\cite{ppprd,ppquad}.

Figure~\ref{pp1} (left) shows uncorrected \yt\ spectra for six \pp\ multiplicity classes averaged over acceptance $\Delta \eta = 2$ and normalized by soft-component density $\bar \rho_s = n_s / \Delta \eta$~\cite{ppquad}. The  bold dotted curve is soft-component L\'evy model $\hat  S_0(y_t)$ with model parameters $T = 145$ MeV and $n = 12.5$~\cite{alicetomspec}. The L\'evy model is slightly modified at lower \pt\ ($< 0.5$ GeV/c) to match data tracking inefficiency there.

Figure~\ref{pp1} (right) shows hard-component data inferred from the left panel via  Eq.~(\ref{a1x})  in the form $H_{pp}(y_t,n_{ch}) /\alpha \bar \rho_s^2$ (thin curves) compared to a fixed Gaussian model function in the form $ \hat H_0(y_t)$ (bold dashed) with centroid $\bar y_t \approx 2.65$ and width $\sigma_{y_t} \approx 0.45$ and with coefficient $\alpha \approx 0.006$ determined by the data-model comparison. Note that from a TCM analysis of \pt\ spectra the hard component (MB jet-fragment distribution) is explicitly determined for direct comparison with {\em measured} dijet properties and constitutes a minority of total hadron production in all cases. In contrast, the PMC ``hard component'' (MPIs) must represent the entire \pt\ spectrum.

The \pp\ spectrum TCM can be extended to \pa\ collisions assuming that {\em dijet production is unchanged} in the more-complex \pa\ system. To do so requires determination of \pa\ centrality in the form of participant-nucleon (N) number $N_{part}$ and \nn\ binary-collision number $N_{bin}$ to isolate individual \pn\ collisions. The mean number of binary collisions per participant pair is $\nu \equiv 2 N_{bin} / N_{part}$. For the \ppb\ centrality study in Ref.~\cite{tomglauber} centrality was inferred from the systematics of \mmpt\ vs \nch\ data~\cite{alicempt}.

\begin{figure}[t]
	\includegraphics[width=3.3in,height=1.65in]{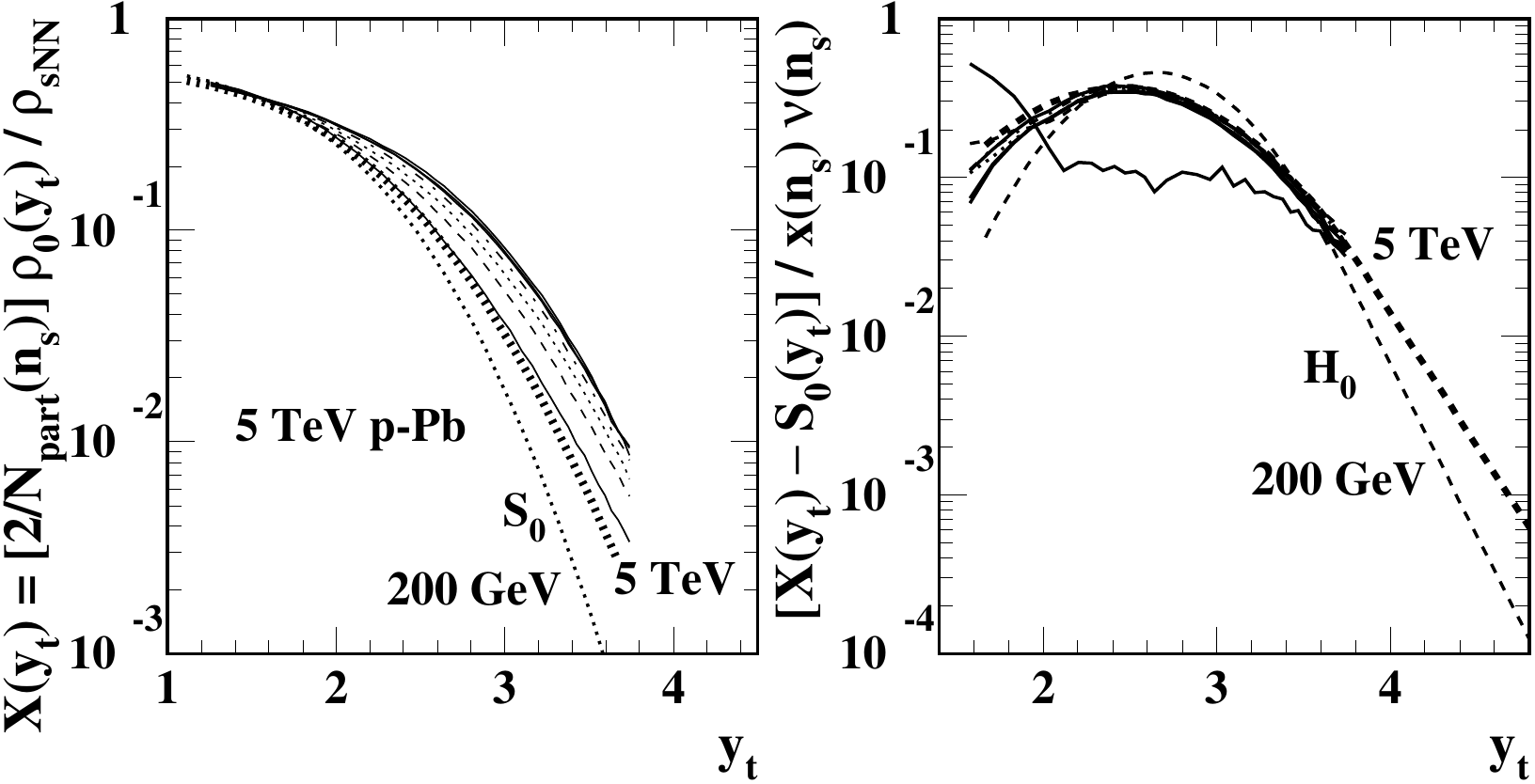}
	\caption{\label{qppb}
		Left: Corrected identified-pion spectra for 5 TeV \ppb\ collisions from Ref.~\cite{alicepionspec} transformed to \yt\ with Jacobian $p_t m_t / y_t$ and normalized by TCM values for $N_{part}$ and $\bar \rho_{sNN}$ from Ref.~\cite{tommpt} (7 thinner curves). $\hat S_0(y_t)$ (bold dotted) is the universal soft-component model.
		Right: Difference $X(y_t) - \hat S_0(y_t)$ normalized by $x(n_s) \nu(n_s) = \alpha \bar \rho_{sNN} \nu(n_s)$ using TCM values from Ref.~\cite{tommpt} (thin curves). The bold dashed curve is the hard-component model $\hat H_0(y_t)$ with exponential tail. Thinner dotted and dashed curves denote 200 GeV models.
	} 
\end{figure}

Figure~\ref{qppb} (left) shows identified-pion spectra from 5 TeV \ppb\ collisions~\cite{alicepionspec}. The published spectra have been multiplied by $2\pi$ to be consistent with $\eta$ densities as in Fig.~\ref{pp1} and transformed to \yt\ with Jacobian $p_t m_t / y_t$. The spectra are then normalized by soft-component density $\bar \rho_s = (N_{part}/2) \bar \rho_{sNN}$ as reported in Table~II of Ref.~~\cite{tomglauber}, except that an additional factor 0.8 is applied to $\bar \rho_s$ values to reflect the pion fraction of soft hadrons. Normalized spectra $X(y_t)$ are then compared with spectrum soft-component model $\hat S_0(y_t)$ (bold dotted curve): a L\'evy model with parameters $T = 145$ MeV and $n = 8.3$ appropriate for 5 TeV \pp\ collisions as reported in Ref.~\cite{alicetomspec}. The 200 GeV soft component is included for comparison.

Figure~\ref{qppb} (right) shows  difference $X(y_t) - \hat S_0(y_t)$ normalized by $x(n_s) \nu(n_s) = \alpha \bar \rho_{sNN} \nu(n_s)$ with TCM values reported in Ref.~\cite{tommpt} and Table~II of Ref.~\cite{tomglauber}.  The result should be directly comparable to the \pp\ spectrum hard-component model in the form $\hat H_0(y_t)$ with model parameters  $(\bar y_t,\sigma_{y_t},q) = (2.65,0.59,3.9)$ for 5 TeV \pp\ collisions as reported in  Ref.~\cite{alicetomspec}. The bold dashed curve is $\hat H_0(y_t)$ with $(\bar y_t,\sigma_{y_t},q) \rightarrow (2.45,0.605,3.9)$. A shift to lower fragment momenta for pions is expected based on  Fig.~7 (left) of Ref.~\cite{eeprd}: pion FFs are softer than kaon FFs are softer than proton FFs. The 200 GeV hard-component model (for unidentified hadrons as in Fig.~\ref{pp1}) is included for comparison.
The overall TCM description is well within point-to-point data uncertainties except for the lowest centrality class (solid curve) where the large deviation is expected based on Ref.~\cite{alicetomspec}. The TCM description of \ppb\ spectra {\em assumes linear superposition} of \pn\ collisions within \ppb\ collisions. However, it also describes realistically the changing properties of \pn\ collisions depending on an applied \ppb\ \nch\ condition. 

In summary, the \pt\ spectrum TCM is an accurate representation of data from several collision systems based on two simple QCD-based hadron production mechanisms.  Evolution of the TCM (and data) from 200 GeV \pp\ to 5 TeV \ppb\ is fully consistent with smooth $\log(\sqrt{s})$ dependences expected for QCD phenomena, as reported in Refs.~\cite{alicetomspec,tommpt,tomglauber}. The {\em independent} energy evolution of soft and hard spectrum TCM components is  notable.

\subsection{Two-particle correlations} \label{2partcorr}

For a self-consistent data description the TCM for \pt\ spectra should have a corresponding description for two-particle correlations: both $(p_{t},p_{t})$ or $(y_{t},y_{t})$ correlations and angular correlations on $(\eta,\phi)$ formulated in terms of difference variables $\eta_\Delta$ and $\phi_\Delta$ defined in Sec.~\ref{fbcorr}.

\begin{figure}[h]
  \includegraphics[width=1.65in,height=1.4in]{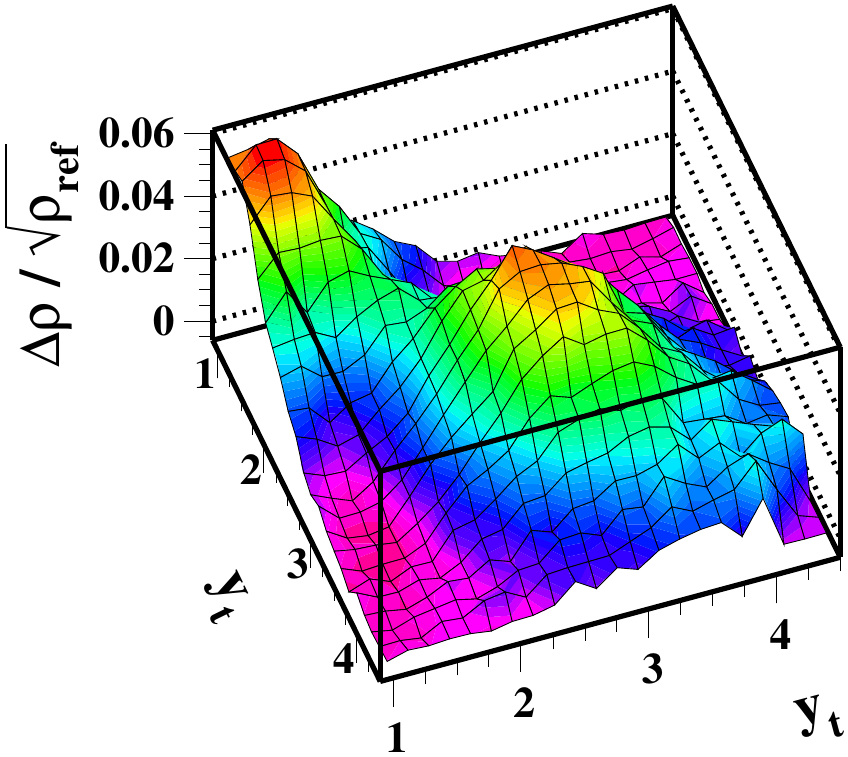}
 \includegraphics[width=1.65in,height=1.4in]{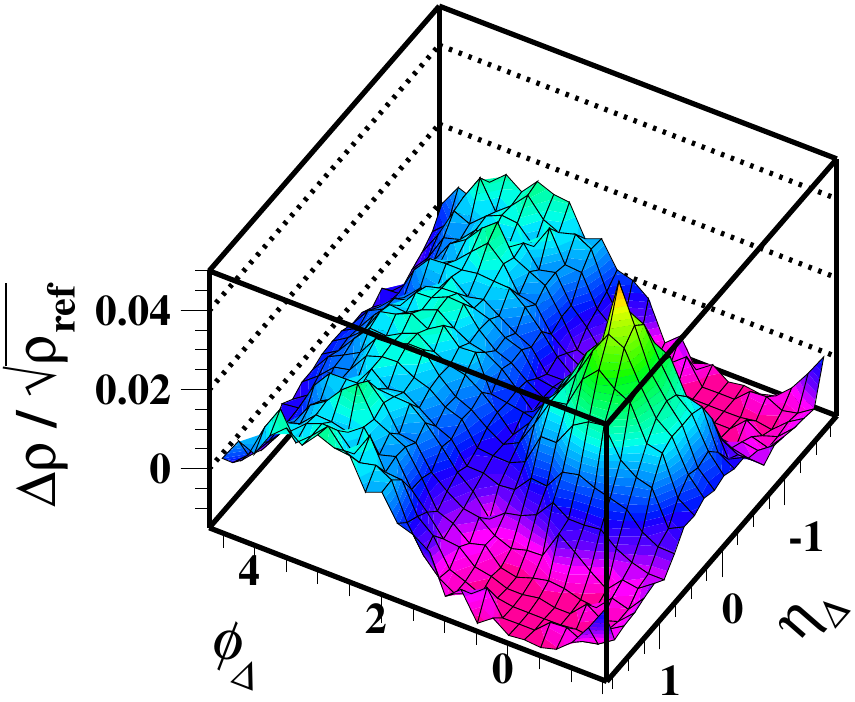}
 \caption{\label{ppcorr} (Color online)  Two-particle correlations on $(y_t, y_t)$ and $(\eta,\phi)$~\cite{porter2,porter3}.
Left: Minimum-bias correlated-pair density on 2D transverse-rapidity space $(y_{t},y_{t})$ from 200 GeV \pp\ collisions showing soft (smaller \yt) and hard (larger \yt) components as peak structures.
Right:  Correlated-pair density on 2D angular difference space $(\eta_\Delta,\phi_\Delta)$. Although hadrons are selected with $p_t \approx 0.6$ GeV/c ($y_t \approx 2.15$) features expected for dijets are still observed: (i) same-side 2D peak representing intrajet correlations and (ii) away-side 1D peak on azimuth representing interjet (back-to-back jet) correlations. 
 }  
 \end{figure}

Figure~\ref{ppcorr} (left) shows correlations on $(y_{t},y_{t})$ from 200 GeV NSD \pp\ collisions for $p_t \in [0.15,6]$ GeV/c ($y_t \in [1,4.5]$)~\cite{porter2,porter3}. Two peaked features are identified as TCM soft and hard components as follows. The lower-\yt\ peak falls mainly below 0.5 GeV/c ($y_t < 2$) and consists exclusively of unlike-sign (US) pairs. Corresponding angular correlations consist of a narrow 1D peak on $\eta_\Delta$ centered at the origin. The combination suggests {\em longitudinal} fragmentation of low-$x$ gluons to {\em charge-neutral} hadron pairs closely spaced on $\eta$ and consistent with spectrum soft component $S_{pp}(y_t,n_{ch})$ in Eq.~(\ref{a1x}). The higher-\yt\ peak extends mainly above 0.5 GeV/c with mode near $p_t = 1$ GeV/c ($y_t \approx 2.65$) and is consistent with  hard component $H_{pp}(y_t,n_{ch})$ in Eq.~(\ref{a1x}) and Fig.~\ref{pp1} (right).

Figure~\ref{ppcorr} (right) shows angular correlations for the same collision system with the condition $p_t \approx 0.6$ GeV/c ($y_t \approx 2.15$), i.e.\ near the {\em lower} boundary of the $(y_t,y_t)$ hard component in the left panel. Despite the low hadron momentum the observed angular correlations exclude a contribution from the soft component (see Sec.~\ref{notr}), exhibiting only structure expected for jets: a same-side (SS, $|\phi_\Delta| < \pi/2$) 2D peak representing {\em intra}\,jet  correlations and an away-side (AS, $|\phi_\Delta - \pi| < \pi/2$) 1D peak representing  {\em inter}\,jet (back-to-back jet) correlations. The SS peak is dominated by US pairs while the AS peak has US $\approx$ LS, consistent with fragmentation of back-to-back {\em charge-neutral} gluons. Note that 2D angular correlations from MB dijets are directly related to interpretations of UE systematics vs a trigger \pt\ condition and the so-called pedestal effect as discussed further in Sec.~\ref{trpedestal}.

\subsection{Hard/soft ratio trends vs $\bf n_s$}

The relation $\bar \rho_h \approx \alpha \bar \rho_s^2$ plays a key role in understanding the nature of \pp\ collisions, especially in the context of evaluating the PMC. It is therefore important to consider the evidence for and accuracy of that relation in terms of ratios $n_h / n_s$ and $\bar P_t / n_s$ vs soft-component density $\bar \rho_s$.

\begin{figure}[h]
  \includegraphics[width=1.62in]{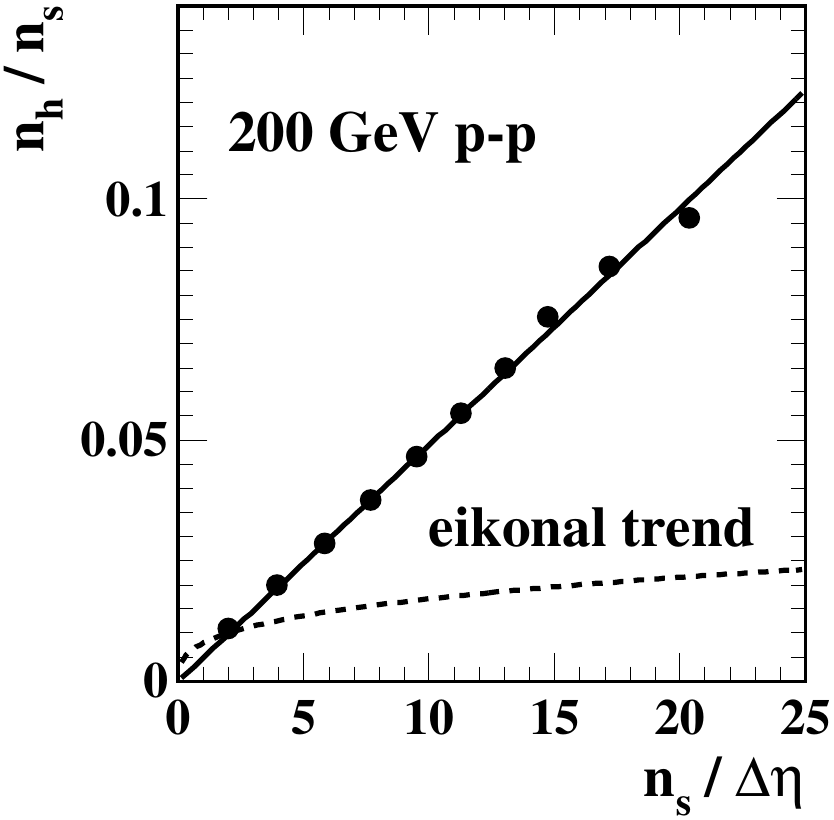}
  \includegraphics[width=1.68in]{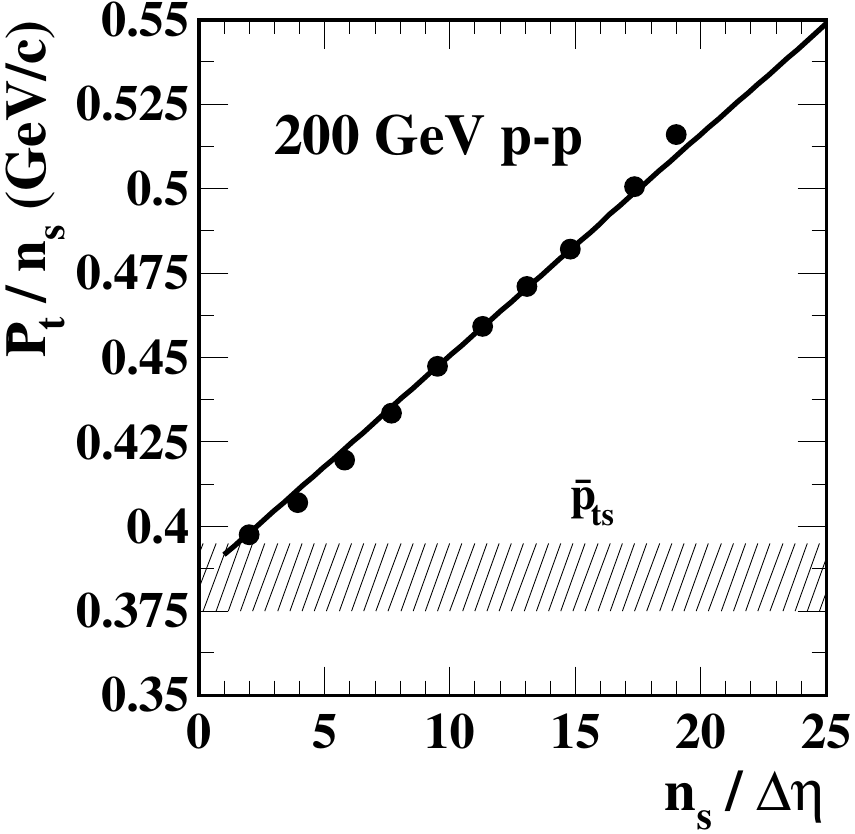}
\caption{\label{ppcomm} 
Left: 
Hard/soft multiplicity ratio $n_h / n_s$ (points) vs soft component $n_s$ consistent with a linear trend (line)~\cite{ppprd,tomphenix}.  Assuming that $n_s$ represents the density of small-$x$ participant partons (gluons) and $n_h$ represents dijet production by parton scattering, an eikonal model of \pp\ collision geometry (analogous to the Glauber model of \aa\ collisions) would predict an $n_s^{1/3}$ trend for the  ratio (dashed curve).
Right: 
The $\bar p_t \rightarrow \bar P_t / n_s$ trend predicted by the \pp\ spectrum TCM (line) and as determined by direct spectrum integration (points). The hatched band represents the uncertainty in soft component $\bar p_{ts}$ from extrapolating \pt\ spectra to zero momentum.
 }  
 \end{figure}

Figure~\ref{ppcomm} (left) shows ratio $n_h / n_s = \bar \rho_h / \bar \rho_s$ vs $\bar \rho_s = n_s / \Delta \eta$ for ten multiplicity classes from Ref.~\cite{ppprd,tomphenix}. $\bar \rho_h$ is the \yt\ integral of hard-component $H(y_t)$ appearing in the form $H(y_t) / \alpha \bar \rho_s^2$ in Fig.~\ref{pp1} (right). The linear trend for $n_h/n_s$ establishes the relation $\bar \rho_h \propto \bar \rho_s^2$. Soft-component density $\bar \rho_s$ may be interpreted as a proxy for the density of low-$x$ gluons released from projectile nucleons in a \pp\ collision. \pp\ spectrum data then reveal that the number of mid-rapidity dijets $\propto \bar \rho_h$ varies {\em quadratically} with number of participant gluons. But for an eikonal collision model the number of gluon-gluon binary collisions should vary as the dashed curve representing $\bar \rho_h \propto \bar \rho_s^{4/3}$, as assumed for a geometric Glauber model and the PMC. These \pp\ data appear inconsistent with the eikonal model. The noneikonal quadratic trend remains accurate over a $\bar \rho_s$ range corresponding to 100-fold increase in dijet production or scattered-parton density.

Figure~\ref{ppcomm} (right) shows $\bar P_t / n_s$ (a ratio of integrated quantities)  vs soft-component mean density $\bar \rho_s$ described by a constant term (soft component $\bar p_{ts}$) plus linear term (hard component $\propto \bar \rho_s$). The solid line is derived from the spectrum $\bar \rho_0(p_t)$ TCM of Eq.~(\ref{a1x}) with $\bar P_t = \Delta \eta \int_0^\infty dp_t p_t^2 \bar \rho_0(p_t)$. Given evidence in  this section, $n_h$ and $\bar P_t$ variations with $\bar \rho_s$ appear to be determined entirely by  a jet-related hard component, and the $\bar P_t$ hard component varies as $\bar P_{th} \propto \bar \rho_s^2$. As noted, the linear ratio trends  (quadratic relation between $n_h$ or $\bar P_{th}$ and $\bar \rho_s$) extend accurately over a $\bar \rho_s$ range corresponding to 100-fold increase of dijet production (and hence scattered-parton density). That result conflicts with a conjectured CR mechanism and consequent parton density (MPI number density) dependence of FFs (Sec.~\ref{color}).

\section{$\bf MPIs~vs$ Minimum-bias dijets}  \label{mpimbjets}

As summarized in Sec.~\ref{mpi} the PMC is based on assumptions that 
(a) almost all hadrons arise from jets (MPIs) -- soft hadron production is negligible, 
(b) the jet (MPI) spectrum extends to $p_t \rightarrow 0$, 
(c) each inelastic \pp\ collision therefore includes at least one MPI, 
(d) MPI production is controlled by \pp\ centrality as described by a geometric Glauber model based on the eikonal approximation and 
(e) parton fragmentation to jets involves a CR mechanism tuned to accommodate certain \mmpt\ data.
This section compares those assumptions to manifestations of MB dijets in several data formats.

\subsection{Systematics of MB dijets from ISR and $\bf Sp\bar pS$} \label{jetspecc}

Reconstructed-jet data available from the ISR and Sp\=pS in the mid eighties are reviewed and described by a simple parametrization inferred recently from \pp\ spectrum and correlation data in the context of the TCM~\cite{jetspec2}. A survey of various manifestations of MB dijets in several data formats is reported in Ref.~\cite{mbdijets}.

\begin{figure}[h]
    \includegraphics[width=1.65in,height=1.6in]{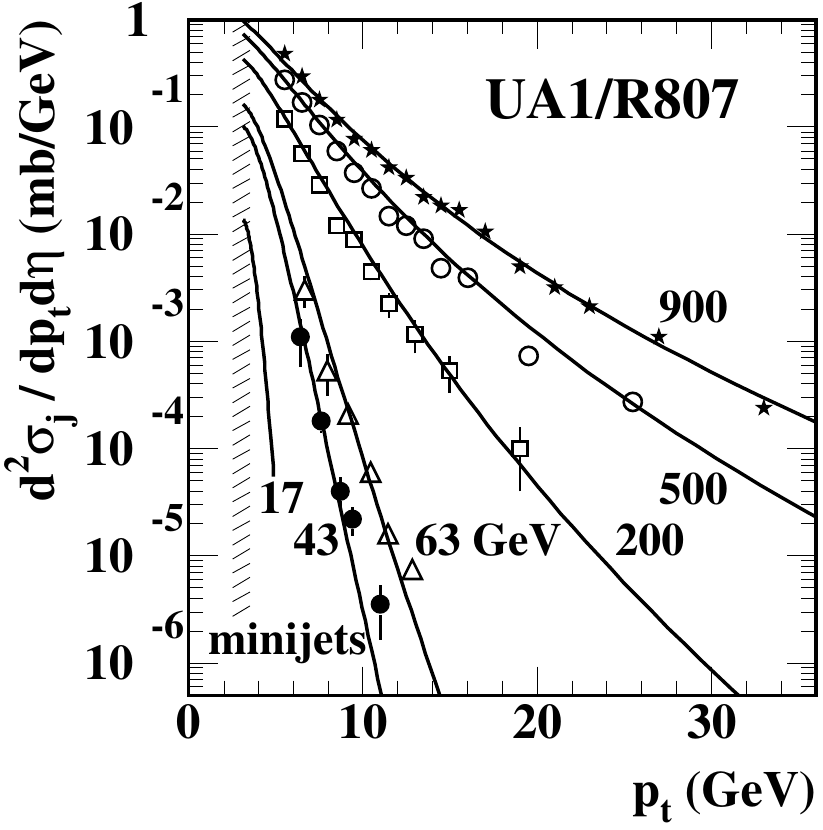}
   \includegraphics[width=1.65in,height=1.6in]{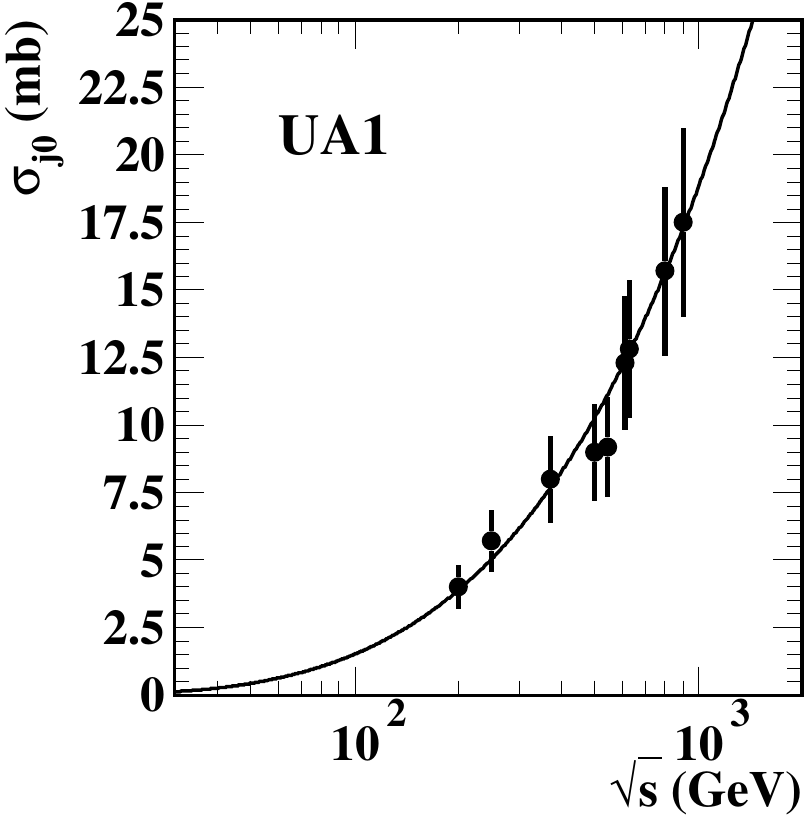}
  \caption{\label{ua1data}
  Left: Inclusive jet cross sections (points) from ISR~\cite{isrfirstjets} and Sp\=pS~\cite{ua1jets} collisions at five energies extending down to 5 GeV/c jet momentum. The curves are from Eq.~(\ref{curious}). The 17 GeV curve is a model extrapolation applicable to \pbpb\ collisions at the SPS.
  Right: Jet (jet-event) total cross sections for  $E_t > 5$ GeV within $|\eta| < 2.5$ from Ref.~\cite{ua1jets}. The curve is described by Eq.~(\ref{jeteta}) (second line).
   }   
   \end{figure}

 Figure~\ref{ua1data} (left) shows MB jet spectra for five \pp\ collision energies from the ISR (43 and 63 GeV~\cite{isrfirstjets}) and Sp\=pS (200, 500 and 900 GeV~\cite{ua1jets}) plotted conventionally on jet (parton) $p_t$. Those innovative analyses provided the first access to very low jet energies. The solid curves through data are defined by Eq.~(\ref{curious}) below. 
 
Figure~\ref{ua1data} (right) shows UA1 total cross sections for MB jet production. The cross sections represent \pp\ events that include at least one jet with $E_t > 5$ GeV within $|\eta| < 2.5$. The curve passing through data is described by Eq.~(\ref{jeteta}) (second). Point-to-point deviations are small compared to  $\pm 20$\% systematic uncertainties (error bars).

To simplify jet spectrum parametrization certain logarithmic rapidity variables are defined in terms of the pion mass. The jet (parton) rapidity is $y_{max} \equiv \ln(2 E_{jet}/m_\pi)$~\cite{eeprd}, with $E_{jet} \rightarrow p_t$ for plotted jet spectra, and the beam rapidity is  $y_b \equiv \ln(\sqrt{s} / m_\pi)$. The conditional jet (scattered-parton) spectrum for a given collision energy $\sqrt{s}$ is denoted by  $d^2\sigma_j /dy_{max}d\eta \equiv S_p(y_{max}|y_{b})$. Systematic analysis of available jet production data leads to a simple parametrization based on quantities $y_{b0} \equiv \ln(\sqrt{s_0} / m_\pi)$ with $\sqrt{s_0} \approx 10$ GeV and $y_{max0} =  \ln(2E_{cut} / m_\pi)$.  Differences $\Delta y_b = y_b - y_{b0}$ and $\Delta y_{max} = y_{b} - y_{max0}$ are defined, with normalized differential jet (parton) rapidity $u = (y_{max} - y_{max0}) / \Delta y_{max}$. 


Section~\ref{tcmstory} established that hard-component density $\bar \rho_h$ (and presumably dijet production as $d\sigma_j/d\eta$) for 200 GeV \pp\ collisions scales with the soft-component density as $\bar \rho_h \propto \bar \rho_s^2$ due to the noneikonal nature of \pp\ collisions.  Given that relation and  $\bar \rho_s  \approx 0.81  \Delta y_b$ near mid-rapidity at and above ISR energies~\cite{alicetomspec} jet spectra near midrapidity should scale vertically as $d\sigma_j/d\eta  \propto (\Delta y_b)^2$. Based on systematics of FFs defined on $y_{max}$ reported in Ref.~\cite{eeprd} jet rapidity as $y_{max} - y_{max0}$ is rescaled horizontally by factor $\Delta y_{max}$ to normalized rapidity $u$. 
Jet spectrum data then collapse to a single locus consistent with a Gaussian {\em if} parameter $y_{max0}$ corresponds to $E_{cut} \approx 3$ GeV.

\begin{figure}[h]
     \includegraphics[width=1.65in,height=1.63in]{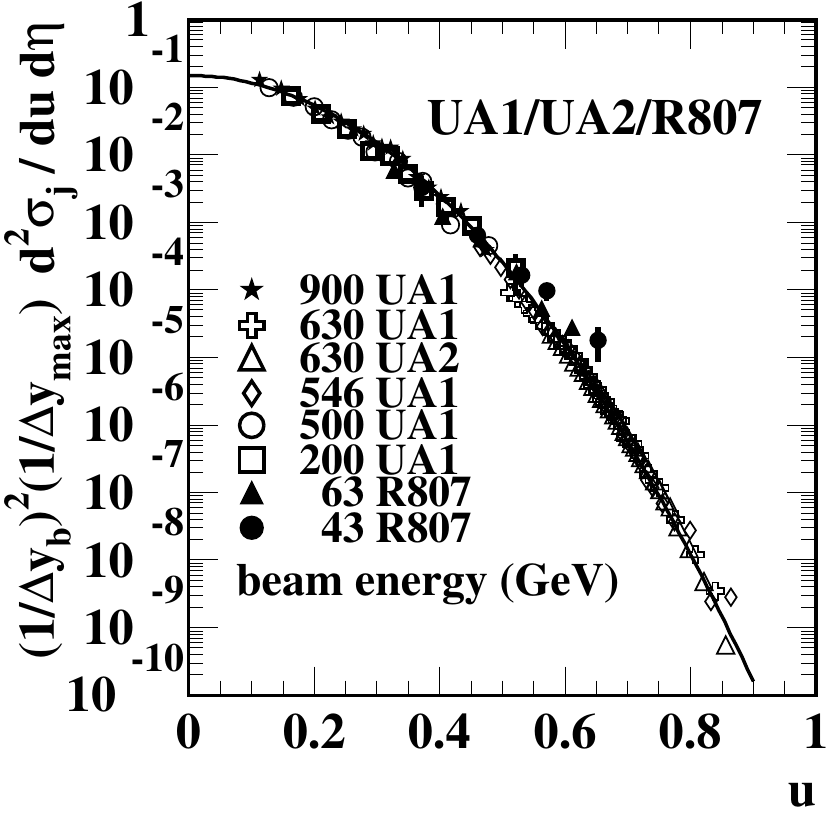}
   \includegraphics[width=1.65in,height=1.6in]{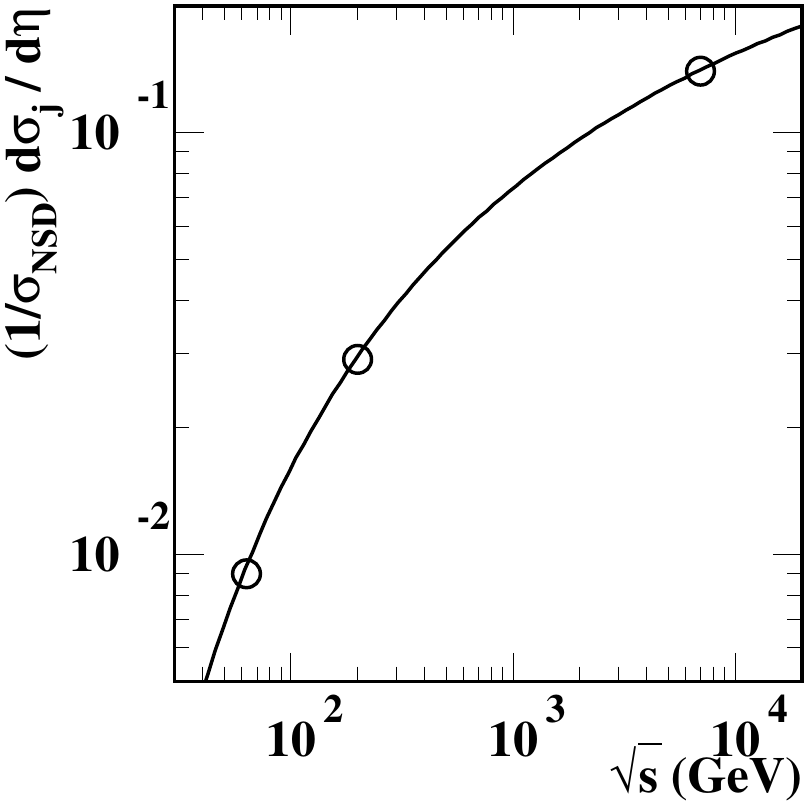}
   \caption{\label{rescale}
   Left: Rescaled jet spectra for several energies~\cite{isrfirstjets,ua1jets,ua2jets} plotted vs normalized fragment rapidity $u$. The data fall on a common curve $0.15 \exp(-25.5 u^2)$. 
 Right: The energy trend of the dijet frequency or dijet $\eta$ density per NSD \pp\ collision $f_{NSD} = (1/\sigma_{\rm NSD}) d\sigma_j /d\eta$ (solid curve) is inferred from a $\sigma_{NSD}(\sqrt{s})$ parametrization in Ref.~\cite{jetspec2} and from Eq.~(\ref{jeteta}).
 }   
    \end{figure}

Figure~\ref{rescale} (left) shows data from Fig.~\ref{ua1data} (left) with the jet spectrum (points) rescaled vertically by factor $(\Delta y_{b})^2$ and parton rapidity $y_{max}- y_{max0}$ rescaled horizontally to $u$ by  $\Delta y_{max}$, with $y_{max0} \approx 3.8$ corresponding to $E_{cut} \approx 3.0$ GeV. All jet data for \pp\ collision energies below 1 TeV fall on a common fitted Gaussian $0.15 \exp(-u^2/2\sigma_u^2)$ (solid curve).
The parton spectrum parametrization, conditional on beam rapidity, is then 
 \bea \label{curious}
 \frac{d^2\sigma_j}{dy_{max} d\eta} &=& p_t \frac{d^2 \sigma_j}{dp_t d\eta}
\\ \nonumber
&\approx& 0.052 (\Delta y_b)^2  \frac{1}{\sqrt{2\pi \sigma^2_u}} e^{-u^2 / 2 \sigma^2_u},
 \eea
where $0.052/\sqrt{2\pi \sigma^2_u} \approx 0.15$ and $\sigma_u \approx 1/7$ are determined empirically from the jet  data.%
\footnote{In Ref.~\cite{jetspec2} the coefficient is given incorrectly as 0.026. Only half the Gaussian is relevant to the jet cross section and therefore integrates to  $\sqrt{2\pi \sigma^2_u}/2$. Equations~(\ref{jeteta}) remain correct.} 
As demonstrated by the curves in Fig.~\ref{ua1data} (left) from Eq.~(\ref{curious}) the jet cross section is represented over nine decades by parameters $y_{b0}$, $y_{max0}$, $\sigma_u$ and $\sigma_X$, the last an overall cross-section scale. Endpoints $y_{b0}$ and $y_{max0}$ are related by kinematic limits on charged-hadron jet production from low-$x$ gluons, while $\sigma_u$ and $\sigma_X$ represent PDF and pQCD $\hat \sigma$ details in Eq.~(\ref{jettspec}).

Integrating Eq.~(\ref{curious}) over $u \in [0,1]$ (first line below) and assuming an effective $4\pi$ $\eta$ interval $\Delta \eta_{4\pi} \approx 1.3 \Delta y_b$~\cite{ua1jets} (second line below) gives 
\bea \label{jeteta}
\frac{d\sigma_j}{d\eta} &=& 0.026 (\Delta y_b)^2 \Delta y_{max}
\\ \nonumber
\sigma_{j0} &\approx & 0.034 (\Delta y_b)^3 \Delta y_{max},
\eea
where the second line defines the solid curve in Fig.~\ref{ua1data} (right). It is notable that the ISR spectrum data were not included in determining the model parameters above. The corresponding curves in Fig.~\ref{ua1data} (left) then serve as predictions that describe the ISR jet data well.

Figure~\ref{rescale} (right) shows a predicted collision-energy trend for the jet  $\eta$ density  per NSD \pp\ collision $f_{NSD} =  (1/\sigma_{\text NSD}) d\sigma_j / d\eta$ based on a parametrization of $\sigma_{\text NSD}(\sqrt{s})$~\cite{jetspec2}  and on Eq.~(\ref{jeteta}) (first line). For 200 GeV \pp\ collisions $f_{NSD} \approx 0.028$   corresponds to $d\sigma_j / d\eta \approx 1$ mb, $\sigma_{j0} \approx 4$ mb and $\sigma_{\text NSD} \approx 36$ mb.  Within the STAR TPC acceptance $\Delta \eta = 2$ the fraction of 200 GeV NSD \pp\ collisions with a dijet is about 6\%. 94\% of \pp\ collisions then have no significant jet activity, thus qualifying as {\em soft events} (i.e.\ no MPIs within the acceptance).

Jet spectrum data described by the model of Eq.~(\ref{curious}) as in Figs.~\ref{ua1data} (left) and \ref{rescale} (left) require a parameter $y_{max0}$ corresponding to $E_{cut} \approx 3$ GeV, a value just below the UA1 (mini)jet spectrum 5 GeV lower bound in Fig.~\ref{ua1data} (left). The 5 GeV represents an estimated limitation on eventwise jet reconstruction, not a physical limit to the jet spectrum. 
As demonstrated below, differential MB jet manifestations in hadron spectra and correlations are more sensitive to the lower bound and confirm an effective physical cutoff near 3 GeV. 
It is notable that the $E_{cut}$ parameter for Eq.~(\ref{curious}) has no apparent energy dependence according to data~\cite{jetspec2}, whereas the PMC MPI spectrum cutoff parameter $p_{\perp 0}$ must vary ``like some power of CM energy''~\cite{sjostrand2} (e.g.\ $\sim E_{CM}^\epsilon$, Sec.~\ref{intro}) to accommodate data.

\subsection{MB dijets and $\bf p_t$-spectrum hard components} \label{spechard}

One can describe or predict {\em fragment distributions} (FDs) via a QCD convolution integral that combines accurate parametrizations of measured \pp\ jet spectra as in Sec.~\ref{jetspecc} and measured \pp\ FFs as in Refs.~\cite{eeprd,jetspec2}. FDs are then directly comparable with \pt\ spectrum hard components as inferred within the context of the TCM.

An ensemble-mean FD for MB dijets is defined by the convolution integral
\bea \label{fold1}
\bar D(y_t) &\approx&   \frac{1}{d\sigma_j/d\eta}  \int_0^\infty \hspace{-.07in}  dy_{max}\, D_\text{pp}(y_t|y_{max})\, \frac{d^2\sigma_j}{dy_{max}d\eta},~~
\eea
where $D_\text{pp}(y_t|y_{max})$ represents measured FFs from \pp\ collisions~\cite{eeprd,jetspec2} and  $d^2\sigma_j / dy_{max}d\eta$ is given by Eq.~(\ref{curious}). Assuming that spectrum TCM hard component $H(y_t)$ represents hadron fragments from MB dijets it can be related to $\bar D(y_t)$ by $y_t H(y_t) \approx  f_{NSD}\,\epsilon\bar D(y_t)$, where $f_{NSD}$ is obtained from Fig.~\ref{rescale} (right)  and $\epsilon \approx 0.6$ within detector acceptance $\Delta \eta = 2$~\cite{jetspec}. 
The charge-density hard component for NSD \pp\ collisions can be expressed in terms of the integral on \yt\ of Eq.~(\ref{fold1})
\bea \label{rhohnsd}
\bar \rho_{h,NSD} &=& \int dy_t  y_t H(y_t) 
\\ \nonumber
&=&f_{NSD}\,  \epsilon 2\bar n_{ch,j},
\eea
where $2\bar n_{ch,j}$ is the mean fragment multiplicity per dijet.

\begin{figure}[h]
   \includegraphics[width=1.65in,height=1.63in]{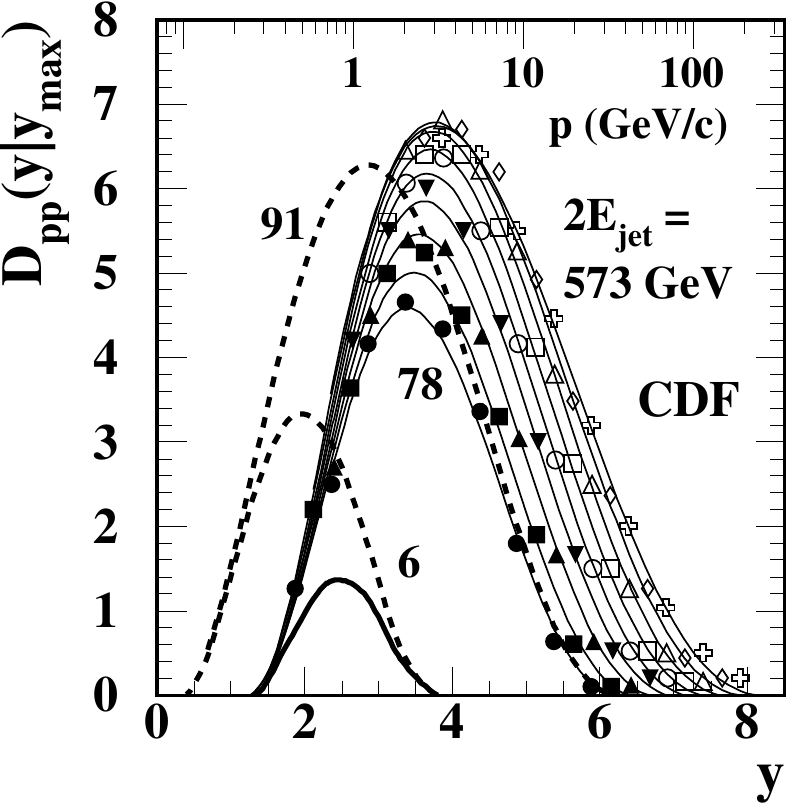}
  \includegraphics[width=1.65in,height=1.63in]{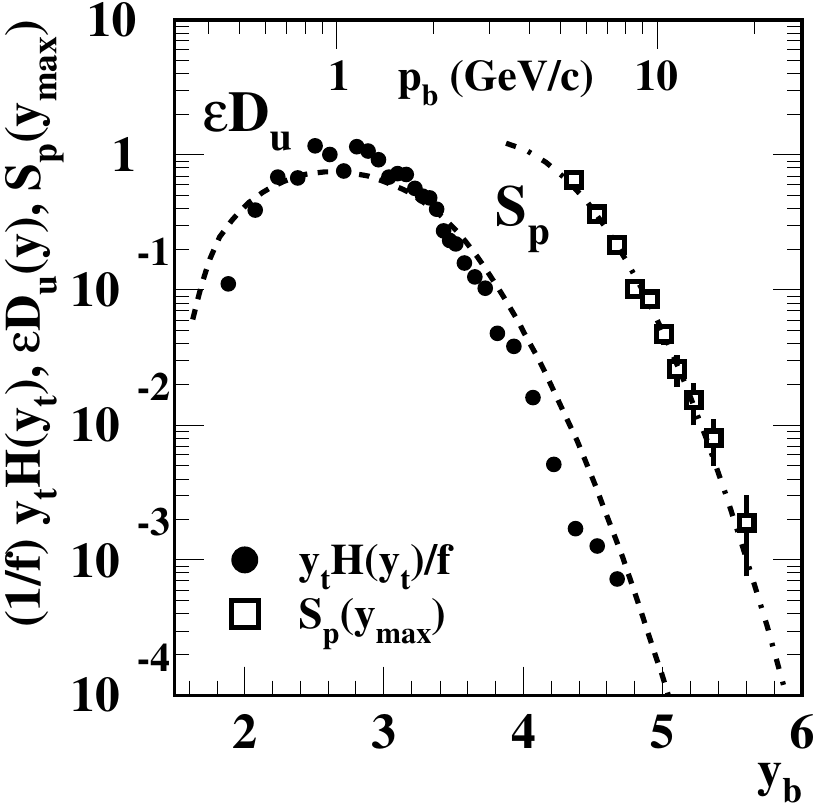}
\caption{\label{ppffs}
Left: Fragmentation functions for several dijet energies (points) from \ppbar\ collisions at 1.8 TeV~\cite{cdfff}. The solid curves represent a \ppbar\ parametrization derived from the \ee\ parametrization in~\cite{eeprd}. The dashed curves show the \ee\ parametrization itself for two energies for comparison.
Right: The spectrum hard component for 200 GeV NSD \pp\ collisions~\cite{ppprd} in the form $y_t H(y_t) / f_{NSD} $ (solid points) compared to calculated mean fragment distribution $\bar D(y_t)$ (dashed) with $y_b = y_t$, $y$ or $y_{max}$~\cite{fragevo}.  The jet spectrum that generated $\bar D(y_t)$ (dash-dotted) is defined by Eq.~(\ref{curious}). The open boxes are 200 GeV \pp\ jet-spectrum data~\cite{ua1jets} transformed to $y_{max}$.
 }  
 \end{figure}

Figure~\ref{ppffs} (left) shows FF data (points) for ten dijet energies from 78 to 573 GeV inferred from 1.8 TeV \ppbar\ collisions (points) using eventwise jet reconstruction~\cite{cdfff}.  The solid curves are the $D_{pp}$ parametrization used in Eq.~(\ref{fold1})~\cite{fragevo}. Comparison with  \ee\ FFs from Ref.~\cite{eeprd} (dashed curves for $2E_{jet} =  6$ and 91 GeV) reveals that a substantial portion of \ee\ dijet FFs at lower fragment momenta may be missing from reconstructed \ppbar\ FFs.

Fig.~\ref{ppffs} (right) shows the corresponding mean FD $\epsilon \bar D(y_t)$ (dashed) described by Eq.~(\ref{fold1}) compared to hard-component data from 200 GeV NSD \pp\ collisions (solid points~\cite{ppprd,ppquad}) in the form $y_t H(y_t) / f_{NSD} $ corresponding to their relation in the text just below Eq.~(\ref{fold1}).  The open boxes are 200 GeV \ppbar\ jet-spectrum data from Ref.~\cite{ua1jets} on $y_{max}$. The dash-dotted curve is $S_p$ from Eq.~(\ref{curious}).

This panel demonstrates that the combination in Eq.~(\ref{fold1}) of a measured jet energy spectrum as in Fig.~\ref{ua1data} (left) and measured \pp\ FFs as in Fig.~\ref{ppffs} (left) accurately describes a measured spectrum hard component from 200 GeV \pp\ collisions~\cite{fragevo}, supporting the interpretation that spectrum hard components  represent the full contribution from MB  large-angle-scattered low-$x$ gluons into angular acceptance $\Delta \eta$, at least for 200 GeV \pp\ collisions. The result depends critically on the choice $E_{cut} \approx 3$ GeV. Similar conclusions are obtained for other energies~\cite{alicetomspec}.

\begin{figure}[h]
  \includegraphics[width=1.67in]{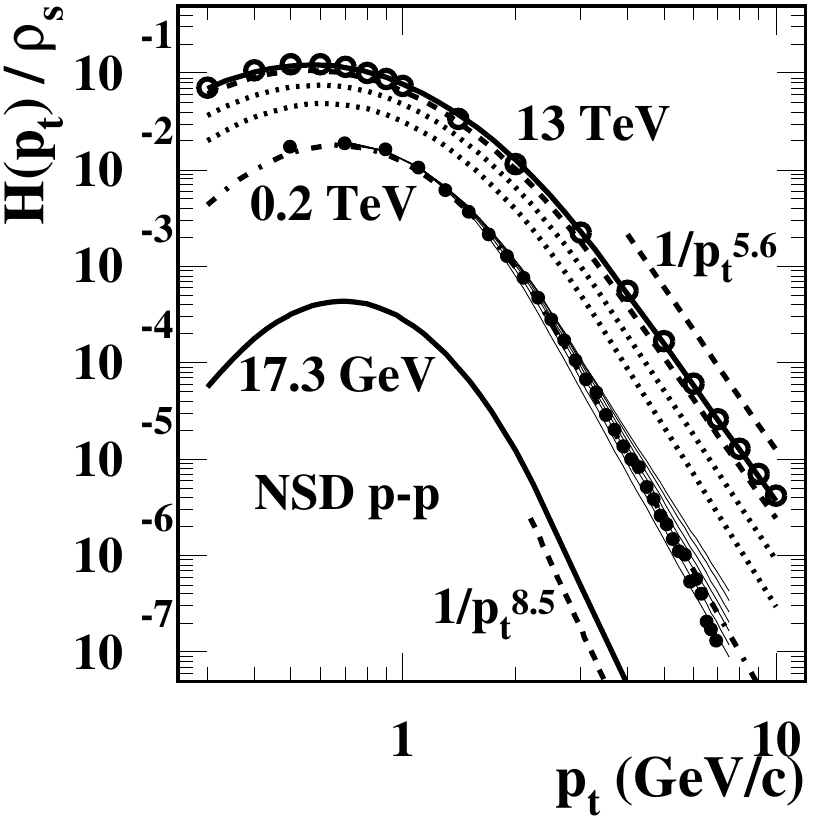}
  \includegraphics[width=1.63in]{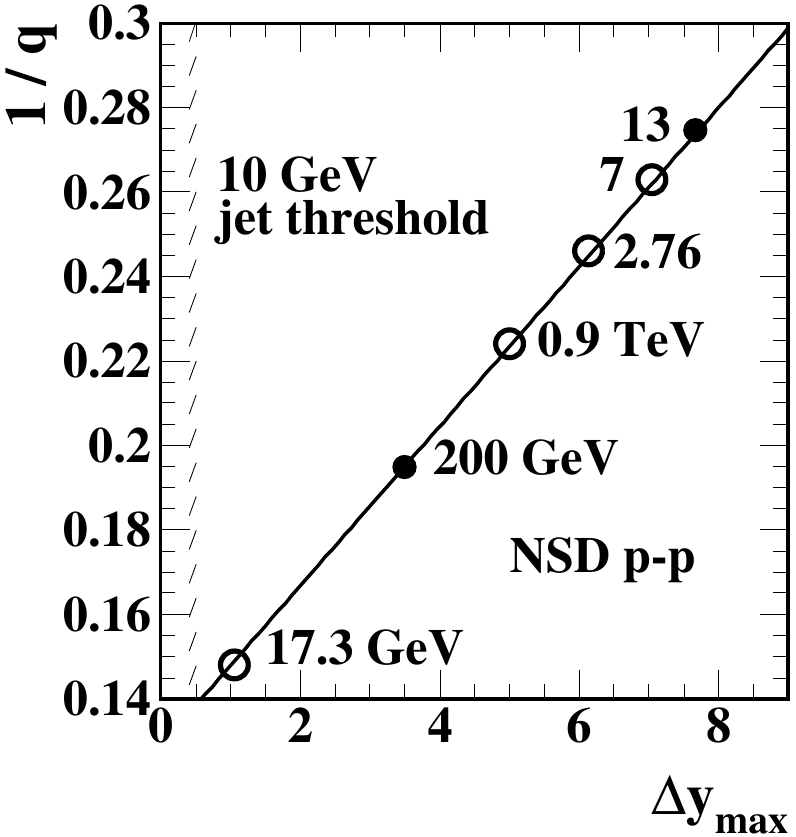}
\caption{\label{enrat3}
Left: A survey of spectrum hard components over the currently accessible \pp\ energy range from threshold of dijet production (10 GeV) to LHC top energy (13 TeV). The curves are determined by TCM parameters for NSD \pp\ collisions from Ref.~\cite{alicespec}. The 200 GeV fine solid curves illustrate \nch\ dependence. The points are from Refs.~\cite{ppquad} (200 GeV) and \cite{alicespec} (13 TeV).
Right: Hard-component exponents plotted in the form $1/q$ determined by analysis of spectrum data (solid points) from Ref.~\cite{ppquad}. The solid line is based on a jet-spectrum parametrization in Ref.~\cite{jetspec2} that also describes ensemble-mean-\pt\ hard-component energy variation~\cite{tommpt}.
 } 
\end{figure}

Figure~\ref{enrat3} (left) shows measured spectrum hard components in the form $H(p_t,\sqrt{s}) / \bar \rho_s(\sqrt{s})$ (points) for 200 GeV and 13 TeV NSD \pp\ collisions representing the spectrum hard component  {\em per soft-component hadron} corresponding (by hypothesis) to dijet production per participant low-$x$ gluon. The curves are TCM model functions in the form $\alpha(\sqrt{s}) \bar \rho_s(\sqrt{s}) \hat H_0(p_t,\sqrt{s})$ with $\hat H_0(p_t,\sqrt{s})$ energy-dependent parameters $(\bar y_t,\sigma_{y_t},q)$~\cite{alicespec}. Isolated hard components establish spectrum energy evolution and its relation to dijet production.  The overall result is a comprehensive description of dijet contributions to \pt\ spectra vs \pp\ energy variation over three orders of magnitude~\cite{alicetomspec}.

 Figure~\ref{enrat3} (right) shows inverse values (solid points) of exponents $q = 5.15$ for 200 GeV and $q = 3.65$ for 13 TeV from Ref.~\cite{alicetomspec} plotted vs quantity $\Delta y_{max} \equiv \ln(\sqrt{s} / \text{6 GeV})$ observed to describe the energy trend for jet spectrum widths $\propto \Delta y_{max}$ from NSD \pp\ collisions assuming a jet spectrum low-energy cutoff near 3 GeV~\cite{jetspec2} as in Fig.~\ref{rescale} (left). Inverse $1/q$ effectively measures the hard-component peak width at larger \yt. The description of 5 TeV \ppb\ spectra in Sec.~\ref{aa1} is based on that relation. Since the \pp\ \pt-spectrum hard component can be expressed as the convolution of a {\em fixed} \pp\ FF ensemble with a collision-energy-dependent jet spectrum~\cite{fragevo}, and the jet-spectrum width trend has an energy dependence $ \propto  \Delta y_{max}$~\cite{jetspec2} the relation $1/q \propto  \Delta y_{max}$ (solid line) should be expected.  That the same relation applies to the ensemble-mean \mmpt\ hard component was established in a separate study~\cite{tommpt} (and see the next subsection).

In summary, direct comparison of measured \pt\ spectrum hard components with measured jet spectra and  FFs combined in a convolution integral confirms an effective lower bound $\approx$ 3 GeV for MB jet spectra and a dijet frequency $f_{NSD} \approx 0.028$ per unit $\eta$ and per NSD event for 200 GeV \pp\ collisions, in contrast to the PMC assumptions of at least one MPI per inelastic \pp\ collision and energy-dependent $p_{\perp0}$~\cite{sjostrand2}. Multiple dijets per unit $\eta$ per \pp\ collision {\em do} occur if a large  \pp\ \nch\ condition is imposed (e.g.\ 100-fold increase for data in Refs.~\cite{ppprd,ppquad}).

\subsection{MB dijets and $\bf \bar p_t$-vs-$\bf n_{ch}$ trends} \label{mpttrends}

The relation of ensemble-mean \mmpt\ vs \nch\ data to the hadron \pt\ spectrum TCM in Sec.~\ref{aa1} and MB dijets in the previous subsection is demonstrated in this subsection. The TCM for ensemble-mean integrated total $\bar P_t$ within acceptance $\Delta \eta$ from \pp\ collisions for given $(n_{ch},\sqrt{s})$ follows from the \pt\ spectrum TCM in Eqs.~(\ref{a1x})
\bea \label{mptsimple}
P_t &=& \Delta \eta \int_0^\infty dp_t\, p_t^2\, \bar \rho_0(p_t) =  P_{ts} +  P_{th}
\\ \nonumber
\bar P_t &=& n_s \bar p_{ts} + n_h \bar p_{th},
\eea
where $\bar p_{ts}$ and $\bar p_{th}$ can be obtained directly from TCM model functions $\hat S_0(p_t)$ and $\hat H_0(p_t)$ in Eq.~(\ref{a1x}) or inferred from \mmpt\ data systematics as described below.
The conventional intensive ratio of extensive quantities
\bea \label{ppmpttcm}
\bar p_t' \equiv \frac{\bar P_t'} {n_{ch}'} &\approx & \frac{\bar p_{ts} + x(n_s) \bar p_{th}(n_s)}{\xi + x(n_s)}
\eea
conflates two simple TCM trends and in effect partially cancels MB dijet manifestations apparent in the form of $x(n_s) \equiv \bar \rho_h / \bar \rho_s = n_h / n_s \approx \alpha \bar \rho_s$. Primes indicate the effect of a \pt\ acceptance cutoff.  Alternatively, the ratio
\bea \label{niceeq}
\frac{n_{ch}'}{n_s} \bar p_t'   \approx \frac{ \bar P_t}{n_s} &= & \bar p_{ts} + x(n_s) \bar p_{th}(n_s)
\\ \nonumber
&\approx& \bar p_{ts} + \alpha(\sqrt{s})\, \bar \rho_s \, \bar p_{th}(n_s,\sqrt{s})
\eea
preserves the simplicity of Eq.~(\ref{mptsimple}) and provides a convenient basis for precise tests of the TCM hypothesis, for instance in Fig.~\ref{ppcomm} (right).

\begin{figure}[h]
   \includegraphics[width=1.65in,height=1.6in]{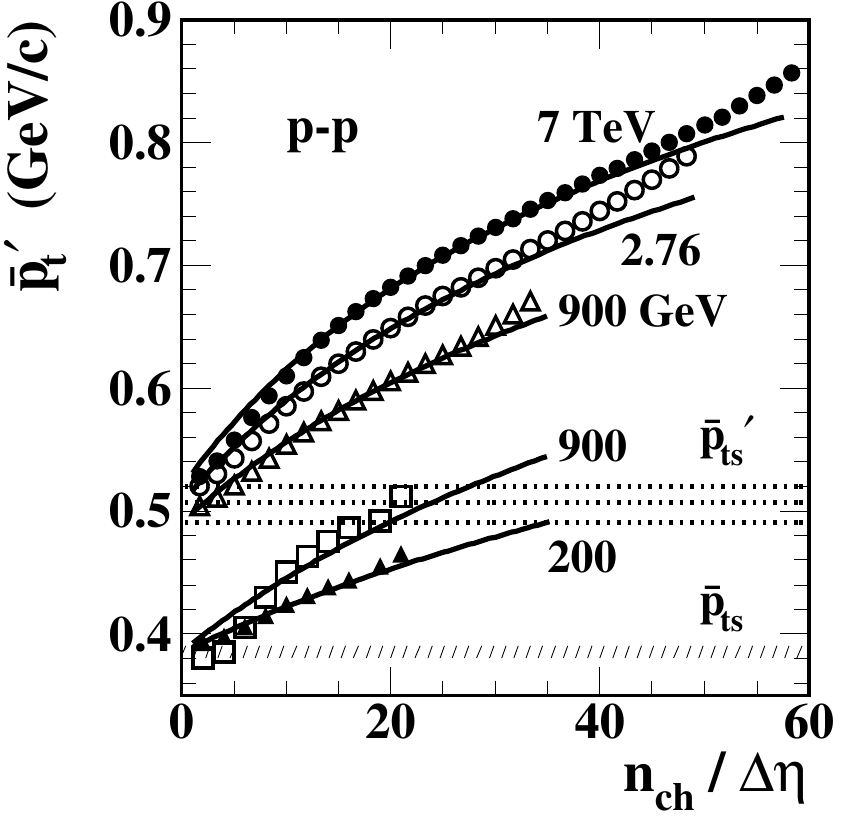}
   \includegraphics[width=1.65in,height=1.6in]{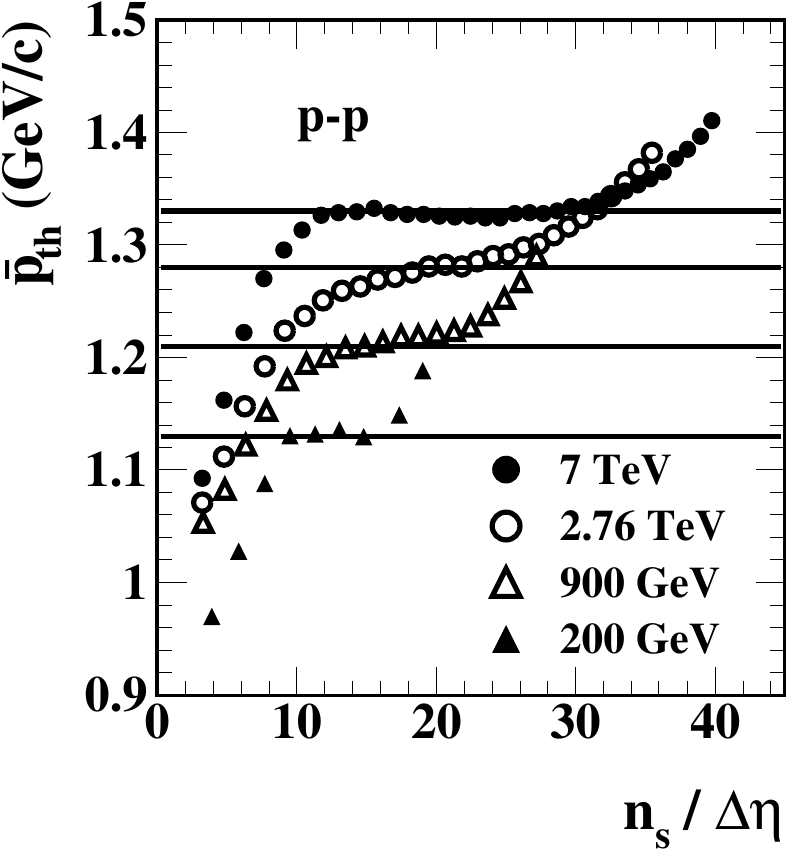}
 \caption{\label{alice5a}
 Left: \mmpt\ vs \nch\ for several collision energies. The upper group of points is from Ref.~\cite{alicempt}. The lower 900 GeV data from UA1 derived from a ``power-law'' spectrum model~\cite{ua1mpt} fall significantly above the TCM for that energy (solid curve) but are consistent with the TCM form with amplitude adjusted. The 200 GeV STAR data are spectrum integrals from Ref.~\cite{ppprd}.
 Right: Hard components  $\bar p_{th}(n_s)$ (points) isolated from data at left per Eq.~(\ref{niceeq}). The horizontal lines represent mean values $\bar p_{th0}(\sqrt{s})$ extracted from \pp\ spectra (Fig.~\ref{enrat3}, left). The  $\bar p_{th}(n_s)$ data vary significantly about those mean values as expected from results in Ref.~\cite{alicetomspec}.
} 
  \end{figure}

Figure~\ref{alice5a} (left) shows $\bar p_t$ data for four \pp\ collision energies from 
the RHIC  (solid triangles~\cite{ppprd}),
the Sp\=pS  (open squares~\cite{ua1mpt})
and the LHC (upper points~\cite{alicempt}) increasing monotonically with charge density $\bar \rho_0 = n_{ch} / \Delta \eta$. The lower points and curves correspond to full \pt\ acceptance.  For acceptance extending down to zero ($\xi = 1$), $\bar p_t' \rightarrow \bar p_t$ in Eq.~(\ref{ppmpttcm}) should vary between the universal lower limit $\bar p_{ts} \approx 0.40$ GeV/c ($n_{ch} \rightarrow 0$) and  $\approx \bar p_{th0}$ ($n_{ch} \rightarrow \infty$) as limiting cases. For a lower-\pt\ cut $p_{t,cut} > 0$ the lower limit is $\bar p_{ts}' = \bar p_{ts} / \xi$ (upper dotted lines) and the data are systematically shifted upward (upper points and curves).  Solid curves represent the \pp\ $\bar p_t$ TCM Eq.~(\ref{ppmpttcm})~\cite{tommpt}.

Figure~\ref{alice5a} (right) shows the result when, following Eqs.~(\ref{niceeq}), $\bar p_t'$ data in the left panel are multiplied by $n_{ch}' / n_{s}$, $\bar p_{ts} \approx 0.4$ GeV/c is subtracted from the product and the difference is divided by $x(\sqrt{s}) \equiv \alpha(\sqrt{s}) \, \bar \rho_s$ to obtain $\bar p_{th}(n_s,\sqrt{s})$ vs $\bar \rho_s$.  The solid lines represent the $\bar p_{th0}$ derived from TCM spectrum hard components as in Fig.~\ref{enrat3} (left). Values for $\alpha(\sqrt{s})$ from Ref.~\cite{tommpt} are consistent with the correspondence between $\bar p_{th}(n_s,\sqrt{s})$ data (points) and $\bar p_{th0}$ values (lines) as shown.  The $\alpha(\sqrt{s})$ values for various energies and detector systems [parametrized by Eq.~(\ref{a3})] are quantitatively consistent to a few percent given the differences in acceptance $\Delta \eta$ between different detectors. 
Some  \nch\ dependence of $\bar p_{th}$ mean values is  expected based on results from Refs.~\cite{ppprd,alicetomspec} [see the corresponding 200 GeV  \nch\ dependence (thin solid curves) in Fig.~\ref{enrat3} (left)]. 

\begin{figure}[h]
   \includegraphics[width=3.3in,height=1.6in]{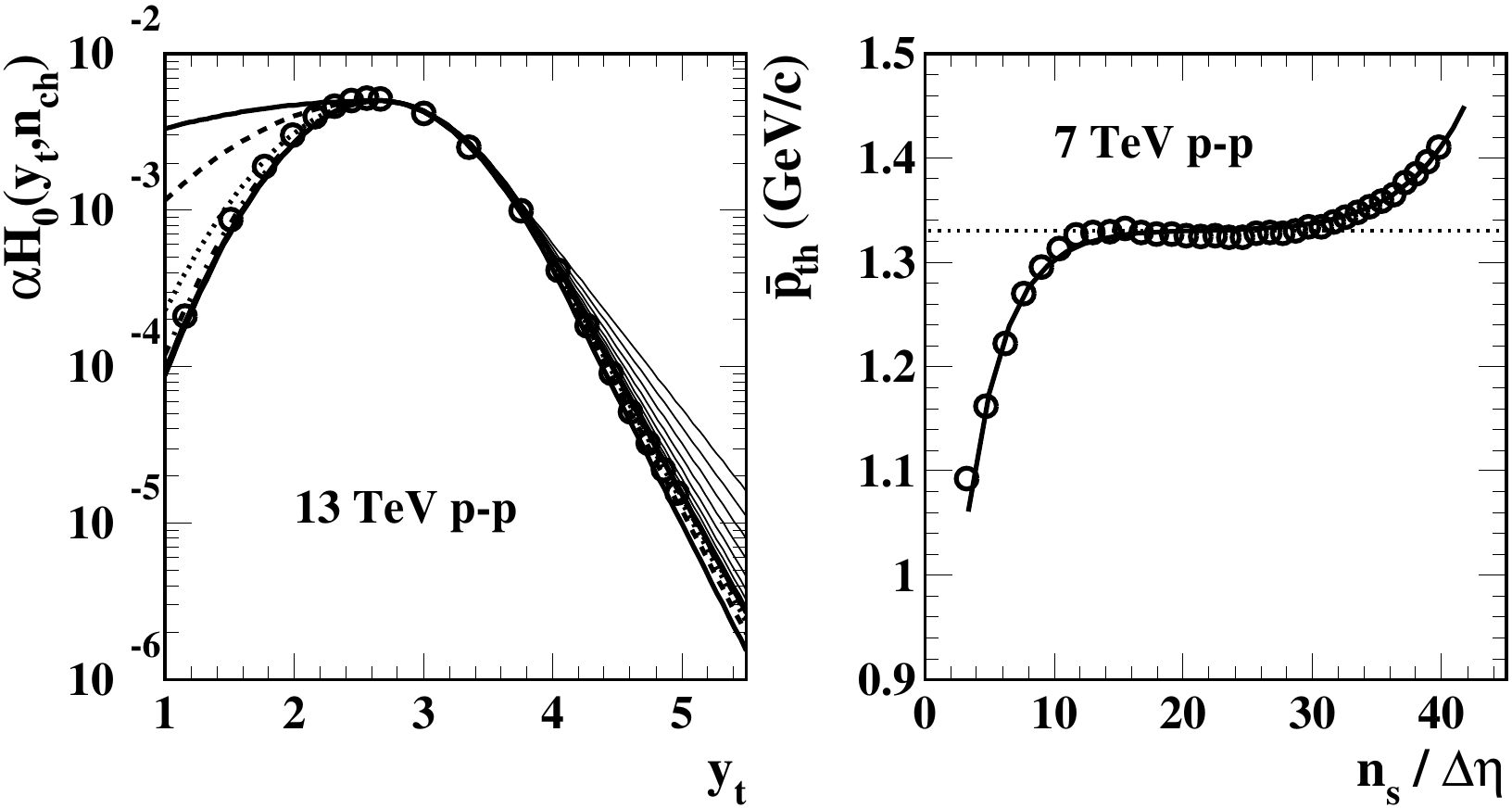}
   \caption{\label{300a}
Left: \pt\ spectrum hard-component models for 13 TeV \pp\ collisions and a range of \nch\ based on a TCM parametrization from  Ref.~\cite{alicetomspec} compared to spectrum data from Ref.~\cite{alicespec} for inelastic \pp\ collisions (points).
 Right: $\bar p_{th}(n_s)$ values inferred from 7 TeV  \pp\ \mmpt\ data as in Ref.~\cite{tommpt} (points) compared to a $\bar p_{th}(n_s)$ trend (curve) inferred from the model functions in the left panel.
 }   
  \end{figure}

Figure~\ref{300a} (left) shows  13 TeV TCM hard-component models for several \pp\ multiplicity classes (curves) as reported in Ref.~\cite{tommpt}. The points are 13 TeV \pt\ spectrum data from Ref.~\cite{alicespec}.  The model-parameter variation for 7 TeV is determined by expressions  interpolated from 13 TeV  to 7 TeV and extrapolated on $\bar \rho_s$ from the rather limited 13 TeV multiplicity range in Ref.~\cite{alicespec}.

Figure~\ref{300a} (right) shows 7 TeV $\bar p_{th}(n_s)$ data from Fig.~\ref{alice5a} (right) (points). The curve is determined by  $\bar p_{th}(n_s)$ values obtained from  spectrum model functions in the left panel. 
The correspondence between \mmpt\ data and TCM is good. In turn, there is quantitative correspondence between \pp\ \pt\  spectrum hard components and {\em measured} jet properties accurate at the percent level as noted in the previous subsections. These detailed TCM results further buttress the conclusion that ensemble-mean $\bar p_t$ variation is completely determined by MB dijets.

In summary, the strong increase of \mmpt\ with \nch\ as in Fig.~\ref{alice5a} (left) is simply described within the TCM by linear variation of relative fractions of {\em fixed} soft and hard spectrum components~\cite{tommpt}. The hard component is derived quantitatively from a measured MB jet spectrum (with effective cutoff near 3 GeV determined by spectrum data) and measured FFs~\cite{fragevo,alicetomspec}. The linear hard/soft ratio variation arises from the noneikonal nature of \pp\ collisions that argues against any role for \pp\ centrality. In contrast, variation of \mmpt\ with \nch\ must be modeled in the PMC by MPIs as the single source of hadrons. To achieve monotonic increase of \mmpt\ an {\em ad hoc} CR mechanism is then required. The CR mechanism in effect alters FFs (modeled by string fragmentation) by decreasing the \nch\ per MPI, supposedly to accommodate an increasing MPI density attributed to increased \pp\ centrality~\cite{ppcent2}. 

\subsection{NBD 1/k and MB jet-spectrum lower limit} \label{1overk}

Further information on the effective lower limit of MB jet spectra can be derived from the systematics of $P(n_{ch})$ distributions, specifically from fitted NBD model parameters $(\mu,k)$.  As noted in Sec.~\ref{pnch} $P(n_{ch})$ distributions for \pp\ collisions up to 1 TeV are well-described by a single NBD distribution. Above that energy and for larger \nch\ a double NBD is required~\cite{aliceppmult}, but for the present argument single-NBD fits closer to the mode are sufficient. In this subsection the energy dependence of NBD parameter $k$ is used to support the conclusion that the MB jet spectrum has an effective lower limit near 3 GeV.

Within a TCM context the \pp\ charge multiplicity trend $n_{ch}(\sqrt{s})$ can be approximated at higher energies as follows~\cite{alicetomspec}: The total charge density near midrapidity is  $\bar \rho_0 \equiv n_{ch} / \Delta \eta = \bar \rho_s + \bar \rho_h$, where  $\bar \rho_s \approx 0.81 \ln(\sqrt{s} / \text{10 GeV})$ (near and above ISR energies) is interpreted to represent participant low-$x$ gluons from proton dissociation. The hard component is  $\bar \rho_h \approx \alpha \bar \rho_s^2$ with $\alpha(\sqrt{s}) \approx O(0.01)$ as described in Ref.~\cite{tommpt} and Eq.~(\ref{a3}). For ISR and Sp\=pS energies $\bar \rho_h \ll \bar \rho_s$ so $\bar \rho_0 \approx \bar \rho_s$. As noted in Sec.~\ref{pnch} Ref.~\cite{guidokno} reports the relation $1/k \approx 0.06 \ln(\sqrt{s} / \text{6 GeV})$ (consistent with $E_{cut} \approx 3$ GeV). From Sec.~\ref{jetspecc} those expressions can be rewritten as $\bar \rho_0 \approx 0.81 \Delta y_b$ and $1/\hat k \approx 0.48 \Delta y_{max}$, where $\hat k \equiv k/\Delta \eta$ is a value averaged over the acceptance (assuming UA5 acceptance $\Delta \eta_{4\pi} \approx 8$~\cite{ua5}).

A number variance $\sigma_n^2 \equiv \overline{n^2} - \bar n^2$ can be interpreted to represent the number of correlated pairs within some acceptance $\Delta \eta$. For a NBD the variance is expressed as
\bea \label{variancen}
\sigma_n^2 &=& \bar n + \bar n^2 / k.
\eea
The first term is the number of self pairs, the Poisson reference $\sigma^2_{n,ref} = \bar n$. The second term, the {\em variance excess}, represents the effective number of correlated pairs. Since $\bar n^2$ is the total number of pairs (given ensemble mean $\bar n$) $1/k$ is the {\em fraction} of those pairs correlated within $\Delta \eta$. While $k$ does increase monotonically with $\Delta \eta$~\cite{guidokno} it is related to a running integral of underlying angular correlations~\cite{inverse,ptscale}. $\hat k$ may vary significantly with $\Delta \eta$.

Returning to \pp\ collisions, the number of correlated pairs represented by an NBD variance {\em excess} $\Delta \sigma^2_{n_{ch}}$ is
\bea \label{varianceexcess}
\Delta \sigma^2_{n_{ch}}  &=&   \sigma^2_{n_{ch}}  - \sigma^2_{n_{ch},ref} = \bar n_{ch}^2 / k 
\\ \nonumber
&\approx & \Delta \eta\, \bar \rho_0^2 / \hat k 
\\ \nonumber
\Delta \sigma^2_{n_{ch}} / \Delta \eta &\approx &  0.3\,  \Delta y_b^2 \, \Delta y_{max}
\eea
based on text above Eq.~(\ref{variancen}). The coefficient 0.3 is an upper limit based on UA5 acceptance $\Delta \eta_{4\pi} \approx 8$. The relevant value could be half that. Equation (\ref{varianceexcess}) (third line) can be compared with the jet cross section in Eq.~(\ref{jeteta}) (first line)  
\bea
d\sigma_j / d\eta &\approx& 0.026 \Delta y_b^2 \Delta y_{max}
\eea
to conclude that at least in terms of energy dependence 
\bea \label{equivalence}
\Delta \sigma^2_{n_{ch}} / \Delta \eta \approx \text{(5 - 10)}\times d\sigma_j / d\eta.
\eea
This result suggests that MB dijets may be the main source of nonPoisson \nch\ fluctuations as manifested by a variance excess for distribution $P(n_{ch})$. The trend for $1/k$ inferred from ISR and Sp\=pS data as reported in Ref.~\cite{guidokno} is consistent with  $1/k \propto \Delta y_{max} \equiv \ln(\sqrt{s} / 2 E_{cut})$ {\em if} $E_{cut} \approx 3$ GeV. Thus, a specific jet spectrum cutoff near 3 GeV  is supported by measured jet spectra, by \pp\ hadron \pt\ spectra and by charge multiplicity fluctuations.

\subsection{Double parton scattering}

The PMC assumption that almost all hadrons arise from MPIs (multiple parton scatters {\em per NSD event}) in high-energy \pp\ collisions appears to conflict with the systematics of MB dijets manifesting as the TCM hard component, dominated by a single dijet per hard event and fraction of hard events $\ll 1$ within an NSD event ensemble and typical acceptance $\Delta \eta$~\cite{pptheory}. However, the TCM description does not exclude double parton scattering (DPS) occurring with some (possibly small) probability, and \pp\ DPS has been demonstrated for example in Ref.~\cite{doubleparton}. In this subsection evidence for DPS and its properties from 1.8 TeV \pp\ collisions are reviewed.


For the DPS study in Ref.~\cite{doubleparton} the applied DPS event trigger is $\gamma$ + 3 jets. Single events with that trigger are compared with pileup event pairs satisfying single-parton-scattering (SPS) triggers: $\gamma$ + jet vs jet + jet. The DPS probability can then be expressed as
\bea \label{pdp}
P_{DP} &=& \frac{\sigma_{DP}}{\sigma_{NSD}} \equiv \frac{m}{2} \frac{\sigma_{NSD}}{\sigma_\text{eff}}\frac{\sigma_{\gamma j}}{\sigma_{NSD}}\frac{\sigma_{jj}}{\sigma_{NSD}},
\eea
where $\sigma_{NSD}$ is a reference NSD cross section, and $\sigma_\text{eff}$ is a defined effective cross section for DPS events.
Factor 1/2 corresponds to Poisson-distributed events and factor $m$ denotes distinguishable (2) vs indistinguishable (1) hard scatters. Experimentally, $m = 2$ is established and the two factors cancel. The corresponding expression for single hard scatters in a pair of pileup events (DI) is
\bea
P_{DI} &=& \frac{\sigma_{DI}}{\sigma_{NSD}} = 2  \frac{\sigma_{\gamma j}}{\sigma_{NSD}}\frac{\sigma_{jj}}{\sigma_{NSD}}
\eea
where factor 2 is the number of ways $\gamma$-jet and dijet processes can be ordered within the DI pair. For an integrated luminosity $\mathcal L t \approx 16/10^{-12} \text{b}$ event numbers $N_{DP} = 7360$ and $N_{DI} = 1060$ were observed, giving an inferred $\sigma_\text{eff}$ cross section~\cite{doubleparton}
\bea \label{eff}
\sigma_\text{eff} &=&  \frac{P_{DI}}{P_{DP}} \, \frac{\sigma_{NSD}}{2}
\\ \nonumber 
&=&\frac{N_{DI}}{N_{DP}} \frac{A_{DP}}{A_{DI}}R_c \, \sigma_{NSD}
\\ \nonumber
&\approx& 1/6.9 \times 1.04 \times 2 \times 50 ~\text{mb} \approx  15~\text{mb}.
\eea

To place this DPS result in a TCM context requires the following numbers for 1.8 TeV \pp\ collisions (interpolated from Ref.~\cite{jetspec2}): $\sigma_{NSD} \approx 50$ mb,
$d\sigma_j / d\eta \approx 5$ mb $\rightarrow d\sigma_{SP} / d\eta$ and $f_{SP}\equiv (1/\sigma_{NSD})d\sigma_{SP} / d\eta \approx 0.1$. It is clear from Fig.~\ref{etatcm} (right) that MB dijets are localized near midrapidity, so a cross-section {\em density} is more appropriate for describing low-energy jets. Expressing Eq.~(\ref{pdp}) in terms of dijet frequency $f$ (defined in Ref.~\cite{ppprd})
\bea
f_{DP} &\equiv& (1/\sigma_{NSD}) d\sigma_{DP} / d\eta
\\ \nonumber
&=& \frac{1}{f_\text{eff}} f_{SP}^2
\\ \nonumber
&\approx&  5 \text{-} 10 \times (0.1)^2
\\ \nonumber
&\approx& 0.05 \text{-} 0.1
\\ \nonumber
d\sigma_{DP}/ d\eta &\approx& 2.5 \text{-} 5 ~\text{mb}
\eea
where $f_\text{eff} = (1/\sigma_{NSD})d\sigma_\text{eff}/ d\eta \approx 0.1 \text{-} 0.2$ is estimated from the result in Eq.~(\ref{eff}). Those numbers suggest that $f_{DP} \ll 1$ describes 1.8 TeV \pp\ collisions. It is then unlikely that MPIs (especially {\em multiple} hard scatters per event) play a dominant role in hadron production from NSD \pp\ collisions near midrapidity. Multiple dijets {\em are} expected for \pp\ events with large \nch\ because of the quadratic dependence of $d\sigma_j / d\eta$ on charge multiplicity.


\section{underlying event $\bf vs$ MB dijets} \label{uembjets}

The UE for high-energy \pp\ collisions is by definition complementary to an eventwise-triggered dijet or pQCD leading-order process~\cite{under5}. Access to the UE is expected via the transverse (azimuth) region or  TR component of the single-particle charge density relative to the trigger (or of 1D azimuth correlations as in Fig.~\ref{minijet}, right), assumed to have no contribution from the triggered dijet. However, the structure of measured MB dijet 2D angular correlations from 200 GeV \pp\ collisions can be used to demonstrate a significant {\em triggered}-dijet contribution to the TR, contradicting UE-related assumptions. 

The relation between an underlying event and TCM descriptions of \pp\ data was considered previously in Ref.~\cite{pptheory} which distinguishes  different responses to \nch\ and $p_{t,trig}$ event selection conditions. An \nch\ condition controls the soft-component charge density $\bar \rho_s$ and therefore the dijet production rate as measured by hard component $\bar \rho_h \propto \rho_s^2$. In contrast, a $p_{t,trig}$ condition selects the fraction of {\em hard events} (at least one jet in the acceptance) vs {\em soft events} (no jet in the acceptance) without changing the soft-component density significantly and therefore without changing MB dijet production, although it does alter (bias) the {\em effective} jet spectrum (see Sec.~\ref{nperpspec}).

\subsection{Conventional underlying-event analysis}

The UE study reported in Ref.~\cite{under5} considers charge-jet evolution and properties of the UE within an acceptance $p_t > 0.5$ GeV/c and $|\eta| < 1$. Trigger (leading) jets (jet in each event with greatest $P_{T1} = \sum_{i \in jet} p_{ti}$) fall in the range $P_{T1} \in [0.5,50]$ GeV/c.  The UE is assumed to consist of beam-beam remnants, initial-state radiation and possibly MPIs (but distinct from the leading jet). The leading-jet axis is an azimuth reference relative to which other hadrons are distributed. The transverse region (TR) subtending $|\phi - \pi/2| < \pi/6$ is said to be  ``very sensitive to the underlying event'' and is assumed to exclude the leading jet (and its partner). The UE accompanying a hard scatter is  said to be ``considerably more active (i.e.\ higher charged particle density and more transverse momentum) than a soft [\ppbar] collision''~\cite{under5}, which defines the term ``activity'' as associated with the UE within the PMC context.

UE-related trends include TR-integrated charge $N_\perp$ vs leading-jet $p_{T1}$ and TR \pt\ spectrum $dN_\perp / dp_t$. UE data are modeled by several Monte Carlos (PYTHIA, ISAJET, HERWIG). ``For PYTHIA we include particles that arise from the soft or semi-hard scattering in multiple parton interactions [MPIs] in the beam-beam remnant component''~\cite{under5}. That strategy is required if no soft component is available. Simulated hard scattering is limited to $p_{t,hard} > 3$ GeV/c. MPIs extend to lower momenta in PYTHIA (1.4 for V6.115 or 1.9 for V6.125 GeV/c). In what follows jet-related correlation structure on $\phi$, $N_\perp$ vs $p_{T1}$ and $dN_\perp / dp_t$ data are reexamined within the context of the TCM, with focus on the PMC.
 
\subsection{MB dijets from $\bf p$-$\bf p$ collisions at 200 GeV} \label{notr}

\begin{figure}[h]
 \includegraphics[width=3.3in,height=1.6in]{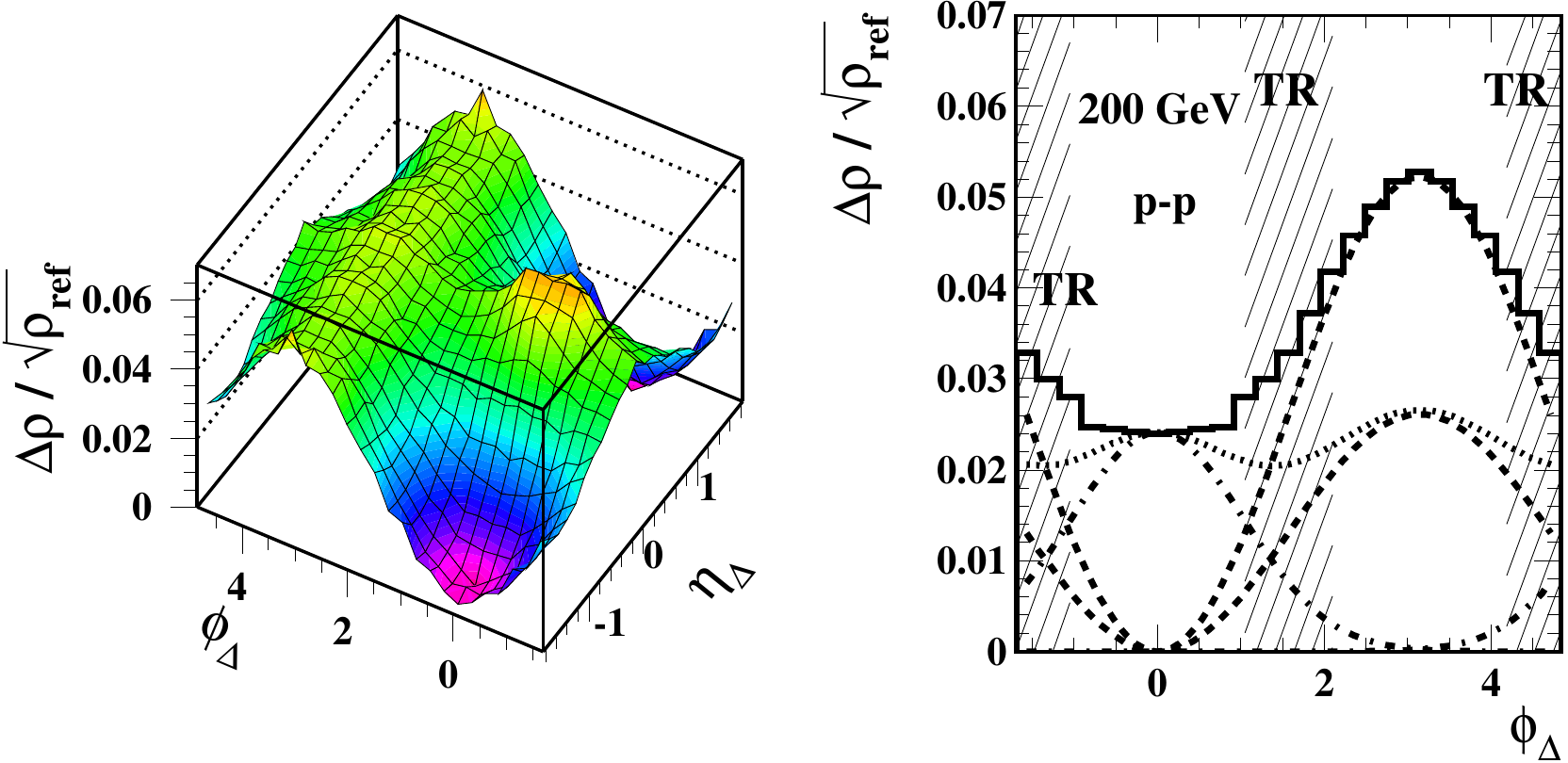}
 \caption{\label{minijet}
 (Color online)
  Left: 
  Minimum-bias jet-related 2D angular correlations from 200 GeV \pp\ collisions~\cite{ppquad}. A same-side 2D peak at the origin is elongated on azimuth. The away-side 1D peak is broad on azimuth and approximated by a dipole $\cos{\phi_\Delta}$ form. These data are corrected for finite $\eta$ acceptance.
  Right: 
  Projection by averaging of jet-related angular correlations onto azimuth. Dash-dotted and upper dashed curves are fitted models for SS and AS peaks. The hatched areas represent the ``transverse region'' (TR) invoked in underlying-event studies. The unhatched areas, denoting ``toward'' ($\phi_\Delta \approx 0$) and ``away'' ($\phi_\Delta \approx \pi$) regions, are conventionally assumed to contain all {\em triggered} dijet structure. The lower dashed curve describes uncorrected pairs and the dotted curve is the actual distribution of jet-related pairs.
 }  
\end{figure}

Figure~\ref{minijet} (left) shows measured 2D angular correlations representing  MB dijets from 200 GeV \pp\ collisions~\cite{ppquad}. No trigger condition is imposed---the distribution represents all combinatoric pairs above a \pt\ acceptance cut at 0.15 GeV/c (accepting 80\% of the soft component and all of the hard component). Contributions from a soft component (1D Gaussian on $\eta_\Delta$), Bose-Einstein (BE) correlations (narrow 2D exponential at origin) and uniform background are subtracted based on 2D model fits that describe data within statistical uncertainties~\cite{ppquad}. The data are corrected for finite $\eta$ acceptance.

For conventional FB analysis (Sec.~\ref{fbcorr}) the quantity $b_{FB}(\eta_\Delta)$ represents a projection of the 2D histogram at left onto 1D $\eta_\Delta$ (symbol  $\Delta \eta$, here denoting a detector acceptance, is also referred to as ``$\eta$ gap''). The 1D projection then superposes several correlation components: (a) the TCM soft component -- a narrow 1D peak on $\eta_\Delta$, (b) the SS 2D jet peak (in projection also a narrow 1D peak on $\eta_\Delta$), (c) the 2D BE peak, (d) the AS 1D jet peak at $\pi$ on azimuth contributing a constant offset to $b_{FB}(\eta_\Delta)$. Each component has its own distinctive charge combination (LS, US or neutral) that helps to identify its source. Within the 1D projection of $b_{FB}(\eta_\Delta)$ there is no possibility to unravel the several production mechanisms.

Figure~\ref{minijet} (right) shows the 2D data projected by averaging onto 1D azimuth (solid histogram). Both the SS 2D peak (dash-dotted) and the AS 1D peak (upper dashed) are broad on azimuth. The AS peak width is approximately $\pi/2$, and the AS periodic peak array~\cite{tzyam} is then approximated by an azimuth dipole extending into the SS region. The SS 2D peak for MB dijets in NSD \pp\ collisions is elongated on azimuth with 2:1 aspect ratio. Thus,  {\em untriggered} SS and AS MB jet peaks are strongly overlapping on azimuth within the TR.  The AS pair structure -- azimuth dipole $\cos(\phi_\Delta - \pi)$ -- is modified by a triangular $\eta$-acceptance correction that overestimates {\em accepted} AS pairs relative to the SS jet peak. The lower dashed curve describes accepted jet-related AS pairs, and the dotted curve is the distribution of all jet-related pairs.


The TR invoked in UE studies and indicated by the hatched regions in the right panel (covering 1/3 of the azimuth acceptance) is conventionally assumed to contain no contribution from a triggered high-$p_t$ (di)jet (if a triggered jet is confined to a cone of radius $R < 1$) and should therefore be particularly sensitive to the UE complementary to the dijet~\cite{under5}. 
Figure~\ref{minijet} reveals that the TR must include a substantial fraction (about 30\%) of the fragment yield from MB dijets. The TR cannot be distinguished from any other part of the MB jet structure and may be dominated by the TCM hard component (MB jet fragments). Since a hard event includes by definition at least one dijet with mean fragment multiplicity $\epsilon 2\bar n_{ch,j}$ the TR in hard events should include a hard-component jet-fragment density corresponding to at least that multiplicity that can be inferred from data. 

Compared to a MB jet sample eventwise triggering of higher-\pt\ jets {\em would} add higher-momentum hadrons successively closer to $\phi_\Delta = 0$ and $\pi$ (e.g.\ within a conventional jet cone with radius $R \leq 1$). However, the trigger condition would not eliminate the MB base in Fig.~\ref{minijet}, common to {\em any} dijet, that should contribute a TCM hard component as {\em part of the triggered jet} to any UE observable  contrary to UE assumptions. That conclusion is supported by \pt\ spectra for transverse multiplicity $N_\perp$ within the TR as discussed in Sec.~\ref{nperpspec}.

Similar issues emerge for trigger-associated (TA) studies of jet-related azimuth correlations where a {\em zero-yield-at-minimum} (ZYAM) assumption is invoked: the jet fragment distribution on azimuth relative to a high-\pt\ trigger hadron is assumed to fall to zero at a minimum in the pair distribution (near or in the TR)~\cite{zyam}. That assumption is challenged in Ref.~\cite{tzyam}. A Bayesian analysis of jet-related azimuth structure reported in Ref.~\cite{tombayes} finds that a MB jet model as in Fig.~\ref{minijet} is required by data.

\subsection{TR multiplicity $\bf N_\perp$ vs $\bf p_{t,trig}$ or $\bf P_j$} \label{trpedestal}


Transverse multiplicity $N_\perp(p_{t,trig})$,  integrated charge within the TR for some trigger \pt\ condition (single-particle $p_{t,trig}$ or jet sum $P_j$), is employed to study UE properties.  In this subsection a parametrization derived from the \pp\ TCM is applied to TR data. TCM energy dependence relevant to $N_\perp$ is summarized in App.~\ref{tcmenergy}.

For quantitative descriptions of $N_\perp(p_{t,trig})$~\cite{pptheory} certain NSD data are required~\cite{alicetomspec,jetspec}: For 200 GeV: $\bar \rho_s \approx 2.43$, $\bar \rho_h \approx 0.006 \bar \rho_s^2 \approx 0.035$ and $\bar \rho_0 \approx 2.5$, with $2\bar n_{ch,j} \approx 2.1$. For 1.8 TeV: $\bar \rho_s \approx 4.2$, $\bar \rho_h \approx 0.011 \bar \rho_s^2 \approx 0.20$ and $\bar \rho_0 \approx 4.4$, with $2\bar n_{ch,j} \approx 3.4$. Soft and hard event fractions $\lambda_x$ are expressed in terms of Poisson jet probabilities $P_0(n_j)$ (no jet) and $[1 - P_0(n_j)]$ (at least one jet), where $n_j(n_{ch})$ is the mean jet number within an acceptance for \pp\ events with multiplicity \nch~\cite{jetspec2}. Also required are running integrals from above $g_x(y_{t,trig})$ of spectrum soft and hard components. Acceptance factors $\gamma_x$ represent the effect of a low-\pt\ acceptance cut at 0.5 GeV/c for CDF data, and factor $\epsilon \approx 0.6$ is the fraction of a dijet within $\Delta \eta = 2$.

An expression for $N_\perp(p_{t,trig})$ derived from the TCM is
\bea \label{nperpeq}
\frac{3}{2}N_{\perp}(p_{t,trig}) &=& \lambda_s(p_{t,trig})g_s(p_{t,trig}) \gamma_s \bar \rho_{s} 
\\ \nonumber
&& \hspace{-.8in} +~ \lambda_h(p_{t,trig})[g_s(p_{t,trig}) \gamma_s \bar \rho_{s}'   + g_h(p_{t,trig}) \gamma_h \epsilon 2\bar n_{ch,j}]
\\ \nonumber
&\rightarrow& \gamma_s \bar \rho_{s}' + \gamma_h \epsilon 2\bar n_{ch,j} ~\text{for}~ p_{t,trig} \rightarrow \infty,
\eea
where $\bar \rho_s'$ is the soft component for hard (or triggered) events and $\bar \rho_s' \approx \bar \rho_s$ is observed (see next subsection). Note  that instead of $\bar \rho_h =  f \epsilon 2 \bar n_{ch,j}$ appearing in the expression for hard events (within square brackets) $f \rightarrow 1$ is assumed for a hard event induced by a \pt\ trigger. A full derivation of Eq.~(\ref{nperpeq}) is presented in Ref.~\cite{pptheory}. For soft  events at 200 GeV with the CDF $p_t$ acceptance $\bar \rho_{s}  \approx 2.43$ and $\gamma_s \approx 0.25$. For hard events $\epsilon 2\bar n_{ch,j}  \approx 1.26$~\cite{ppprd} and $\gamma_h \approx 0.95$. For 1.8 TeV the numbers are $\bar \rho_s \approx 4.2$ and $\epsilon 2\bar n_{ch,j} \approx 2.0$ with limiting value $N_\perp \approx 2.0$. For those TCM results the most-probable jets are assumed to be unbiased, having the same properties for any trigger. See App.~\ref{nperptcm} for further details.

\begin{figure}[h]
  \includegraphics[width=1.65in,height=1.6in]{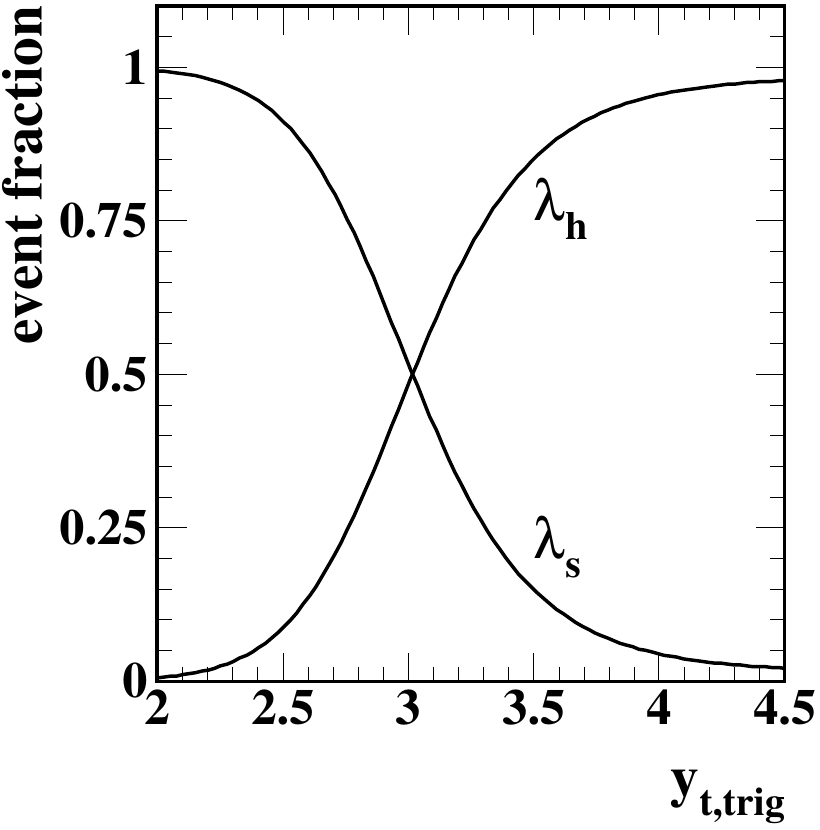}
 \includegraphics[width=1.65in,height=1.63in]{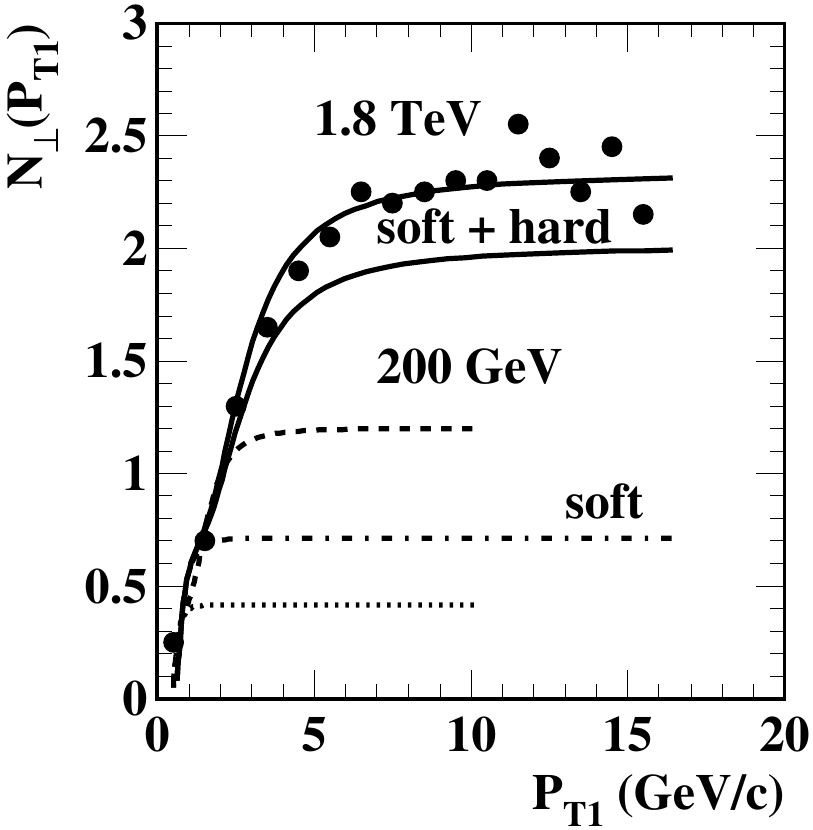}
 \caption{\label{trends2}
  Left: Event fractions $\lambda_x$ for soft (s) and hard (h) 200 GeV \pp\ collisions vs single-particle trigger condition $y_{t,trig}$.
  Right: TR integrated yield $N_\perp$ vs summed trigger (leading) jet momentum $P_{T1}$ data from Ref.~\cite{under5} (points). The curves are TCM results from Eq.~(\ref{nperpeq}) for for 200 GeV (dashed, dotted) and 1.8 TeV (solid, dash-dotted) \pp\ collisions. The two solid curves are explained in Sec.~\ref{nperpspec} and App.~\ref{tcmenergy}.
 }  
\end{figure}

Figure~\ref{trends2} (left) shows soft and hard event fractions $\lambda_s$ and $\lambda_h$ vs trigger condition $y_{t,trig}$ for 200 GeV \pp\ collisions. The fractions become equal for $y_{t,trig} \approx 3$ or $p_{t,trig} = m_\pi \sinh(y_{t,trig}) \approx 1.4$ GeV/c.

Figure~\ref{trends2} (right) shows $N_\perp(P_{T1})$ data for 1.8 TeV \ppbar\ collisions from Ref.~\cite{under5} (points) where $P_{T1}$ in that case refers to summed \pt\ per trigger jet $P_{T1}  \equiv \sum_{i \in \text{jet}} p_{ti}$. The TCM $N_\perp(P_{T1})$ trends (curves) obtained from Eq.~(\ref{nperpeq}) represent running integration from above of soft and hard spectrum components ($g_x$) and a transition from almost all soft events to almost all hard events ($\lambda_x$). The 200 GeV curves stop near 10 GeV/c while the 1.8 TeV curves extend to the limits of data.  The soft + hard curves correspond to the $\bar n_{ch,j}$ estimates above while the soft-only curves correspond to setting $\bar n_{ch,j}$ to zero.  The two solid curves are explained in Sec.~\ref{nperpspec} and App.~\ref{tcmenergy}.

The  200 GeV curves are plotted vs $p_{t,trig}$ for illustration assuming a single trigger particle whereas the 1.8 TeV data correspond to a \pt\ sum for a trigger jet. For a leading-track  (single-particle) trigger the plateau begins near $p_{t,trig} \approx 2.5$ GeV/c, whereas for a leading-track-{\em jet}  trigger the plateau begins near $P_{T1} \approx 5$ GeV/c. To model the difference $p_{t,trig}$ values applied to  Eq.~(\ref{nperpeq}) are converted for plotting to jet-sum approximations by $p_{t,trig} \rightarrow P_{T1} \approx p_{t,trig} + a\, p_{t,trig}^2$ with $a = 0.5$ determined by data. That expression is motivated by the observation that for lower $p_{t,trig}$ a ``jet'' is more likely a single particle sampled from the soft component, with random soft background contribution, whereas for higher $p_{t,trig}$ the trigger particle is more likely associated with a real jet, and multiple correlated jet fragments then contribute to the $P_{T1}$ sum. See the jet-finding algorithm in Ref.~\cite{under5} for details.

The UE is expected to include only MPIs {\em not} identified with the triggered jet, and the elevated plateau (extra UE  ``activity'') is then interpreted to confirm their presence. But that expectation is not realistic for two reasons: (a) a second significant dijet is unlikely (less than 10\% within the CDF acceptance for 1.8 TeV NSD \ppbar\ collisions) except for large-\nch\ events and (b) the TR must include a substantial contribution from any trigger jet as demonstrated in the previous subsection.
Figure~\ref{minijet} (right) confirms that for hard events (those including at least one jet in the acceptance) the TR contains on average 1/3 of the jet fragments from MB dijets, that is, from a triggered dijet selected at random from a MB population. The increase of TR ``activity'' with an applied trigger noted in Ref.~\cite{sjostrand2}%
\footnote{The pedestal effect: ``events with high-$p_\perp$ jets on the average contain more underlying [UE] activity than minimum-bias ones, also {\em well away from {\em[trigger]} jets themselves} [emphasis added].''} actually represents the single trigger dijet in hard events, to the extent that hard events are preferred by the trigger condition, as well as a substantial contribution from the TCM soft component.

These results demonstrate that a TCM description of $p_t$ spectra and minimum-bias angular correlations predicts the general form of the $N_\perp(P_{T1})$ trend with single-particle or jet-sum \pt\ condition. For the CDF \pt\ acceptance ($p_t > 0.5$ GeV/c) the $N_\perp$ increase from zero up to some plateau value includes a smaller contribution from the soft component (projectile nucleon dissociation) and a larger contribution from the hard component (large-angle base of any triggered dijet). However, for a lower acceptance cutoff (e.g.\ $p_t > 0.15$ GeV/c)  the {\em soft} component (actual beam remnants) would dominate $N_\perp(P_{T1})$. Given the conventional UE analysis procedure a ``pedestal'' would result even in the absence of jets.

\subsection{TR $\bf dN_{\perp}/dp_t$ spectrum structure} \label{nperpspec}

TR yield $N_\perp$ can also be characterized in terms of a \pt\ spectrum. In this subsection the TCM is applied to $dN_\perp / dp_t$ spectrum data to demonstrate that the $N_\perp$ soft component has the same universal form inferred from NSD \pp\ collisions, and the hard component is consistent with a trigger condition that prefers \pp\ hard events with at least one MB dijet. A previous analysis of $dN_\perp/ dp_t$ data in Ref.~\cite{pptheory}, lacking detailed information on spectrum structure at higher collision energies, assumed TCM model functions for 200 GeV \pp\ collisions and adjusted certain Eq.~(\ref{a1x}) coefficients to accommodate $dN_\perp / dp_t$ spectrum data. Given new information in Ref.~\cite{alicetomspec} quantitative {\em predictions} are now possible.

\begin{figure}[h]
 \includegraphics[width=1.65in,height=1.6in]{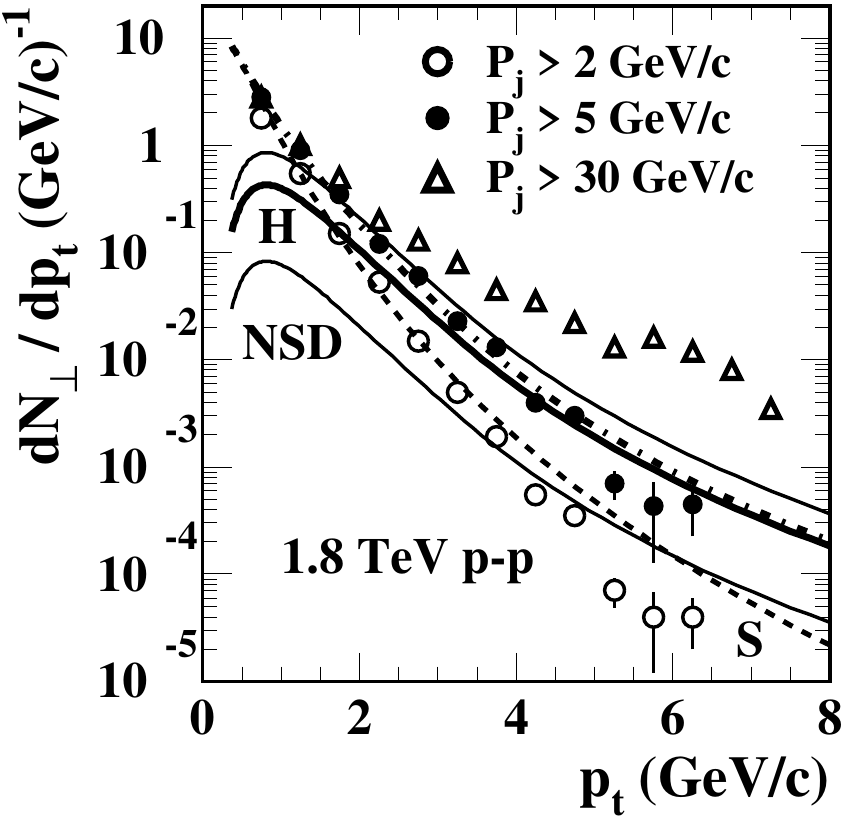}
 \includegraphics[width=1.65in,height=1.6in]{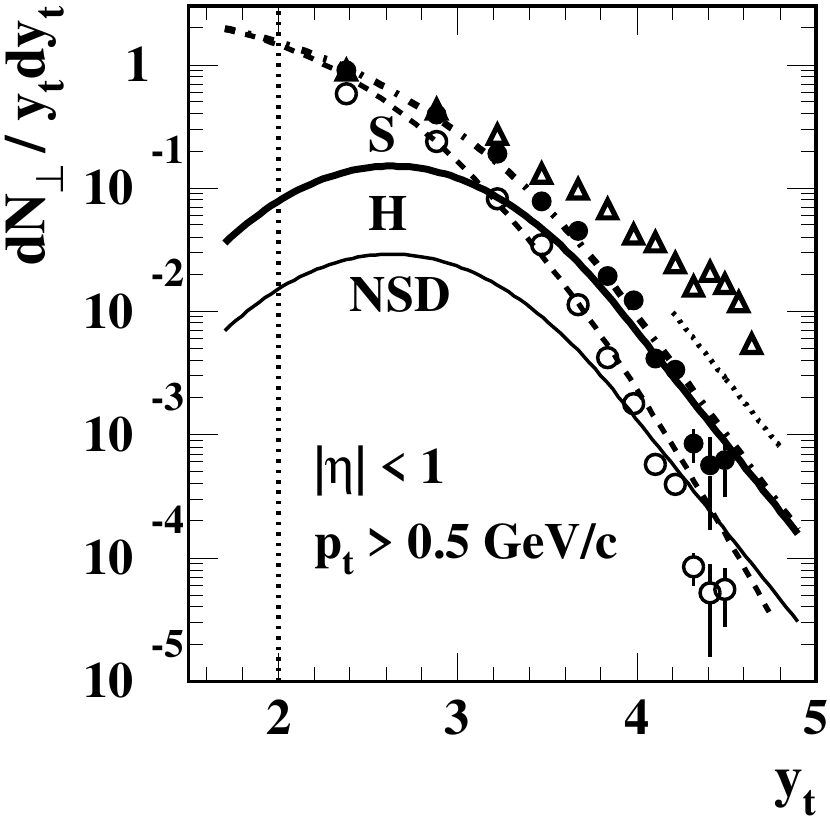}
 \caption{\label{spectrum}
  Left: TR \pt\ spectra $dN_\perp/ dp_t$ for 1.8 TeV \pp\ collisions from Fig.~37 of Ref.~\cite{under5} within $p_t > 0.5$ GeV/c and $|\eta| < 1$ for jet triggers $P_j > 5$ (solid points), $P_j > 2$ (open circles) and for  $P_j > 30$  GeV/c (triangles). Curve $S$ (dashed) is a TCM soft component with data normalization corresponding to soft-component density $\bar \rho_s = n_{s}/\Delta \eta = 4.2$.  NSD $H$ is a 1.8 TeV hard component predicted from $\bar \rho_s = 4.2$ per Refs.~\cite{alicetomspec,jetspec2}. The dash-dotted curve is $S + H$ for $P_j > 5$ GeV/c.
  %
  Right: Data and curves at left transformed to  $dN_\perp/ y_t dy_t$ with Jacobian $m_t / y_t$. The vertical dotted line corresponds to $p_t = 0.5$ GeV/c. The dotted line at right relates to a power-law trend for an underlying MB jet spectrum.
 }  
\end{figure}

Figure~\ref{spectrum} (left) shows $dN_\perp / dp_t$  data for 1.8 TeV \pp\ collisions for specific trigger conditions: summed (highest) jet momentum $P_j > 5$ (solid points), $P_j > 2$  GeV/c (open points) and $P_j > 30$  GeV/c (triangles) from Fig.~37 of Ref.~\cite{under5}.%
\footnote{Because jet spectra are steeply falling imposed lower limit $P_j$ is effectively the same as the mode and mean value. Since $P_{T1}$ or $P_j$ represents the {\em highest} jet momentum it is also an upper limit.} 
The $N_\perp$ spectrum data can be described by Eq.~(\ref{a1x})  given transformation $y_t \rightarrow p_t$ and added factor $p_t$.  Based on the spectrum TCM from Ref.~\cite{alicetomspec} and parametrization of jet energy spectra from Ref.~\cite{jetspec2} the $N_\perp$ spectrum data for 1.8 TeV \pp\ can be predicted by interpolation. Spectrum soft component $S$ (dashed) is described by $\bar \rho_s \approx 4.2$ (consistent with Ref.~\cite{under5}), $T \approx 145$ MeV and $n = 9.0$. NSD hard component $H$ (lowest solid) is described by $\bar \rho_h =f_\text{NSD} \, \epsilon  2 n_{ch,j} \approx \alpha \bar \rho_s^2 \approx 0.20$ with $f_{NSD} \approx 0.1$, $\epsilon = 0.6$, $\bar y_t = 2.63$, $\sigma_{y_t} = 0.55$ and $q = 4.2$. Those values are taken from Refs.~\cite{jetspec2,alicetomspec}. TCM model functions include additional factor 2/3 representing $\Delta \eta = 2$ and TR azimuth acceptance 1/3 of $2\pi$.

The $N_\perp$ TCM has one free parameter: the degree to which a $P_j$ trigger condition alters the {\em effective} MB dijet frequency $f \equiv (1/\sigma_{NSD}) d\sigma_\text{jet}/d\eta$ relative to $f_\text{NSD} \approx 0.1$ or the fraction of hard events $\lambda_h$. The bold solid curve $H$ corresponds to $\lambda_h \rightarrow 0.5$, implying that 50\% of triggered events include at least one dijet within acceptance $\Delta \eta$ (i.e.\ are hard events). It also implies that 50\% are soft events with no significant jet structure despite a ``jet'' trigger. When that $H$ is incorporated into Eq.~(\ref{a1x}) the dash-dotted curve  describing $P_j > 5$ GeV/c data results. The upper solid curve corresponding to $\lambda_h \approx 1$ is consistent with $P_j > 30$ GeV/c data below $p_t  \approx 3$ GeV/c but falls increasingly below the data above that point. 
In contrast,  the $P_j > 2$ GeV/c data are consistent with soft component $S$ alone ($\lambda_h \sim  f \approx 0$), i.e.\ negligible dijet production. Those results can be compared with Fig.~\ref{trends2} (left) plotted in terms of 200 GeV single-particle $y_{t,trig}$. For $P_j > 2$ or $> 5$ GeV/c triggers the data suggest suppression at higher \pt\ due to the trigger condition.

Figure~\ref{spectrum} (right) shows the same data and curves on transverse rapidity $y_t$ with factor $1/y_t$ added. While $p_t$ may be directly measured, details at smaller $p_t$ are obscured compared to $y_t$. The linear power-law trend evident at larger $y_t$ (compare with the dotted line at right) is consistent with an underlying MB jet spectrum~\cite{fragevo,jetspec2}. $dN_\perp / dp_t$ data for the $P_j > 5$ GeV/c trigger condition reveal spectrum structure consistent with a TCM description of triggered events where 50\% (hard events) have at least one MB jet per unit $\eta$ and the complement (soft events) have none. Data for the $P_j > 2$ GeV/c condition suggest that almost all events are soft: The condition that the {\em highest} ``jet'' momentum is $\approx 2$ GeV/c is an effective {\em antijet} trigger. The $P_j > 30$  GeV/c data are consistent both with a higher hard-event fraction ($\lambda_h \approx 1$) and with a biased jet spectrum that deviates substantially from a MB distribution, as might be expected.

The trigger bias illustrated by the $P_j > 30$  GeV/c data spectra in Fig.~\ref{spectrum} refers back to the two solid curves in Fig.~\ref{trends2} (right). The lower solid curve there corresponds to the upper (thin) solid curve in Fig.~\ref{spectrum} (left) assuming no bias of the underlying jet spectrum and $\lambda_h \approx  1$ (all hard events, each including at least a single dijet). The upper solid curve in Fig.~\ref{trends2} (right) corresponds to a 25\% increase of $\epsilon 2 \bar n_{ch,j}$ that accommodates the 1.8 TeV $N_\perp(P_{T1})$ data. That increase can be compared with the difference between the TCM upper thin solid curve in Fig.~\ref{spectrum} (left) and $P_j > 30$ GeV/c data (open triangles).

In summary, application of a jet momentum ($P_j$) trigger has two consequences for $dN_\perp / dp_t$ spectra: (a) the fraction of hard events is altered from untriggered NSD -- greater {\em or lesser} depending on the trigger value -- and (b) the effective jet spectrum is biased in a manner also depending on the trigger value relative to a most-probable jet energy near 3 GeV. The TCM soft- and hard-component curves in Fig.~\ref{spectrum} are predictions based on independent jet- and hadron-spectrum measurements with no adjustment to accommodate $N_\perp$ data. It is especially notable that three distinct trigger conditions retain the same spectrum soft component $S$. Fixed $S$ argues against assumptions that a jet-momentum trigger condition should bias to more-central \pp\ collisions with greater soft or UE charge densities (activity)~\cite{ppcent2}, instead buttresses the conclusion that centrality plays no role in \pp\ collisions~\cite{pptheory,ppquad}. $N_\perp$ \pt\ spectrum behavior appears to confirm a jet spectrum lower bound near 3 GeV (a 2 GeV/c jet trigger effectively excludes jets entirely whereas a 5 GeV/c trigger selects a 50-50 mixture of soft and hard events). The $N_\perp$ \pt\ spectrum also confirms that the trigger jet makes a substantial contribution to the TR (the $N_\perp$ hard component changes dramatically with varying trigger \pt\ condition).

\section{Discussion} \label{disc}

This section considers several issues relating to the PMC:
(a) fluctuations vs correlations vs MB dijets -- relating fluctuation and correlation measurements of several types to the PMC,
(b) direct jet counting from data -- comparing diject production inferred from data to PMC assumptions about MPIs,
(c) a universal TCM soft component that appears to dominate hadron production according to several data manifestations but has no counterpart within the PMC and
(d) the PMC CR mechanism compared to measured FFs and \pt\ spectrum trends.

\subsection{Fluctuations vs correlations vs MB dijets}

Fluctuations and correlations of total \nch\ and total $P_t$ integrated within some angular acceptance $(\Delta \eta,\Delta \phi)$ are intimately related but conventionally treated as independent topics. For instance, the well-studied structure of $P(n_{ch})$ distributions as in Sec.~\ref{pnch} is not considered in parallel with corresponding $P(P_t)$ distributions. In either case distribution structure is strongly influenced by a hard-component contribution from MB dijets~\cite{mbdijets,ptscale} quadratically related to the soft component~\cite{ppprd,alicetomspec,ppquad}. Within an \aa\ context defined by expectations for QGP formation, systematic trends for ensemble-mean $\bar p_t \equiv \bar P_t / n_{ch}$ (radial flow?)~\cite{alicempt} are considered independently of fluctuations in eventwise mean $\langle p_t \rangle \equiv \langle P_t / n_{ch}\rangle$ (temperature fluctuations?)~\cite{aliceptfluct}, although data suggest  that both trends are again dominated by MB dijets as the common element~\cite{tommpt,ptscale}. Forward-backward (FB) \nch\ (but not $P_t$?) correlations on $\eta$ (Sec.~\ref{fbcorr}) are treated in isolation, although FB correlations represent a 1D projection of 2D angular correlations on $(\eta_\Delta,\phi_\Delta)$ (Sec.~\ref{notr}), again dominated by MB dijets~\cite{ppquad} and directly related to \nch\ and $P_t$ fluctuations and correlations~\cite{inverse,ptscale}.

The assumed sources of fluctuations are quite different for PMC and TCM. In a PMC context particle and \pt\  fluctuations result primarily from \pp\ centrality variation (modeled by geometric Glauber MC based on eikonal approximation) and Poisson fluctuations of MPI number.

In a TCM context the relevance of \pp\ centrality is challenged by the noneikonal quadratic relation between soft and hard components -- jet production $\propto \bar \rho_h \propto \bar \rho_s^2$. The soft component dominates, and for NSD \pp\ collisions multiple dijets within a typical $\Delta \eta$ acceptance (MPIs) are unlikely. Hard events with at least one dijet comprise only a few percent of the total at 200 GeV. Instead, the most likely source of \nch\ or $\bar \rho_s$ fluctuations is the splitting cascades within inelastically-scattered projectile protons leading to large fluctuations of the soft component.%
\footnote{``The proton substructure is represented by a parton [splitting] cascade, which at high energies is described by BFKL evolution. The 
fluctuations in this evolution are known to be very large [Ref.~\cite{bfklflucts} of the present paper].'' G. Gustafson, Ref.~\cite{mpiref} p. 43.}
Given $\bar \rho_h \propto \bar \rho_s^2$, dijet production must fluctuate to a greater degree (e.g.\ see Fig.~3 of Ref.~\cite{pptheory}).

How those fluctuations are revealed in statistical measures depends critically on the measure definition~\cite{mbdijets,aliceptflucttom}. For instance, the variance of integrated $P_t$ within some acceptance $\sigma^2_{P_t} = \overline{(P_t - \bar P_t)^2}$, with $\bar P_t$ an ensemble mean, is sensitive to all $P_t$ fluctuations~\cite{clt}. However, conditional variance $\sigma^2_{P_t|n_{ch}} \equiv \overline{(P_t - n_{ch} \hat p_t)^2}$, with $\hat p_t$ a single-particle ensemble mean, is sensitive only to fluctuations in integrated $P_t$ {\em relative to} what would be expected for \nch\ fluctuations with fixed $\hat p_t$ in every event~\cite{ptscale}. Its scale (bin size) dependence can be inverted to reconstruct underlying \pt\ angular correlations~\cite{inverse}. Those angular correlations, e.g.\ from 200 GeV \auau\ collisions with \nn\ collisions as a limit, are dominated by MB dijet structure~\cite{ptscale,ptedep}, a result consistent with Secs.~\ref{2partcorr} and \ref{notr} for \pp\ collisions.  Equation~(\ref{equivalence}) suggests that \pp\ charge (number) \nch\ fluctuations are also associated with MB dijets as the dominant source of two-particle correlations near midrapidity (see Sec.~\ref{notr}), one of several jet manifestations represented by the TCM hard component. 

\subsection{Jet counting and jet-related correlations}

As a response to the PMC assumption that almost all hadron production must result from MPIs it is possible to derive actual jet production rates from several data formats that present a consistent picture of the role of MB dijets and are consistent with the TCM soft component as the dominant mechanism for hadron production. For example, jet-related angular correlations from \aa\ collisions can be analyzed to determine the absolute mean dijet number as a function of \aa\ centrality~\cite{jetspec}. Such estimates are consistent with results from TCM \pt\ spectrum analysis~\cite{fragevo,ppprd,ppquad}, from ensemble-mean \mmpt\ analysis~\cite{tommpt} and from jet spectrum analysis~\cite{jetspec2}. In each case maximum information is derived from diverse data formats and collision systems to develop a consistent TCM.

In contrast, the PMC addresses less-differential data formats (e.g.\ the pedestal effect, $P(n_{ch})$, FB correlations) that do not include actual jet counting. A restricted data set is accommodated with a combination of jet spectrum lower cutoff $p_{\perp0}$ and CR mechanism (variable FFs), both adjusted to fit data. The TCM soft component that is required by several data trends (e.g.\ spectrum~\cite{ppprd} and angular-correlations~\cite{ppquad} \nch\ dependence) is excluded, leading to a requirement for unrealistically large dijet production at very low jet energies (below 2 GeV~\cite{under5}) where fragmentation to jets is unmeasured and unlikely. The lower limit on jet energies is not $\lambda_\text{QCD}$ but the density of hadronic final states. For a jet spectrum terminating near 3 GeV~\cite{jetspec2} the lowest-energy ``jets'' consist of charge-neutral hadron (pion) pairs~\cite{porter2,porter3}. Jet production must then be strongly inhibited below that point.

\subsection{TCM soft component as true underlying event}

The UE is conventionally defined as complementary to a triggered high-\pt\ dijet and is by hypothesis expected to include beam-beam remnants (projectile-nucleon dissociation products), initial-state radiation and MPIs~\cite{under5}. The UE is therefore ill-defined because the triggered dijet is conventionally misrepresented (illustrated for instance in Sec.~\ref{notr}) in that any dijet includes a substantial contribution to the TR region not typically acknowledged. In contrast, the TCM soft component is well-defined phenomenologically, with consistent manifestations in yields, spectra and correlations, and is equivalent to ``beam remnants'' within the context of the PMC. What should be the established reference (soft component) is confused with the ill-defined UE, whereas what should be the object of study -- high-\pt\ jets -- is adopted as the reference.

As noted in Sec.~\ref{models} soft events (``diffractive events'' with no MPI) are assumed to be rare.
``Such events [with no apparent hard scatter] are associated with nonperturbative low-$p_\perp$ physics, and are simulated by exchanging a very soft gluon between the two colliding hadrons, making the hadron remnants colour-octet objects rather than colour-singlet ones''~\cite{sjostrand2}. The possibility that one or two color {\em singlets} (e.g.\ soft or hard Pomerons) are exchanged~\cite{pomerons,levin} is not considered.
The statement ``Translated into modern terminology, each cut pomeron corresponds to the exchange of a soft gluon, which results in two `strings' being drawn between the two beam remnants''~\cite{sjostrand2} implies that part of the soft component is excluded from the beam remnants, whereas the statement ``For PYTHIA we include particles that arise from the soft or semi-hard scattering in multiple parton interactions [MPIs] in the beam-beam remnant component''~\cite{under5} implies that beam remnants as simulated by the PMC include MPIs as well as some unspecified fraction of the projectile dissociation component.

Within the PMC model the ``hard component'' (i.e.\ MPIs) must represent the entire \pt\ spectrum. As a consequence ``The charged particle $p_\perp$ spectrum is underestimated at low $p_\perp$ scales''~\cite{sjostrand}. In contrast, the TCM soft component is required by a broad array of data and describes the low-\pt\ portion of hadron spectra within data uncertainties, as demonstrated in Figs.~\ref{pp1} and \ref{qppb}. The spectrum soft component has a universal form with fixed slope parameter $T \approx 145$ MeV corresponding to fixed \mmpt\ soft component $\bar p_{ts} \approx 0.4$ GeV/c consistent with all presently available \mmpt\ data~\cite{tommpt}.

As demonstrated in Sec.~\ref{nperpspec} an imposed jet \pt\ trigger may bias the higher-\pt\ hadron distribution (hard component) but does nothing to influence the lower-\pt\ soft component, which qualifies as the true ``underlying event'' complementary to all dijet production. That a $p_t$ trigger may prefer hard events but {\em does not change the soft-component multiplicity} implies both that soft/hard ratio in triggered events is no different from NSD and that \pp\ centrality is not a relevant degree of freedom, contradicting Ref.~\cite{ppcent2} and the PMC. That all jet-triggered events maintain the same \mmpt\ is also explained: the ratio of TCM soft vs hard components in Eq.~(\ref{ppmpttcm}) remains fixed.

\subsection{Color reconnection and MB dijets}
 
The PMC must include a CR mechanism to accommodate \mmpt\ vs \nch\ data~\cite{alicempt}, as noted in Sec.~\ref{mpttrends}. The CR mechanism is in effect equivalent to FFs strongly dependent on scattered-parton (MPI) density.  Within a TCM context FFs, such as $D_{pp}$ in Eq.~(\ref{fold1}), are approximately independent of the \pp\ collision system as implied by TCM analysis of \pp\ spectrum data over a large collision-energy interval~\cite{alicetomspec}.  \mmpt\ variation with \nch\ is interpreted to arise from noneikonal quadratic increase of dijet production with increasing \nch~\cite{ppquad,tommpt}.

According to assumptions supporting the PMC (Sec.~\ref{color}) each MPI should contribute ``the same (semi)hard $p_\perp$ kick''~\cite{sjostrand} (e.g.\ some $\bar p_{t0}$) to integrated $\bar P_t$ while the ``hadrons per string'' (jet fragment multiplicity) would decrease with increasing $\bar n_{MPI}$. That hypothesis can be represented by rearranging Eq.~(\ref{ppmpttcm}) (with $\xi \rightarrow 1$) describing \mmpt\ data within their statistical uncertainties
\bea
\bar p_t \equiv \frac{\bar P_t}{n_{ch}} &\approx&  \frac{\bar p_{ts} + x(n_s) \bar p_{th}(n_s)}{1 + x(n_s)} \rightarrow \frac{\bar p_{t0} \, \bar n_{MPI}} {n_{ch}( \bar n_{MPI})}
\\ \nonumber
2 \bar n_{ch,j} &\approx& \frac {n_{ch}(\bar n_{MPI})}{\bar n_{MPI}} \approx  \frac{\bar p_{t0}}{\bar p_{ts}} \cdot \frac{1 + \alpha \bar \rho_s}{1+ \alpha \bar \rho_s\, \bar p_{th0}/\bar p_{ts} },
\eea
where the second line gives the PMC mean fragment multiplicity per MPI via $\bar p_{ts} \approx 0.4$ GeV/c from Fig.~\ref{alice5a} (left), $\bar p_{th0} \approx 1.2$ GeV/c from Fig.~\ref{alice5a} (right), $\alpha \approx O(0.01)$ from Eq.~(\ref{a3}) and $\bar p_{t0}$, a free parameter within the CR model. The CR-related expression then implies that the mean jet fragment yield $2 \bar n_{ch,j}$ should decrease asymptotically by factor 1/3 as \nch\ (and dijet multiplicity) increases.

That relation contradicts the systematics of jet fragmentation wherein a fixed mean fragment multiplicity per dijet is determined by $2 \bar n_{ch,j} = \int dy_t \bar D(y_t)$ with $\bar D(y_t)$ given by measured jet properties combined as in Eq.~(\ref{fold1}). Data indicate that $2 \bar n_{ch,j}$, described by Eq.~(\ref{nchj}), changes with \pp\ collision energy but not with \nch\ or $\bar \rho_s$ and therefore does not depend on parton density (MPI number?), which contradicts the PMC CR model. From spectrum data $2 \bar n_{ch,j} = \int dy_t y_t H(y_t) / \epsilon f_{NSD}$ actually {\em increases} slightly with \nch\ increase corresponding to 100-fold increase of dijet production (Fig.~\ref{300a} and Ref.~\cite{alicetomspec}).

\section{Summary}  \label{summ}

The PYTHIA Monte Carlo (PMC) model of high-energy \pp\ collisions, motivated by certain data features emerging from the super proton-antiproton synchrotron (Sp\=pS) program, includes several basic assumptions: (a) almost all hadrons arise from multiparton interactions (MPIs) described by perturbative QCD (pQCD), (b) the scattered-parton (jet) spectrum extends down to zero \pt\ (or jet energy), (c) a color reconnection (CR) mechanism controls parton fragmentation to jets and (d) \pp\ centrality, modeled by a geometric Glauber model based on the eikonal approximation, controls hadron production.

The two-component (soft + hard) model (TCM) of hadron production, introduced concurrently with the PMC, provides an alternative description of hadron production in A-B collisions. The TCM was inferred inductively from measured data trends and describes a broad array of collision systems and data formats accurately.  Whereas the PMC is a one-(hard)-component model the TCM soft component appears to be {\em required} by data, e.g.\ \pt-spectrum, $\eta$-density and two-particle-correlation data for 200 GeV \pp\ collisions. In any collision system the TCM soft component represents the majority of hadrons.

The dijet-related TCM hard component exhibits a noneikonal quadratic dependence on the soft component interpreted to represent participant low-$x$ partons (gluons). The quadratic trend remains accurate at the  percent level over a soft-component range corresponding to 100-fold increase in dijet production. The noneikonal trend suggests no dependence on \pp\ centrality: each participant parton in one projectile proton may interact with {\em any} participant in the partner proton -- all or nothing. 

Differential spectrum analysis reveals that the TCM model (including soft component) describes \pt\ spectra accurately down to very low \pt, whereas the PMC (hard component only) model fails at lower \pt. The \pt-spectrum TCM hard component is predicted by convoluting jet spectra having a lower bound near 3 GeV with fragmentation functions independent of parton density (i.e.\ no CR) and consistent with measurements. 

Detailed analysis of 2D angular correlations reveals that any dijet from a MB ensemble must make a substantial contribution  to the  ``trans'' region (TR) (including $\phi = \pi/2$ relative to the trigger direction) of azimuth angular correlations that is assumed to be especially sensitive to the underlying event (UE) for jet-triggered events. Imposing a jet trigger (e.g.\ jet \pt\ sum) results in contributions of the triggered jet to the TR, contradicting a basic PMC assumption. The trigger condition does not change the soft component, arguably the real ``underlying event'' for any collision conditions, which also appears inconsistent with any connection to \pp\ centrality dependence.

In conclusion, a variety of data manifestations is inconsistent with the basic assumptions of the PMC, whereas the TCM is consistent with a large body of jet measurements and provides a simple and accurate representation of hadron production in a variety of collision systems. These results have implications for other Monte Carlos derived from the PMC, such as HIJING and AMPT.

\begin{appendix}

\section{TCM Hard-component collision-energy dependence} \label{tcmenergy}

This appendix combines MB dijet collision-energy dependence in Sec.~\ref{jetspecc} and TCM hadron \pt\ spectrum collision-energy trends in Secs.~\ref{aa1} and \ref{spechard} to support UE analysis in  Secs.~\ref{trpedestal} and \ref{nperpspec} that requires additional TCM elements. The object is  self-consistent quantitative representation of single-particle yields and \pt\ spectra vs $N_\perp$ yields and spectrum

\subsection{p-p spectra and MB dijets}

The TCM hard component (various manifestations) is related to QCD jets by the following expressions
\bea \label{a1}
\bar \rho_h(n_{ch},\sqrt{s}) &=& \alpha(\sqrt{s}) \bar \rho_s^2 
\\ \nonumber
&=& f(n_{ch},\sqrt{s}) \, \epsilon 2 \bar n_{ch,j}(\sqrt{s}),
\eea
where the first line is derived from \pp\ \pt\ spectrum analysis~\cite{ppprd,ppquad} and the second line is derived from jet analysis~\cite{jetspec2}. From spectrum studies $\bar \rho_s \approx 0.81 \Delta y_b$~\cite{alicetomspec} with $\Delta y_b = \ln(\sqrt{s} / \text{10 GeV})$. From jet systematics and a measured NSD cross section trend~\cite{jetspec2} the per-NSD-event dijet $\eta$ density is
\bea \label{a2}
f_{NSD}  &\equiv& \frac{1}{\sigma_{NSD}} \left[\frac{d\sigma_j}{d\eta}\right]_{NSD}   \hspace{-.1in} \approx 0.037 \frac{\Delta y_{max} \Delta y_b^2}{32 + \Delta y_b^2 },~~
\eea
where $\Delta y_{max} \equiv \ln(\sqrt{s} / \text{6 GeV})$ and $\sigma_{NSD} \approx  0.85 (32 + \Delta y_b^2) $.
The trend of $\alpha(\sqrt{s})$ values inferred from \pp\ \pt\ spectra~\cite{alicetomspec} is given by
\bea \label{a3}
\alpha \approx 0.02 \frac{\Delta y_{max}^2 }{32 + \Delta y_b^2 }.
\eea
Combining Eqs.~(\ref{a1}), (\ref{a2}) and (\ref{a3}) gives~\cite{alicetomspec}
\bea \label{nchj}
 2 \bar n_{ch,j}(\sqrt{s}) &\approx& 0.60 \Delta y_{max}
\eea
and Eq.~(\ref{a1}) becomes
\bea
\bar \rho_h  &\approx& 0.013 \frac{\Delta y_{max}^2 \Delta y_b^2}{32 + \Delta y_b^2 }.
\eea

Table~\ref{tcmparams} presents NSD values from those relations  evaluated for $\sqrt{s} = 200$ and 1800 GeV assuming accepted dijet fraction $\epsilon \approx 0.6$ for acceptance $\Delta \eta \approx 2$.

\begin{table}[h]
	\caption{TCM and jet parameters for NSD \pp\ collisions at two energies. Must rescale otherwise based on above relations.
	}
	\label{tcmparams}
	\begin{center}
		\begin{tabular}{|c|c|c|c|c|c|c|c|} \hline
		$\sqrt{s}$ (GeV) &$\Delta y_b$ & $\Delta y_{max}$ & $\bar \rho_{sNSD}$ &  $\alpha$  & $f_{NSD}$ & $2\bar n_{ch,j}$  & $\bar \rho_{hNSD}$  \\ \hline
		   200 & 3.0 & 3.5 & 2.43 & 0.006 & 0.028  & 2.1    & 0.035  \\ \hline
		   1800 & 5.2 & 5.7 & 4.20 & 0.011 &  0.097 & 3.4   &  0.20 \\ \hline
		\end{tabular}
	\end{center}
\end{table}

The expressions above estimate NSD TCM parameter values for any $\sqrt{s}$. For \pp\ collision ensembles other than NSD (e.g.\ with some \nch\ condition imposed) corresponding parameter values can be obtained by scaling. For instance $f(n_{ch}) = (\bar \rho_s / \bar \rho_{sNSD})^2 f_{NSD}$ reflects the noneikonal quadratic dependence of MB dijet production on TCM soft component $\bar \rho_s$ (i.e.\ $\propto \Delta y_b^2$). On the other hand, $\alpha(\sqrt{s})$ and $2 \bar n_{ch,j}(\sqrt{s})$ do not appear to depend on \pp\ \nch, and their energy dependence is controlled by $\Delta y_{max}$ which measures the width of MB jet spectra~\cite{jetspec2}. The lack of \nch\ dependence for mean dijet fragment multiplicity $2 \bar n_{ch,j}(\sqrt{s})$  suggests that parton fragmentation does not depend on scattered-parton density (over a 100-fold variation~\cite{ppprd,ppquad}) as implied by the CR mechanism described in Refs.~\cite{sjostrand,sjostrand2}.
Imposition of a \pt\ condition (single-particle $p_{t,trig}$ or jet sum $P_j$) on events requires a different treatment as reported in Ref.~\cite{pptheory} and summarized briefly in Sec.~\ref{trpedestal}

\subsection{TR $\bf N_\perp$ predictions from the TCM} \label{nperptcm}

The results of the previous subsection can  be used to predict TR $N_\perp$ trends.
The TR pedestal effect is related to a varying \pt\ ($P_{T1}$) condition placed on \pp\ collisions. If hadron production in the TR were unbiased (no condition) one expects the TCM spectrum description
\bea
\frac{3}{2} \frac{dN_\perp}{p_tdp_t} &=& \bar \rho_s  \hat S_0(p_t) + \bar \rho_h  \hat H_0(p_t).
\eea
Integrating that expression over the CDF \pt\ acceptance ($p_t > 0.5$ GeV/c) gives [using unit-integral TCM $\hat S_0(p_t)$ and $\hat H_0(p_t)$ models interpolated to 1.8 TeV~\cite{alicetomspec}]
\bea
\frac{3}{2}N_\perp &=& 0.25 \bar \rho_s  + 0.95 \bar \rho_h .
\eea
If a $P_{T1}$ condition is imposed such that only hard events are selected (but with NSD soft and hard components) then in Eq.~(\ref{a1}) $\bar \rho_h \rightarrow \epsilon 2 \bar n_{ch,j}$, i.e.\ $f \rightarrow 1$. In that case
\bea
N_\perp &=& \frac{2}{3} (0.25\, \bar \rho_s  + 0.95\, \epsilon 2 \bar n_{ch,j})
\\ \nonumber
&\rightarrow& 0.70 + 1.29 \approx 2.0
\eea
using NSD values from Table~\ref{tcmparams}. That asymptotic value corresponds to the lower solid curve in Fig.~\ref{trends2} (right) assuming that the TCM hard component is unbiased by the imposed \pt\ condition. Figure~\ref{spectrum} makes clear that the spectrum in the TR {\em is} biased by the trigger condition, with an effective increase of $ 2 \bar n_{ch,j}$ by factor 1.25. Note that for 100\% \pt\ acceptance the saturation value becomes $N_{\perp} \rightarrow  0.7 / 0.25 + 1.25 \cdot 1.29 / 0.95 \approx 2.8 + 1.7 = 4.5$. That is, fixed soft component $S$ actually dominates $N_{\perp}$.

\section{TCM description of $\bf \rho_0(\eta)$ distributions} \label{etadist}
 
 TCM  $\eta$ densities $\bar \rho_0$, $\bar \rho_s$ and $\bar \rho_h$ appearing in Eqs.~(\ref{a1x}) are mean values of differential densities $\rho_x(\eta)$ averaged over some acceptance $\Delta \eta$. The hadron density near midrapidity is represented by joint density $ \rho_0(y_t,\eta;n_{ch})$. Based on results from spectrum analysis in Ref.~\cite{ppprd}  the basic TCM decomposition is given by the first line of
 \bea \label{eq15}
 \rho_0(y_t,\eta;n_{ch}') &=& S(y_t,\eta;n_{ch}') + H(y_t,\eta;n_{ch}')
 \\ \nonumber
 &\approx&   \rho_{s0}(n_{ch}') S_0(\eta)  \hat S_0(y_t)
 \\ \nonumber
 &+&  \rho_{h0}(n_{ch}') H_0(\eta) \hat H_0(y_t),
 \eea 
 where $n_{ch}'$ (with prime) is an uncorrected multiplicity within $\Delta \eta$.
 The second line  invokes factorization of soft and hard components.  Soft component $S(y_t,\eta;n_{ch}')$ is assumed factorizable within some limited acceptance $\Delta \eta$. Hard component $H(y_t,\eta;n_{ch}')$ may include significant $\eta$-$y_t$ covariances but factorization is also assumed here. Unit integral (denoted by carets) spectrum models $\hat S_0(y_t)$ and $\hat H_0(y_t)$ are as defined in Ref.~\cite{ppprd} over the full \yt\ acceptance.  Model functions $S_0(\eta)$ and $H_0(\eta)$ on $\eta$ are newly defined below, and the $\rho_{x0}$ represent soft and hard hadron densities at $\eta = 0$ ({\em not} averaged over $\Delta \eta$).
 
 Integrating Eq.~(\ref{eq15}) over a \pt\ or \yt\ acceptance with nonzero lower limit (e.g.\ $p_{t,cut} \approx 0.15$ GeV/c) leads to
 \bea \label{rho0eta}
 \rho_0'(\eta;n_{ch}') &\approx&   \rho_{s0}'(n_{ch}') S_0(\eta)  
 +  \rho_{h0}(n_{ch}')  H_0(\eta)
 \\ \nonumber
 \rho_0'(\eta;n_{ch}',\Delta \eta) & \approx &  \bar  \rho_{s}'(n_{ch}',\Delta \eta) \tilde S_0(\eta;\Delta \eta)  
 \\ \nonumber
 &+& \bar  \rho_{h}(n_{ch}',\Delta \eta)  \tilde H_0(\eta;\Delta \eta)~\text{within $\Delta \eta$}
 \eea 
 where $\tilde X_0(\eta;\Delta \eta) \equiv X_0(\eta) / \bar X_0(\Delta \eta)$ are unit integral on $\Delta \eta$, $\bar \rho_x$ are charge densities averaged over $\Delta \eta$ and it is assumed that the \pt\ acceptance cut does not affect the hard component. The goal of analysis is to infer TCM models $\tilde S_0(\eta;\Delta \eta)$ and $\tilde H_0(\eta;\Delta \eta)$ from $\rho_0'(\eta;n_{ch}') $ data.
 
\begin{figure}[h]
  \includegraphics[width=1.65in,height=1.6in]{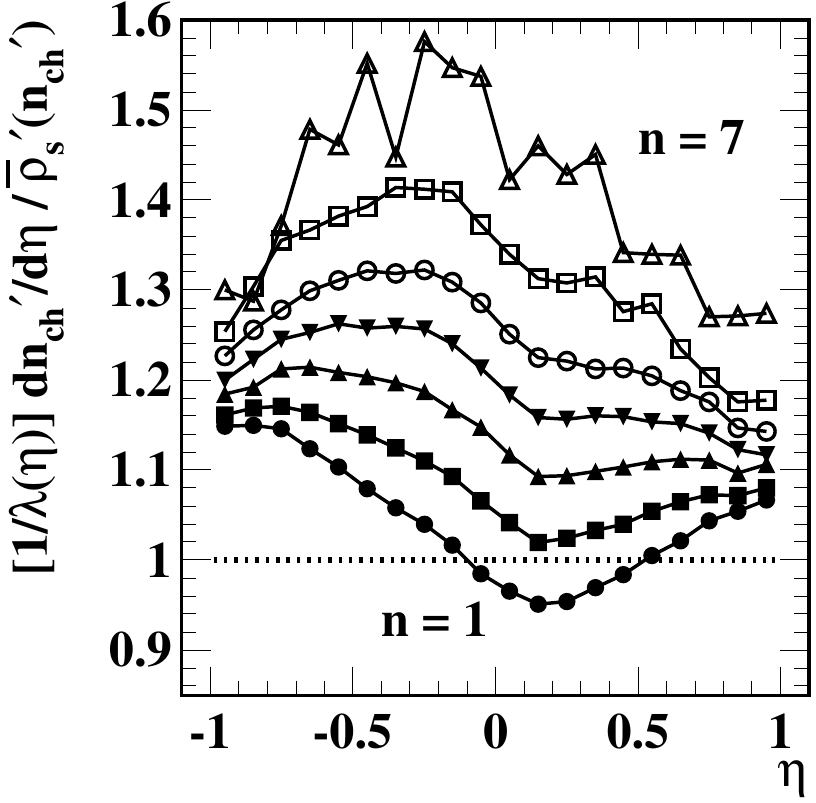}
  \includegraphics[width=1.65in,height=1.6in]{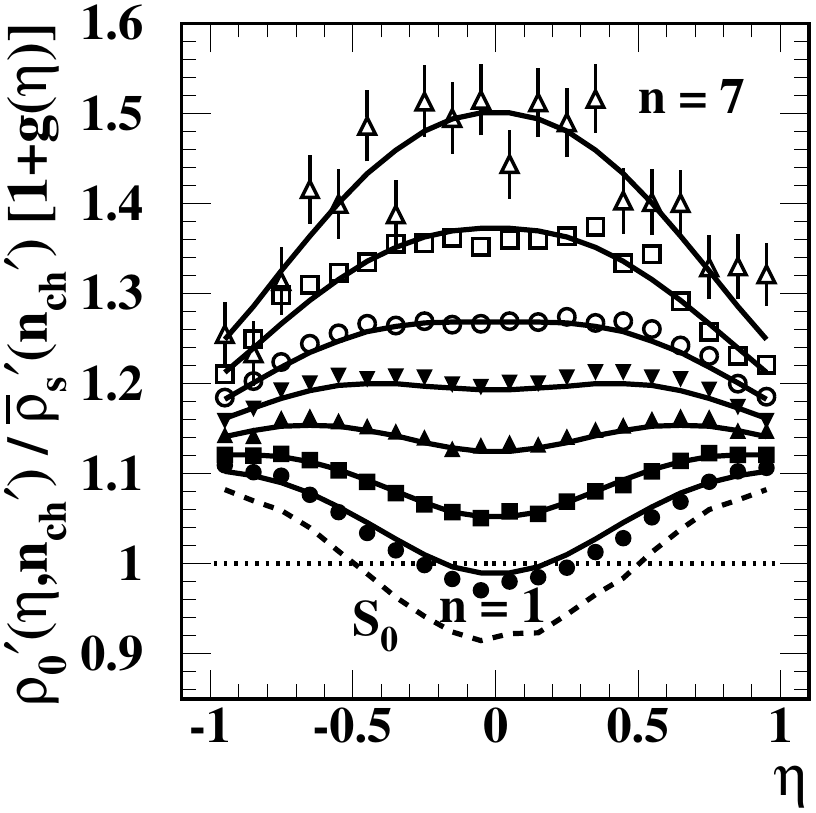}
  \caption{\label{dndeta1}
   Left: Uncorrected $\eta$ densities within $\Delta \eta = 2$ for seven multiplicity classes denoted by index $n$~\cite{ppquad}. The curves connecting data points guide the eye.
   Right:  Corrected $\eta$ densities within $\Delta \eta = 2$ for seven multiplicity classes. The dashed curve is normalized soft-component model $\tilde S_0(\eta) \equiv S_0(\eta)/ \bar S_0(\Delta \eta)$ from Eq.~(\ref{s00}). The solid curves are Eq.~(\ref{rho0eta}) with TCM elements defined in Eqs.~(\ref{h00}) and (\ref{s00}).
  }   
 \end{figure}

 Figure~\ref{dndeta1} (left) shows measured density distributions $dn_{ch}'/d\eta$ for seven \nch\ classes of 200 GeV \pp\ collisions normalized by uncorrected soft component  $\bar \rho_s' = n_s' / \Delta \eta$  inferred from $n_{ch}'$ values assuming $\alpha = 0.006$~\cite{ppquad,alicetomspec}. Corrected charge density $\bar \rho_0$ varies from 1.76 to 18.8 corresponding to a 100-fold increase in dijet production.
 An $\eta$-symmetric  inefficiency $\lambda(\eta) \leq 1$ deviates from unity only for two outer  bins at the ends of the $\eta$ acceptance. The plotted data are corrected for  $\lambda(\eta)$, but a common $\eta$-asymmetric distortion $1 + g(\eta)$ remains due to different tracking efficiencies in two halves of the TPC detector.
 
 Figure~\ref{dndeta1} (right) shows asymmetry-corrected renormalized densities $\rho_0'(\eta;n_{ch}') /  \bar \rho_{s}'(n_{ch}')[1 + g(\eta)]$ (points). Solid curves through data are described by Eq.~(\ref{rho0eta}) (lower line) with soft and hard model functions defined below. Dashed curve $\tilde S_0(\eta)$ is defined by Eq.~(\ref{s00}), the soft-component limit to the ratio $\rho_0'(\eta) / \bar \rho_s'$ as $n'_{ch} \rightarrow 0$. Data are  modeled by fixed $\tilde S_0$ plus hard component $\propto \bar \rho_s$.
 
 An iterative analysis method is described in detail in Ref.~\cite{ppquad}. Briefly, differences between adjacent pairs of multiplicity classes in Fig.~\ref{dndeta1} (right)  are used to obtain estimates of $ (\bar \rho_h / \bar \rho_s') \tilde H_0(\eta;\Delta \eta)$ per Eq.~(\ref{rho0eta}) (second line). The ensemble of data differences is used to determine the common hard-component model
 \bea \label{h00}
 \tilde H_0(\eta;\Delta \eta) \equiv \frac{H_0(\eta)}{\bar H_0(\Delta \eta)} &=& 1.47 \exp[-(\eta / 0.6)^2/2].~~~
 \eea
 With a provisional hard-component model defined the soft-component model is estimated from data as follows.
 
\begin{figure}[h]
  \includegraphics[width=1.65in,height=1.6in]{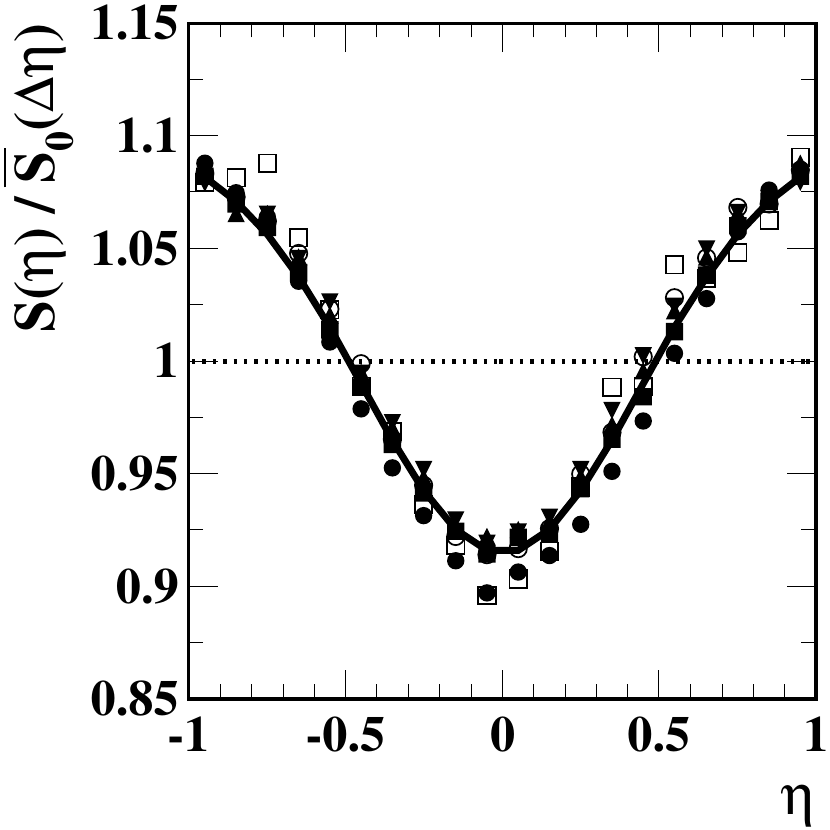}
  \includegraphics[width=1.65in,height=1.6in]{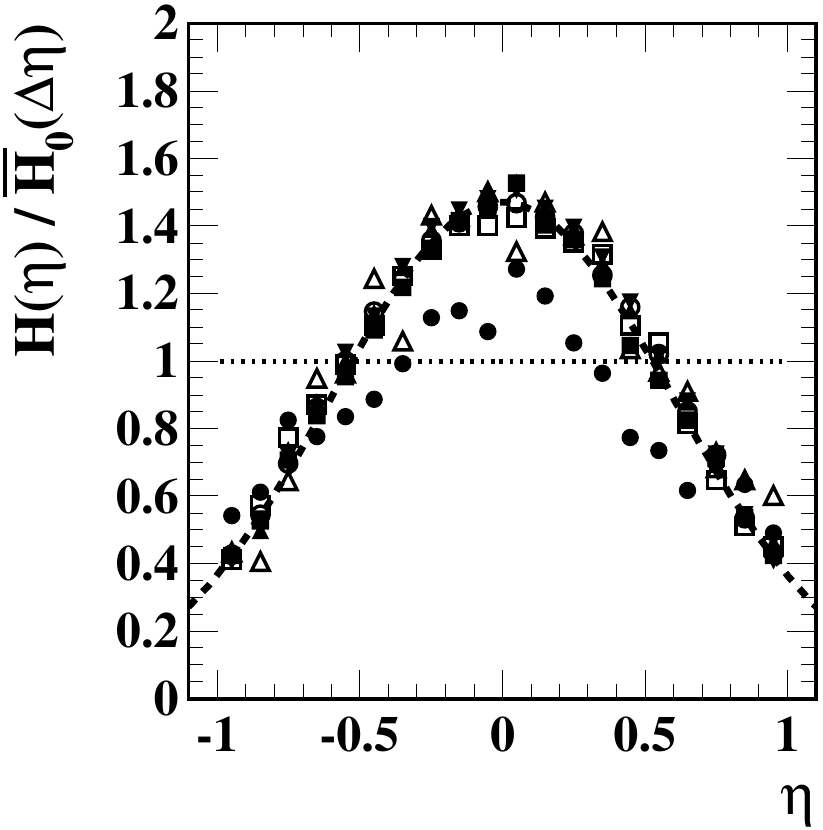}
  \caption{\label{etatcm}
   Left:  Data soft-component estimates as in Eq.~(\ref{ss}). The solid curve is normalized soft-component model $\tilde S_0(\eta)$ as defined by Eq.~(\ref{s00}).
   Right:  Data hard-component estimates as in Eq.~(\ref{hc01}). The solid curve is normalized hard-component model $\tilde H_0(\eta)$ as defined by Eq.~(\ref{h00}). The data for $n = 1$ (solid dots) are significantly low compared to the common trend.
  }  
 \end{figure}

 Figure~\ref{etatcm} (left) shows the soft-component estimator
 \bea \label{ss}
 \frac{S_n(\eta)}{\bar S_0(\Delta \eta)} &\equiv& \rho_0'(\eta)_{n} /  \bar  \rho_{s,n}' - (\bar \rho_{h,n} / \bar \rho_{s,n}') \tilde H_0(\eta;\Delta \eta),~~~
 \eea
 for each multiplicity class $n$ with $\tilde H_0(\eta;\Delta \eta)$ as defined in Eq.~(\ref{h00}).  The inferred soft-component model for $\Delta \eta = 2$ (solid curve) is defined by
 \bea \label{s00}
 \tilde S_0(\eta;\Delta \eta)\hspace{-.02in} \equiv\hspace{-.02in} \frac{S_0(\eta)}{\bar S_0(\Delta \eta)}\hspace{-.05in} &=&\hspace{-.05in} 1.09 - 0.18 \exp[-(\eta / 0.44)^2/2].~~~~~
 \eea
 The form of the soft component appears to be stable over a large \nch\ interval. Small data deviations from the model are consistent with statistical uncertainties. The minimum at $\eta = 0$ is expected given  the Jacobian for $\eta \leftrightarrow y_z$, where an approximately uniform density on $y_z$ is expected within a limited $\Delta y_z$ acceptance about $y_z = 0$.
 
 Figure~\ref{etatcm} (right) shows the hard component re-estimated from $\rho_0'(\eta;n_{ch}')$ data using an alternative method that assumes model $\tilde S_0(\eta;\Delta \eta)$ from Eq.~(\ref{s00})
 \bea \label{hc01}
 \frac{H_n(\eta)}{\bar H_0(\Delta \eta)} &\equiv& \frac{\rho_0'(\eta)_{n} /   \bar \rho_{s,n}' - \tilde S_0(\eta;\Delta \eta)} {\bar \rho_{h,n} / \bar \rho_{s,n}'},
 \eea
 which substantially reduces statistical noise in  the differences. The dashed curve is the hard-component model $\tilde H_0(\eta;\Delta \eta)$ defined by Eq.~(\ref{h00}) demonstrating the overall self-consistency of the TCM. 
 The hard-component density in the right panel suggests that hadron fragments from MB dijets are strongly peaked near $\eta = 0$, localized mainly within $\Delta \eta = 2$ and consistent with the dominant dijet source being low-$x$ gluons corresponding to small $y_z$ or $\eta$. The functional form on $\eta$ is also consistent with a peaked  ``gluon-gluon source'' component predicted by Ref.~\cite{georg}. Its  $N_{\rm ch}^{gg}\propto \ln^3(s_{NN}/s_0)$ collision-energy dependence can be compared with Eq.~(\ref{jeteta}) (first line) below. 
 
 In summary, $\bar \rho_0(\eta)$ density distributions present another case where TCM soft and hard components are accurately separable, are quite different (for understandable reasons) and do not share a common production mechanism (i.e.\ MPIs) as assumed for the PMC.

\end{appendix}


\end{document}